%% file: v2-02-20-2026.tex
\documentclass[11pt]{article}
\RequirePackage{amsthm,amsmath}

\input{myoperator.tex}

\usepackage{amsgen,amsmath,amstext,amsbsy,amsopn,amssymb,latexsym,float}
\usepackage{array}

\usepackage{multirow}
\usepackage{graphicx}
\usepackage{amssymb}
\usepackage{comment}
\usepackage{hyperref,xr}
\usepackage{mathtools}
\usepackage[round]{natbib}
\usepackage{algpseudocode} 
\usepackage{algorithm} 
\usepackage{makecell}
\usepackage{mathbbol}
\usepackage{subcaption}
\usepackage{enumerate}
\usepackage{soul}
\usepackage{MnSymbol} 
\usepackage{tikz}       
\usetikzlibrary{arrows.meta,calc,patterns}
\usepackage{pgfplots}   
\usepackage{authblk}
\usepackage{multibib}
\usepackage[dvipsnames,table,dvipsnames*, svgnames*, hyperref]{xcolor}

\newcommand{\RR}{\mathbb{R}}
\newcommand{\Prob}{\mathbb{P}}

\newcommand{\Cov}{{\rm Cov}}
\newcommand{\Var}{{\rm Var}}
\newcommand{\Ical}{\mathcal{I}}

\newcommand{\m}{{[m]}}

\newcommand{\op}{{\rm op}}
\newcommand{\mstar}{{[m^*]}}
\newcommand{\err}{{\rm err}_{n,p}(M;\alpha_0)}
\newcommand{\errml}{{\rm err}_{n,p}(M;{\alpha}_{T_0})}
\newcommand{\len}{{\rm\bf Length}}
\renewcommand{\T}{\intercal}

\renewcommand{\hat}{\widehat}
\renewcommand{\tilde}{\widetilde}
\renewcommand{\l}{\left}
\renewcommand{\r}{\right}
\renewcommand{\P}{\mathbb{P}}

\algrenewcommand{\algorithmiccomment}[1]{\hfill\(\triangleright\) \textit{#1}}

\theoremstyle{plain}
\newtheorem{theorem}{Theorem}
 
\newtheorem{lemma}[theorem]{Lemma} 
 
\theoremstyle{definition}
 
\newtheorem{assumption}{Assumption} 
\theoremstyle{remark}

\DeclareMathOperator*{\argmin}{arg\,min}

\newcites{app}{Appendix References}

\title{Perturbed Double Machine Learning: \\ Nonstandard Inference Beyond the Parametric Length}

\author[1]{Mengchu Zheng}
\author[1,*]{Matteo Bonvini}
\author[2,*]{Zijian Guo}

\affil[1]{Department of Statistics, Rutgers, The State University of New Jersey, USA}
\affil[2]{Center for Data Science, Zhejiang University, China}

\date{\today}

\begin{document}

\maketitle

\def\thefootnote{*}\footnotetext{Correspondence to Matteo Bonvini (\nolinkurl{mb1662@stat.rutgers.edu}) and Zijian Guo (\nolinkurl{zijguo@zju.edu.cn}).} 

\setcounter{footnote}{0}
\renewcommand{\thefootnote}{\arabic{footnote}}
\interfootnotelinepenalty=10000  
\setlength{\skip\footins}{1em}   

\maketitle

\begin{abstract}
We study inference on a low-dimensional functional $\beta$ in the presence of infinite-dimensional nuisance parameters. Classical inferential methods are typically based on Wald intervals, whose large-sample validity rests on asymptotic negligibility of nuisance error; for example, influence-curve based estimators (Double/Debiased Machine Learning, DML) are asymptotically Gaussian when nuisance estimators converge faster than $n^{-1/4}$. Although such negligibility can hold even in nonparametric classes, it can be restrictive. To relax this requirement, we propose Perturbed Double Machine Learning, which ensures valid inference even when nuisance estimators converge slower than $n^{-1/4}$. Our proposal is to (i) inject randomness into the nuisance estimation step to generate perturbed nuisance models, each yielding an estimate of $\beta$ and a Wald interval, and (ii) filter out perturbations whose deviations from the original DML estimate exceed a threshold. For Lasso nuisance learners, we show that, with high probability, at least one perturbation yields nuisance estimates sufficiently close to the truth, so the associated estimator of $\beta$ is close to an oracle with known nuisances. The union of retained intervals delivers valid coverage even when the DML estimator converges slower than $n^{-1/2}$. The framework extends to general machine-learning nuisance learners, and simulations show coverage when state-of-the-art methods fail.
\end{abstract}


\section{Introduction}

In many domains, e.g., causal inference \citep{kennedy2024semiparametric} and machine learning \citep{kandasamy2014influence}, the relevant inferential targets can be expressed as summaries (functionals) of the data generating distribution. 
Research on functional estimation dates back decades; see, e.g., \cite{bickel1988estimating, birge1995estimation, laurent1996efficient, Laurent:1997:EIF, Vaart1998, robins2008higher, robins2009quadratic, robins2017minimax, bickel1993efficient, newey1990semiparametric, tsiatis2006semiparametric, hines2022demystifying, kennedy2024semiparametric}, among others. Particularly, semiparametric efficiency theory offers principled guidelines on how to infer parameters that depend on unknown quantities, the so-called nuisances components, that need to be estimated despite not being of immediate interest. 

A key feature of functional estimation is that, under certain conditions, and when the parameter of interest is “sufficiently smooth” in the data generating distribution, one can construct influence-function-based estimators that converge to the truth faster than the rate at which the nuisance estimators converge to their corresponding targets.
Throughout, we refer to this general estimation strategy as Double/Debiased Machine Learning (DML), borrowing the terminology from the influential work by \cite{Chernozhukov2018}, which recently popularized these methods.

A straightforward approach to inference in this context is based on the assumption that the nuisances are estimated accurately enough so that the estimator is asymptotically linear and the Wald interval is valid in large samples. As DML estimators exhibit second-order dependence--- involving products or squares--- on the nuisance errors, this condition allows for nonparametric nuisance estimation under structural conditions; e.g, one may assume nuisance convergence faster than $n^{-1/4}$-rates. However, when the nuisances are of high complexity, asymptotic linearity can be an heroic assumption. {For example, if the nuisances are not sufficiently smooth or sparse,} the functional minimax rate of convergence {is} slower than $n^{-1/2}$ \citep{robins2009semiparametric, bradic2019minimax, balakrishnan2023fundamental, jin2024structure}.

Several authors have studied inference in settings where standard conditions for asymptotic linearity fail. A broad strategy is to construct estimators whose leading bias depends more weakly on nuisance estimation error than does that of the DML estimator; higher-order influence function theory is a prominent example \citep{robins2017minimax}. Yet, \cite{balakrishnan2023fundamental} and \cite{jin2024structure} have shown that, in a strong sense, improvements over DML can only be attained under conditions on the data distribution that are not fully exploited by the DML estimator. We provide a detailed review of existing results on inference beyond standard Wald intervals in Section \ref{section:lit_review} of the Supplement. 

To the best of our knowledge, the construction of confidence intervals that are agnostic to nuisance-model specification remains largely unexplored when the estimator converges at a rate slower than $n^{-1/2}$ or has a non-Gaussian limiting distribution. In this work, we make progress toward this goal by adding a perturbation-and-filtering step to DML; we refer to this approach as 	\textit{Perturbed DML}. The perturbation step repeatedly injects noise into the fitting process of the nuisance models, yielding a collection of perturbed DML estimators and corresponding Wald intervals. A properly filtered union is then taken as the final confidence set. Under suitable conditions, this set retains coverage without being overly conservative, even when the parameter of interest is not estimated at the $n^{-1/2}$  rate.

\subsection{Results and Contributions}

We introduce a novel inferential approach when the nuisance estimation error might not be negligible and the parameter of interest not estimable at the parametric rate $n^{-1/2}$. To highlight our methodology, we focus on a projection parameter that reduces to the linear coefficient in a partially linear model when partial linearity holds \citep{vansteelandt2022assumption}. Our procedure builds upon the DML framework by augmenting it with a perturbation-and-filtering layer.

\begin{figure}[ht]
    \centering
    \includegraphics[width=\linewidth]{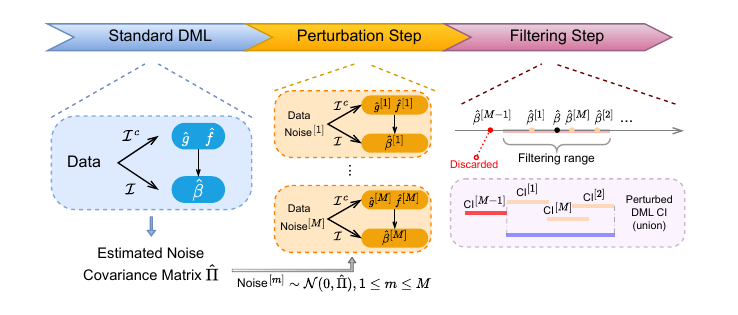}
    \caption{\small Workflow of the \textit{Perturbed DML} procedure.}
    \label{fig:workflow}
\end{figure}

In the perturbation step, we inject simulated noise into nuisance fitting by subtracting the simulated noises from the response variables. Repeating this process yields a collection of perturbed nuisance estimators. Intuitively, among many perturbations, at least one noise draw nearly cancels the true noise, leading to nuisance estimates close to the truth. Under the condition that the perturbation step has produced at least one valid, yet unidentifiable Wald interval, it is natural to consider the union of all intervals indexed by the perturbed nuisance models as the final confidence set. However, this union interval can be potentially conservative in practice. To address this issue, we propose a filtering step that discards those intervals whose corresponding estimates deviate excessively from the unperturbed  DML estimate. 
The filtering step ensures that the union does not yield an overly conservative confidence set while retaining the valid interval with high probability. Figure \ref{fig:workflow} presents a workflow of the proposed Perturbed DML procedure.

\vspace{-4mm}
Theorem \ref{thm: mstar lasso} formalizes this intuition rigorously in settings where the nuisance functions are high-dimensional linear models. 
When these models are consistently estimated by Lasso, we show that the proposed interval attains nominal coverage even when the DML estimator converges slower than $1/\sqrt{n}$ and might not be asymptotically Gaussian. With an appropriate filtering threshold, the interval also achieves the minimax expected length established in Theorem~1 of \cite{cai2017confidence}. Theorem~\ref{thm: coverage lasso} summarizes our coverage and length guarantees in the high-dimensional linear setting.

We extend our perturbation-and-filtering approach to settings where the nuisances are estimated by general machine learning methods. The perturbation and filtering steps remain as above, with the Lasso replaced by a different method. However, deriving theoretical guarantees in this general setting is more challenging. In Supplementary Section \ref{sec:general theory},  Theorem \ref{thm:mstar ml beta} provides an informal justification for employing our approach in the more general case based on an isoperimetric inequality on the Gaussian density \citep{bobkov1997isoperimetric,cousins2018gaussian}. Furthermore, our simulations show that the Perturbed DML approach applied to settings where the nuisances are estimated by generalized {additive models and} XGBoost \citep{chen2016xgboost} yields valid confidence sets in settings where the standard DML method fails to achieve nominal coverage. Moreover, in both simulations and real data analysis, compared to an oracle benchmark that has access to the empirical bias and standard error, Perturbed DML delivers confidence sets that are not excessively wide.

\section{Semiparametric Estimators and Inference Challenges}
\label{sec: model}

To better illustrate the method, we focus on the following estimand, though our procedures can be applied to a variety of functionals:
\begin{align}
\beta = \frac{\E\{\Cov(Y_i, D_i \mid X_i)\}}{\E\{\Var(D_i \mid X_i)\}} = \frac{\E(Y_iD_i) - \E\{f(X_i) g(X_i)\}}{\E[\{D_i - f(X_i)\}^2]},
\label{eq: causal estimand}
\end{align}
where $Y_i \in \RR$ is an outcome, $D_i \in \R$ is a treatment of interest, and $X_i \in \RR^p$ denotes the baseline covariates. We let $g(X_i) = \E(Y_i \mid X_i)$ and $f(X_i) = \E(D_i \mid X_i)$, which are the two key nuisance functions entering the definition of $\beta$. Our goal is to construct a confidence interval for $\beta$ having access to $n$ independent and identically distributed (i.i.d.) copies of $\mathcal{O}_i = \{Y_i, D_i, X_i\} \sim P$. Inference for $\beta$ is a well-studied problem as it arises naturally when considering the partially linear model,  
\begin{equation}\label{eq:PLR}
    \E(Y_i \mid D_i, X_i) = \psi D_i + h(X_i),
\end{equation} 
for some function $h$ only depending on $X_i$. If partial linearity holds, then $\beta$ defined in \eqref{eq: causal estimand} equals the homogeneous treatment effect $\psi$ of $D$ on $Y$ (identified under no-unmeasured-confounding). However, $\beta$ in \eqref{eq: causal estimand} remains well-defined even under model misspecification. For example, when $D$ is a binary, $\beta$ can be interpreted as a variance-weighted average treatment effect without reference to \eqref{eq:PLR}.  See \cite{vansteelandt2022assumption} for an in-depth discussion on this parameter and related estimands. 

\begin{figure}[ht]
    \centering
    \includegraphics[width=\linewidth]{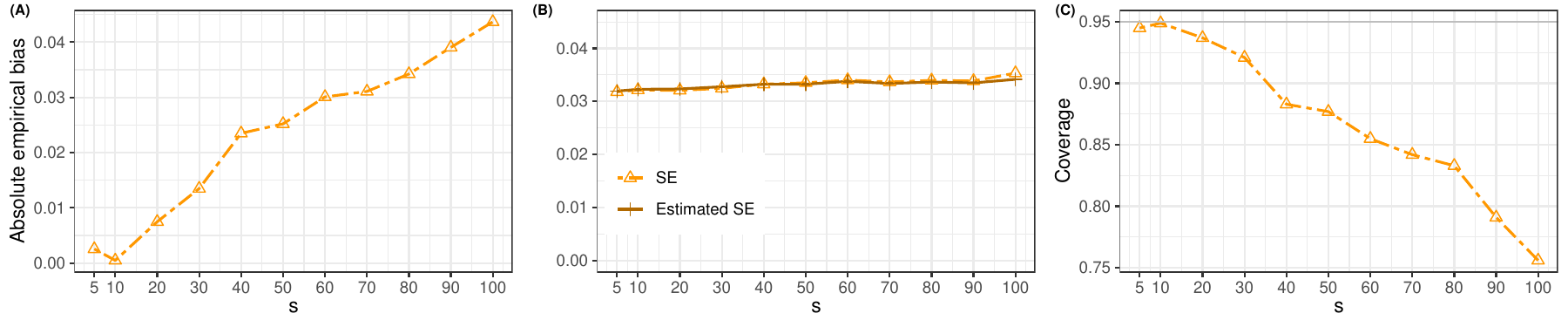}
    \caption{\small DML with high-dimensional sparse linear models where the sparsity level $s$ ranges from 5 to 100. (A): Absolute empirical bias of the DML estimator. (B): Estimated and empirical standard errors. (C): Empirical coverage of Wald CIs. Results are calculated based on 1000 simulations.}
    \label{fig: dml problem sparse}
\end{figure}

We start by reviewing the semiparametric efficient DML estimator for estimating $\beta$ defined in \eqref{eq: causal estimand} by modeling the data distribution in a nonparametric way, using the well-known estimator based on the (unique) influence function of $\beta$, and then highlight the associated inference challenges; see Figure \ref{fig: dml problem sparse}. These challenges arise when the nuisance models $f(\cdot)$ and $g(\cdot)$ are estimated at slow convergence rates, in which case the estimator fails to attain the usual $1/\sqrt{n}$ convergence rate. 

In the following discussion, and in the rest of the paper (unless specified otherwise), we assume that the nuisance functions $f(\cdot)$ and $g(\cdot)$ are estimated on an auxiliary training sample  $\mathcal{I}^c$. We further leverage these nuisance estimators together with the main sample $\mathcal{I}$ to estimate $\beta$.  For simplicity, we assume that both samples $\mathcal{I}$ and $\mathcal{I}^c$ are of size $n$. We expect all the arguments made in this paper to apply when cross-fitting is performed. 

Let $\widehat{g}$ and $\widehat{f}$ denote estimators of $g$ and $f$ constructed using observations from $\Ical^c$. Define the estimator
\begin{align}
\widehat\beta = \frac{\sum_{i \in \mathcal{I}} \{Y_i - \widehat{g}(X_i)\}\{D_i - \widehat{f}(X_i)\}}{\sum_{i \in \mathcal{I}}\{D_i -\widehat{f}(X_i)\}^2}. \label{eq:dml estimator cross-fitted general}
\end{align}
This estimator has been analyzed by various authors; see, e.g., \cite{vansteelandt2022assumption, balakrishnan2023fundamental, kennedy2024semiparametric}. We now outline a common approach to analyze its properties. Define
\begin{align*}
    \varphi(O_i; \beta) = \frac{\{Y_i - g(X_i)\}\{D_i - f(X_i)\} - \{D_i - f(X_i)\}^2 \beta}{\E\{\Var(D_i \mid X_i)\}},
\end{align*}
and $\widehat\varphi(O_i; \beta)$ to be equal to $\varphi(O_i; \beta)$ except that, in the numerator, $f$ and $g$ are replaced by $\widehat{f}$ and $\widehat{g}$, respectively. The quantity $\varphi(O_i; \beta)$ is the influence function of $\beta$.

By a direct calculation, we have that
\begin{align}\label{eq:beta_decomposition}
    \widehat\beta - \beta = Z_n + T_n  + S_n, \textrm{ with } Z_n  = \frac{1}{n}\sum_{i \in \Ical} \varphi(O_i; \beta), \ 
    T_n = \E\{\widehat\varphi(O_i; \beta) - \varphi(O_i; \beta) \mid \mathcal{I}^c\}.
\end{align}
The first term $Z_n$ is a central limit term that converges to $N\left(0, \Var\{\varphi(O_i; \beta)\}\right)$ when scaled by $\sqrt{n}$. The last term $S_n$ is a collection of empirical process and higher-order terms, which are asymptotically negligible as long as $\widehat{f}$ and $\widehat{g}$ are consistent in $L_2$; see, e.g., Lemma 2 in \cite{kennedy2018sharp}. 
Importantly, we refer to the second term $T_n$ as ``the nuisance bias term" throughout this paper, which evaluates to
\begin{equation}\label{eq:Tn}
\begin{aligned} 
    |T_{n}| &= C\cdot  \left| \E\left[[\widehat{f}(X_i) - f(X_i)][\widehat{g}(X_i) - g(X_i)] \mid \mathcal{I}^c\right] - \beta \E\left[[\widehat{f}(X_i) - f(X_i)]^2 \mid \mathcal{I}^c\right]\right| \\
     &\leq C(r_f r_g + |\beta|r_f^2)
\end{aligned}
\end{equation}
where $C = 1 / \E\{\Var(D_i \mid X_i)\}$, and $r_f=\left(\E\left[\{\widehat{f}(X_i) - f(X_i)\}^2 \mid \mathcal{I}^c\right]\right)^{\frac{1}{2}}$ and $r_g=\left(\E\left[\{\widehat{g}(X_i) - g(X_i)\}^2 \mid \mathcal{I}^c\right]\right)^{\frac{1}{2}}$ are the root-mean-square errors for estimating $f$ and $g$, respectively. 
The above upper bound in \eqref{eq:Tn} implies that a sufficient condition for the negligibility of $T_n$ is that $r_f$ and $r_g$ converge to zero faster than $n^{-1/4}$. 

When the dominant term is $Z_n$ and the nuisance bias component $T_n$ is negligible, a Wald-type confidence interval can readily be computed as 
\begin{equation}
\left[\widehat{\beta}-z_{\alpha/2}\widehat{\textrm{SE}}(\widehat{\beta}), \ \widehat{\beta}+z_{\alpha/2}\widehat{\textrm{SE}}(\widehat{\beta})\right],
\label{eq: wald}
\end{equation}
where $z_{\alpha/2}$ is the $\alpha/2$ upper quantile of the standard normal and 
\begin{equation}\label{eq:se for beta hat}
    \widehat{\textrm{SE}}(\widehat{\beta}) = \sqrt{\frac{n^{-1}\sum_{i \in \mathcal{I}} (\widehat\epsilon_{i} - \widehat\beta \widehat\delta_{i})^2 \cdot \widehat{\delta}_{i}^2}{n\l(n^{-1}\sum_{i \in \mathcal{I}} \widehat{\delta}_{i}^2\r)^2}},\quad\text{with}\quad \widehat\delta_i = D_i - \widehat{f}(X_i),\;\; \widehat\epsilon_{i} = Y_i - \widehat{g}(X_i). 
\end{equation}
The construction of $\widehat\beta$ is agnostic with respect to $\widehat{f}$ and $\widehat{g}$, and thus it is amenable to the use of black-box machine learning algorithms (cf. \cite{balakrishnan2023fundamental}). 

To illustrate the main idea, we consider high-dimensional nuisance models $g(x)=x^\top\eta$ and $f(x)=x^\top\gamma$, with the sparse vectors $\eta$ and $\gamma$ estimated by the following Lasso estimators, yielding $\widehat g(x)=x^\top\widehat\eta$ and $\widehat f(x)=x^\top\widehat\gamma$,

\begin{equation}\label{eq: lasso optimization problems}
\begin{aligned}
& \widehat\eta = \argmin_{u \in \R^p} \frac{1}{2 n} \sum_{i\in \Ical^c} u^\intercal X_iX_i^\intercal u - \frac{1}{n} \sum_{i\in \Ical^c}u^\intercal X_iY_i + \lambda_{\eta} \|u\|_1, \\
& \widehat\gamma = \argmin_{u \in \R^p} \frac{1}{2n} \sum_{i\in \Ical^c}u^\intercal  X_iX_i^\intercal u - \frac{1}{n} \sum_{i\in \Ical^c}u^\intercal X_iD_i + \lambda_{\gamma} \|u\|_1, 
\end{aligned}
\end{equation}
where $\lambda_\eta>0$ and $\lambda_\gamma>0$ are penalty parameters selected via cross-validation. Note that the estimators in \eqref{eq: lasso optimization problems} are equivalent to those obtained from the regularized least-squares loss; removing the constant terms $\frac{1}{2n}\sum Y_i^2$ and $\frac{1}{2n}\sum D_i^2$ from the regularized least-squares losses yields the objective functions in \eqref{eq: lasso optimization problems}.


\noindent{\bf Inference Challenges.} The asymptotic validity of the Wald interval in \eqref{eq: wald} relies on $T_n$ in \eqref{eq:Tn} being $o_P(n^{-1/2})$. When the nuisance models $f$ and $g$ are high-dimensional  sparse linear models, and in virtue of sample splitting, this translates to 
\begin{align}\label{eq:bound_Tn_linear}
T_n \propto (\widehat\gamma - \gamma)^\intercal \E(X_iX_i^\intercal) (\widehat\eta - \eta) - \beta (\widehat\gamma - \gamma)^\intercal \E(X_iX_i^\intercal) (\widehat\gamma - \gamma) = o_P(n^{-1/2}).
\end{align} 
Suppose that $\eta$ and $\gamma$ are $s_\eta$- and $s_\gamma$-sparse vectors in $\R^p$, then it is well-known that $\|\widehat\gamma - \gamma\|^2 = O_P(s_\gamma \log p / n)$ and $\|\widehat\eta - \eta\|^2 = O_P(s_\eta \log p / n)${; see Theorem 7.2 in \cite{bickel2009simultaneous}.} This further leads to a high probability upper bound:
\begin{equation}
|T_n|\leq \rho_n = c^*\l(s_\gamma + \sqrt{s_\eta s_\gamma}\r) \frac{\log p}{n}
\label{eq: Tn bound}
\end{equation}
where $c^*$ is a constant independent of $n$ and $p$. Together with the requirement $T_n=o_P(n^{-1/2})$, the upper bound \eqref{eq: Tn bound} leads to $(s_\gamma + \sqrt{s_\eta s_\gamma}) \log p \ll \sqrt{n}$ as a sufficient condition for the validity of the Wald interval in this setting (assuming the operator norm of $\E(X_iX_i^\intercal)$ is bounded). This condition highlights that, if $s_\eta$ and $s_\gamma$ are sufficiently large relative to $n$, the coverage of the Wald interval is expected to degrade. Using simulated data, the following example demonstrates how the Wald confidence interval fails to achieve the desired coverage for a relatively dense model. 

\noindent {\bf Example 1.}
We evaluate the finite-sample performance of $\widehat\beta$ defined in \eqref{eq:dml estimator cross-fitted general} (with $K=2$ cross-fitting and five splits) under a sparse linear nuisance specification. We generate data such that $\E(Y_i \mid X_i, D_i) = \beta D_i + h(X_i)$ where $h(X_i)=X_i^\T \mu$ and $\E(D_i \mid X_i)=X_i^\T\gamma$. The nuisance coefficients $\mu$ and $\gamma$ are $s$-sparse with partially overlapping supports and the nonzero entries alternate in sign. The generation of $X_i$ and noises are detailed in Section \ref{sec:simulation}. We fix $n=1000$ and $p=500$, while we vary the sparsity level $s$ from 5 to 100.  
{As shown in Figure \ref{fig: dml problem sparse}, the absolute empirical bias of $\widehat\beta$ grows rapidly with $s$ due to the growing magnitude of the nuisance bias $T_n$ while the standard error is accurately estimated across various $s$. When the true nuisance models become too dense, the nuisance bias $T_n$ is larger than the order of $n^{-1/2}$, resulting in the undercoverage of the Wald CI.}

\section{Perturbed DML: High-Dimensional Linear Models}
\label{sec: method section}

In this section, we describe the two key steps of the proposed approach: perturbation and filtering. 
We illustrate the main idea by focusing on the high-dimensional sparse linear models $g(x) = x^\intercal \eta$ and $f(x) = x^\intercal \gamma$, where $\eta$ is $s_\eta$-sparse and $\gamma$ is $s_\gamma$-sparse. We will discuss the general scenario with the use of machine learning models in Section \ref{sec: method general setting}. 

\noindent{\bf Step 1: Perturbation.} We outline our procedure and rationale for injecting noise into the nuisance Lasso optimizations using the data $\Ical^{c}$, which are further used to construct a perturbed DML estimator using the data $\Ical$. We adopt the setup from Section \ref{sec: model} and write $\epsilon_i = Y_i - g(X_i)$ and $\delta_i = D_i - f(X_i)$, and define the $p$-dimensional vectors $\xi = n^{-1/2}\sum_{i \in \Ical^c} X_i \epsilon_i$ and $\kappa = n^{-1/2} \sum_{i \in \Ical^c} X_i \delta_i$. 
With the decomposition $X_i Y_i = X_iX_i^\intercal \eta + X_i \epsilon_i$ and $X_i D_i = X_i X_i^\intercal \gamma + X_i \delta_i$, we write the nuisance Lasso optimization in \eqref{eq: lasso optimization problems} as
\begin{equation}\label{eq: lasso optimization problem unperturbed}
\begin{aligned}
& \widehat\eta = \argmin_{u \in \R^p} \frac{1}{2 n} \sum_{i\in \Ical^c} u^\intercal X_iX_i^\intercal u - u^\intercal \left(\frac{1}{n} \sum_{i\in \Ical^c}X_i X_i^\intercal \eta + n^{-1/2} \xi\right) + \lambda_{\eta} \|u\|_1,  \\
& \widehat\gamma = \argmin_{u \in \R^p} \frac{1}{2 n} \sum_{i\in \Ical^c} u^\intercal X_iX_i^\intercal u - u^\intercal \left(\frac{1}{n} \sum_{i\in \Ical^c}X_i X_i^\intercal \gamma + n^{-1/2} \kappa\right) + \lambda_{\gamma} \|u\|_1.
\end{aligned}
\end{equation}
As the main randomness in the above optimization arises from $\xi$ and $\kappa$, an oracle with access to them could remove them from the objective functions so that their minimizers would recover $\eta$ and $\gamma$ given suitably chosen $\lambda_\eta$ and $\lambda_\gamma$. Building on this observation, we propose perturbing \eqref{eq: lasso optimization problem unperturbed} by subtracting off artificial noise sampled from distributions mimicking those of $\xi$ and $\kappa$, respectively. Specifically, we generate $M$ independent copies of $\xi$ and $\kappa$ as, for $1\leq m\leq M$,
\begin{equation}\label{eq: resample xi kappa}
\xi^\m \sim \mathcal{N}( \mathbf{0}, \hat\Sigma+\nu I ) \quad \text{and}\quad \kappa^\m \sim \mathcal{N}(\mathbf{0}, \hat{\Lambda}+\nu' I),
\end{equation}
with $\hat\Sigma := \frac{1}{n}\sum_{i\in\Ical^c} (Y_i - X_i^\T\hat{\eta})^2 X_i X_i^{\intercal}$ and $\hat{\Lambda} := \frac{1}{n}\sum_{i\in\Ical^c} (D_i - X_i^\T\hat{\gamma})^2 X_iX_i^{\intercal}$ where $\widehat\eta$ and $\widehat\gamma$ are the Lasso estimates from \eqref{eq: lasso optimization problems} based on the data from $\Ical^c$. We choose $\nu=\min_{1\leq j\leq p} \hat{\Sigma}_{j,j}>0$ and $\nu'=\min_{1\leq j\leq p} \hat{\Lambda}_{j,j}>0$ in \eqref{eq: resample xi kappa} to make sure that the covariance $\widehat{\Sigma}+\nu I$ and $\widehat\Lambda+\nu' I$ are positive definite, even in the high-dimensional regime with $p>n$.

The motivation for the artificial noise generating distribution in \eqref{eq: resample xi kappa} is that, for a fixed covariates' dimension $p$ and by the central limit theorem, $\xi \indist \mathcal{N}_p(\mathbf{0}, \E(\epsilon_i^2 X_i X_i^\intercal))$ and $\kappa \indist \mathcal{N}_p(\mathbf{0}, \E(\delta_i^2 X_i X_i^\intercal))$, as $n\to\infty$. However, it is actually non-essential that the injected noise is Gaussian nor that it has a distribution close to that of the true noise. Our proposal's validity rests on the assumption that its distribution is sufficiently diffuse so that the true noise lies within its support with non-negligible probability.

Next, given $\xi^\m$ and $\kappa^\m$, we solve the perturbed Lasso optimization problems 
\begin{equation}    
\begin{aligned}\label{eq: lasso optimization problem eta m}
    & \widehat\eta^\m = \argmin_{u \in\RR^p} \frac{1}{2n} \sum_{i \in \Ical^c} u^{\intercal} X_i X_i^\intercal u -  u^{\intercal}\left\{\frac{1}{n}\sum_{i \in \Ical^c} X_iY_i - n^{-1/2}\xi^\m\right\}  + \lambda^\m_\eta \|u\|_1, \\
    & \widehat\gamma^\m = \argmin_{u\in\RR^p} \frac{1}{2n} \sum_{i \in \Ical^c} u^{\intercal} X_i X_i^\intercal u -  u^{\intercal} \left\{\frac{1}{n}\sum_{i \in \Ical^c}X_iD_i - n^{-1/2} \kappa^\m\right\} +\lambda^\m_\gamma \|u\|_1,
\end{aligned}
\end{equation}
where $\lambda_\eta^\m>0$ and $\lambda_\gamma^\m>0$ are positive tuning parameters whose selection is discussed at the end of this section. 
Notice that the expressions $n^{-1}\sum_{i \in \Ical^c}X_iY_i - n^{-1/2} \xi^\m$ and $n^{-1}\sum_{i \in \Ical^c}X_iD_i - n^{-1/2} \kappa^\m$ in \eqref{eq: lasso optimization problem eta m} are equal to 
$$n^{-1}\sum_{i \in \Ical^c}X_iX_i^\T\eta + n^{-1/2}(\xi - \xi^\m)\quad \text{and} \quad n^{-1}\sum_{i \in \Ical^c}X_iX_i^\T\gamma + n^{-1/2}(\kappa - \kappa^\m).$$ 

After solving the perturbed Lasso problems in \eqref{eq: lasso optimization problem eta m} $M$ times, we obtain a collection of estimates of $\eta$ and $\gamma$, which we use to construct estimates of $\beta$ on sample $\Ical$:
\begin{equation}\label{eq: dml estimator m}
    \hat{\beta}^\m =\frac{\sum_{i\in\Ical} (D_i-X_i^{\intercal}\hat{\gamma}^\m)(Y_i - X_i^{\intercal}\hat{\eta}^\m)}{\sum_{i\in\Ical}(D_i-X_i^{\intercal}\hat{\gamma}^\m)^2}.
\end{equation}
Compared to $\widehat{\beta}$ defined in \eqref{eq:dml estimator cross-fitted general}, each $\widehat\beta^\m$ simply replaces the Lasso nuisance estimators $\widehat\eta$ and $\widehat\gamma$ with the perturbed Lasso estimators $\widehat\eta^\m$ and $\widehat\gamma^\m$, respectively. The key to our analysis is the observation that, for $M$ being sufficiently large, there should be an index $m^*$ such that $\xi^\mstar$ and $\kappa^\mstar$ are close to $\xi$ and $\kappa$, respectively. In turn, this means that $\widehat\eta^\mstar$ and $\widehat\gamma^\mstar$ would be respectively close to $\eta$ and $\gamma$ and the perturbed estimator $\hat{\beta}^\mstar$ would nearly recover the following oracle estimator computed using the true nuisance functions, 
\begin{align}
    \hat{\beta}^{\rm ora} = \frac{\sum_{i\in\Ical} (D_i-X_i^{\intercal}{\gamma})(Y_i - X_i^{\intercal}{\eta})}{\sum_{i\in\Ical}(D_i-X_i^{\intercal}{\gamma})^2}. \label{eq: betaHat ora}
\end{align}
Under (essentially) no assumptions, the oracle estimator $\widehat\beta^{\rm ora}$ is $\sqrt{n}$-CAN with influence function equal to $\varphi(O; \beta)$.
In this light, a perturbed estimator $\hat{\beta}^\mstar$ sufficiently close to $\hat{\beta}^{\rm ora}$ can be used to center a Wald interval ${\rm CI}^\mstar$ with asymptotic nominal coverage.

\begin{figure}[ht]
    \centering
    \includegraphics[width=0.5\linewidth]{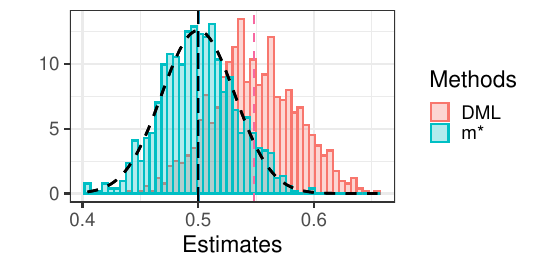}
    \caption{Empirical distributions of $\hat{\beta}^\mstar$ and $\hat{\beta}$ in Example 1 with $s=150$ and $M=500$, where the dashed curve represents the distribution $\widehat\beta^{\rm ora}$, namely $N(\beta,n^{-1}\Var\{\varphi(O_i;\beta)\}$. }
    \label{fig:mstar_lasso}
\end{figure}

In the following Theorem \ref{thm: mstar lasso}, we provide a rigorous justification of the above discussion. In particular, we establish that, when $M$ is sufficiently large, such a $\hat{\beta}^\mstar$ exists with high probability and its uncertainty is governed by the usual influence-function term $Z_n$ defined in \eqref{eq:beta_decomposition}. The corresponding Wald interval centered at $\hat{\beta}^\mstar$ thus achieves the asymptotic nominal coverage. 
Figure \ref{fig:mstar_lasso} empirically illustrates the existence of such a $\hat{\beta}^\mstar,$ where we set $M=500$ and $m^* = \argmin_{1\leq m\leq M}|\hat{\beta}^\m - \hat{\beta}^{\rm ora}|$. {The Figure is based on data generated as in Example 1 with $s = 150$ across $1000$ simulations. It can be seen that the distribution of $\hat{\beta}^\mstar$ nearly matches the asymptotic distribution of $\hat{\beta}^{\rm ora}$.} 

In practice, however, it is impossible to identify such $\hat{\beta}^\mstar$ {since $\hat{\beta}^{\rm ora}$ is unknown}. We therefore construct the Wald interval for each perturbation. 
For a given significance level $\alpha\in(0,1/2)$, we budget the significance level $\alpha_0$ (e.g. $\alpha_0=\alpha/10$) to account for not being able to recover $\xi$ and $\kappa$ and use the remaining significance $\alpha'=\alpha-\alpha_0$ to build the Wald type confidence interval centered at $\hat{\beta}^\m$:
\begin{equation}\label{eq: dml ci m}    
\textrm{CI}^\m = \left[\hat{\beta}^\m-z_{\alpha'/2}\hat{\textrm{SE}}(\hat{\beta}), \ \hat{\beta}^\m + z_{\alpha'/2}\hat{\textrm{SE}}(\hat{\beta})\right].
\end{equation}
Since the DML's estimated standard error aims to quantify the variability of $Z_n$, we simply take $\hat{\rm SE}(\hat{\beta})$ defined in \eqref{eq:se for beta hat} to construct each Wald interval. In constructing $\hat{\textrm{SE}}(\hat{\beta})$ in \eqref{eq:se for beta hat}, we use the estimated residuals $\hat{\epsilon}_i = Y_i-X_i^\T\hat{\eta}$ and $\hat{\delta}_i = D_i - X_i^\T\hat{\gamma}$, where $\hat{\eta}$ and $\hat{\gamma}$ are unperturbed nuisance estimators defined in \eqref{eq: lasso optimization problem unperturbed}.

Finally, we notice that there could be different, equally valid strategies for injecting the noise in the nuisance fitting step, as done, for instance in Section \ref{sec: method general setting}; depending on the fitting procedure employed, some may be more natural than others.

\noindent{\bf Step 2: Filtering Perturbed DML Estimators.} 
By injecting noise into the nuisance estimation procedure, we obtain a collection of estimates of $\beta$ denoted by $\hat{\beta}^\m$, for $1\leq m\leq M$. For a large $M$, we should expect that there exists at least one $m^*$ such that $\widehat\beta^\mstar$ is close to $\widehat\beta^{\rm ora}$. Since it is impossible to identify which particular perturbation will achieve the goal, taking the union of all $M$ intervals as the final confidence interval guarantees asymptotic coverage. However, because $M$ should be large enough so that the probability of obtaining at least one valid interval is sufficiently large, taking an unfiltered union can be conservative and potentially wide, though it need not be overly wide in practice. We tackle this problem by proposing a filtering procedure that screens out clearly unpromising perturbations, so that the length of the resulting union is controlled while retaining at least one valid interval with high probability.

The main idea is to filter out those Wald intervals whose center $\hat{\beta}^{(m)}$ deviates substantially from the original $\hat{\beta}$. The rationale is based on the following decomposition. For the perturbation $m^{\star}$ with $\widehat{\beta}^{[m^{\star}]}$ lying closest to $\hat{\beta}^{\rm ora}$, the deviation satisfies $$ \bigl|\hat{\beta}^{[m^{\star}]} - \hat{\beta}\bigr| \leq \bigl|\hat{\beta}^{[m^{\star}]} - \hat{\beta}^{\rm ora}\bigr| + \bigl|\hat{\beta}^{\rm ora} - \hat{\beta}\bigr|. $$
Because $\hat{\beta}^{(m^{\star})}\approx \hat{\beta}^{\rm ora}$, the dominant term in the above decomposition is typically $\bigl|\hat{\beta}^{\rm ora} - \hat{\beta}\bigr|$, which is close to the nuisance-bias term $|T_n|$ and is therefore upper bounded by $\rho_n$ as in \eqref{eq: Tn bound}. Moreover, $\bigl|\hat{\beta}^{(m^{\star})} - \hat{\beta}^{\rm ora}\bigr|\to 0$ as $M$ becomes sufficiently large; this term, together with the higher-order error in approximating $\bigl|\hat{\beta}^{\rm ora} - \hat{\beta}\bigr|$ by $|T_n|$, can be bounded by $c\rho_n + \widehat{\textrm{SE}}(\hat{\beta})$ for any $c>0$ (e.g., $c=0.01$) with high probability. Hence, we obtain \begin{align} \bigl|\hat{\beta}^{(m^{\star})} - \hat{\beta}\bigr| &\leq \bigl|\hat{\beta}^{(m^{\star})} - \hat{\beta}^{\rm ora}\bigr| + \bigl|\hat{\beta}^{\rm ora} - \hat{\beta}\bigr| \leq 1.01 \cdot \rho_n + \widehat{\textrm{SE}}(\hat{\beta}). 
\label{eq: filtering reason}
\end{align}

Motivated by \eqref{eq: filtering reason} and the rationale of retaining $m^*$ in the filtering set, we propose the following perturbed DML interval
 \begin{align}
\label{eq:filtered union CI}
  \text{CI} = \cup_{m \in \mathcal{M}} \text{CI}^\m, \quad \text{with}
 \end{align}
 \begin{equation}
     \mathcal{M} = \l\{1\leq m\leq M: |\hat{\beta}^\m - \hat{\beta}| \leq 1.01 \cdot \rho_{n} + \widehat{\textrm{SE}}(\widehat{\beta}) \r\}.
\label{eq: filtering 1}
 \end{equation}
{Strictly speaking, ${\rm CI}$ may not be a continuous interval. However, since in practice it is most often so, we will refer to it as an interval.} In practice, choosing $c^*$ and the right sparsity coefficients $s_\gamma$ and $s_\eta$ in $\rho_n$ is non-trivial. Instead, we propose to filter out the intervals corresponding to the $100 \cdot \pi^* \%$ largest $|\widehat\beta^\m - \widehat\beta|$ and modify \eqref{eq: filtering 1} as follows: 
 \begin{align} 
 \label{eq: filtering 2}
 \mathcal{M} = \left\{1 \leq m \leq M: \ |\widehat\beta^\m - \widehat\beta| \leq q^*\right\},
 \end{align}
where $q^*$ is the empirical $\pi^*$-quantile of $\{|\widehat\beta^\m - \widehat\beta|\}_{1\leq m\leq M}$.
In simulations, we find that the procedure is rather insensitive to $\pi^*$ {if $\pi^*\geq 0.95$}; so we take $\pi^* = 0.95$ unless otherwise specified. 
We further compare the confidence intervals constructed with the above two filtering sets using simulated data in Section \ref{sec: compare ci with bias bound}.

\begin{figure}[htp!]
    \centering
    \includegraphics[width=\linewidth]{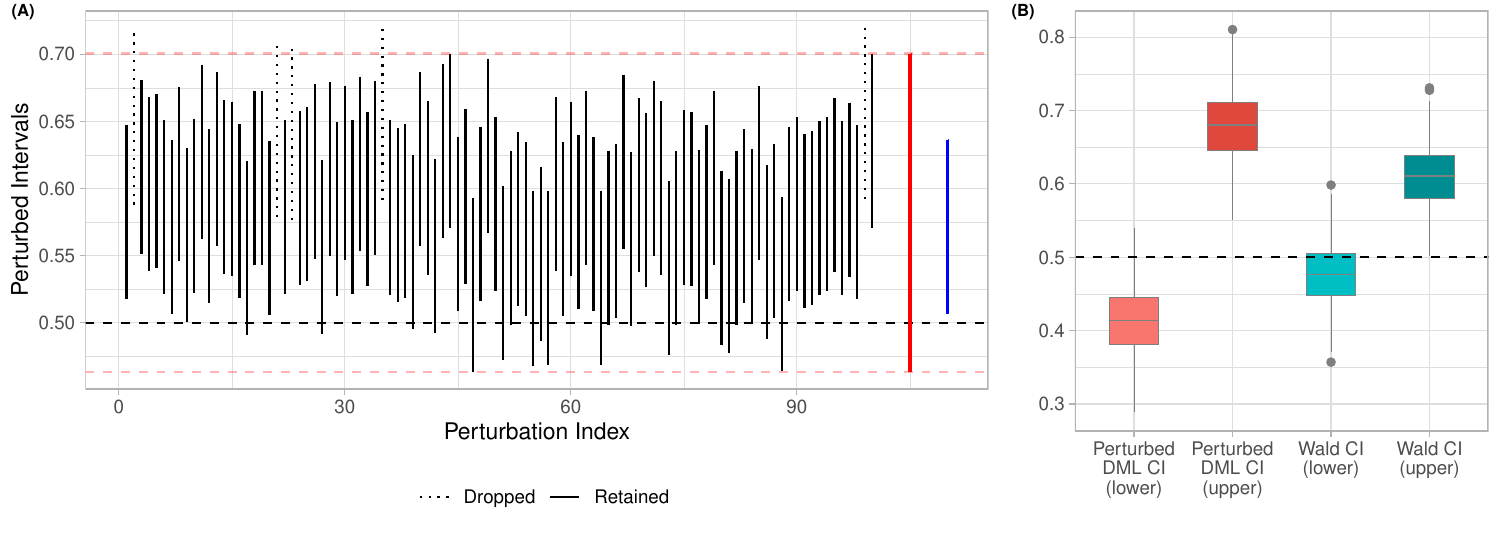}
    \caption{\small Illustration of filtering and aggregation using Example 1 with $n=1000$, $p=500$, $s=100$, $M=100$ and $\pi^*=0.95$. 
    {(A): Illustration of 100 perturbed intervals from a single simulation, where the red and blue segments are the Perturbed DML CI in \eqref{eq:filtered union CI} and the Wald CI in \eqref{eq: wald}. (B): Boxplots of lower and upper limits of the Perturbed DML CIs and Wald CIs across 500 simulations; boxes indicate the 25th to 75th percentiles. The black dashed line denotes the true parameter $\beta=0.5$.}}
    \label{fig:union ci}
\end{figure}

We illustrate the construction of our union interval in Figure \ref{fig:union ci}. We simulate data as in Example 1 with $n=1000$, $s=100$. We implement our procedure with $M=100$, which is smaller than our proposed default value $M=500$, for illustration clarity, and choose $\pi^* = 0.95$ as the filtering cutoff. 
On Panel (A) of Figure \ref{fig:union ci}, the true value of $\beta$ is given by the black dashed line ($\beta = 0.5$); the black solid lines represent all intervals $\rm CI^\m$ of which we take the union to obtain the final interval (red solid line) while the vertical black dotted lines represent the intervals that are filtered out. The original Wald interval centered at $\widehat\beta$ is given in blue and can be seen to fail to cover $\beta$ in this simulation. Panel (B) of Figure \ref{fig:union ci} displays boxplots of the lower and upper limits of the proposed and Wald CIs across 1000 simulations. The proposed CIs' limits remain concentrated and do not vary excessively compared to Wald CIs'. In particular, the upper edge of the box (75\% quantile) for DML lower limits lies above the true $\beta$, which implies that more than 25\% of Wald CIs miss $\beta$ from the below. This corresponds to the fact that the Wald CI attains only 0.714 coverage in this setting, whereas the proposed CI achieves the coverage of 0.964.

\noindent{\bf Selection of Tuning Parameters.}
Our proposal requires choosing the following set of tuning parameters: the number of perturbations $M$, the filtering proportion $\pi^*$ {used in defining the filtering set $\mathcal{M}$ in \eqref{eq: filtering 2}}, and the tuning parameters $\lambda_\eta^\m$ and $\lambda_\gamma^\m$ for each perturbed optimization. 
Through extensive simulations reported in Supplementary Section \ref{sec: sensitivity to tuning}, we found that,  as long as $M \geq 500$ and $\pi^* \geq 0.95$,  the finite-sample performance of our procedure is rather insensitive to the choice of $M$ and $\pi^*$. Based on this numerical exploration, we set $M=500$ and $\pi^*=0.95$ as default values {throughout this paper}.

A natural way to choose the tuning parameters $\lambda_\gamma^\m$ and $\lambda_\eta^\m$ is via cross-validation. However, since our procedure requires solving $M$ perturbed Lasso optimizations, cross-validation can be rather time consuming without any modifications. To address this, we restrict the candidate parameters to be of the form $r \cdot \widehat\lambda_\eta$ and $r \cdot \widehat\lambda_\gamma$, where $\widehat\lambda_\eta$ and $\widehat\lambda_\gamma$ are the tuning parameters of the original Lasso optimizations chosen by cross-validation. We then choose $r$ from a small set, e.g., $r = \{0.1, 0.2, \ldots, 1\}$, by cross-validation. The reason why we propose restricting the search of the optimal $r$ to values less than 1 is as follows. {In the Lasso theory \citep{bickel2009simultaneous}, the optimal penalty parameters are closely tied to the noise levels in the response variable: a smaller noise level requires a smaller penalty to maintain the optimal convergence rate. Since the validity of our procedure relies on the high probability event that at least one injected random term $\xi^\m$ nearly cancels the true term $\xi$, the noise level is expected to decrease. } In this sense, it is natural to expect that $\widehat\lambda_\eta^\mstar \ll \widehat\lambda_\eta$, suggesting taking $r \leq 1$. The same reasoning applies to choosing $\lambda_\gamma^\m$.

\section{Perturbed DML with General Machine Learning}
\label{sec: method general setting}
In this section, we generalize the perturbation-based approach to settings where generic machine learning methods are employed to estimate the nuisances, with a theoretical justification provided in Supplementary Section \ref{sec:general theory}.

We consider the general models $Y_i = g(X_i) + \epsilon_i$ and $D_i = f(X_i) + \delta_i$, where $\E(\epsilon_i \mid X_i) = 0$ and $\E(\delta_i \mid X_i) = 0$ and $g(\cdot)$ and $f(\cdot)$ are unknown functions that can be consistently estimated by machine learning algorithms. We use $\widehat{g}$ and $\widehat{f}$ to denote the machine learning prediction models trained using observations on sample $\Ical^c$: 
\begin{align}
    \widehat{g} = \argmin_{h \in \mathcal{G}} \frac{1}{n}\sum_{i \in \Ical^c}\{Y_i - h(X_i)\}^2 \quad \text{and} \quad \widehat{f} = \argmin_{h \in \mathcal{F}} \frac{1}{n}\sum_{i \in \Ical^c} \{D_i - h(X_i)\}^2, \label{eq:general optimization}
\end{align}
where $\mathcal{G}$ and $\mathcal{F}$ denote the considered function classes. Using $\widehat{g}$ and $\widehat{f}$, one can construct the unperturbed, influence-function based estimate of $\beta$ on $\Ical$ (Section \ref{sec: model}).

The perturbation step is conceptually similar to that described in Section \ref{sec: method section}. The goal is to create a collection of perturbed nuisance models $\widehat{g}^\m$ and $\widehat{f}^\m$, for $1 \leq m \leq M$. We first generate perturbed noises $\epsilon_i^\m$ and $\delta_i^\m$ using the estimated covariance as detailed below, then inject them into the optimizations \eqref{eq:general optimization}: 
\begin{equation}\label{eq:general optimization perturbed}
\begin{aligned}
    \widehat{g}^\m &= \argmin_{h \in \mathcal{G}} \frac{1}{n}\sum_{i \in \Ical^c}\{Y_i - \epsilon_i^\m - h(X_i)\}^2, \\
    \widehat{f}^\m &= \argmin_{h \in \mathcal{F}} \frac{1}{n}\sum_{i \in \Ical^c}\{D_i - \delta_i^\m - h(X_i)\}^2. 
\end{aligned}
\end{equation}
Specifically, conditioning on the observed data, one may generate the {i.i.d.} bivariate noise vector $(\epsilon_i^\m, \delta_i^\m)$, for $i \in \Ical^c$, following
\begin{align}\label{eq: resample noise}
    \begin{pmatrix} \epsilon_i^{[m]} \\ \delta_i^{[m]}\end{pmatrix} \sim \mathcal{N}_2\l(\mathbf{0}, \hat{\Pi}\r), \quad \textrm{with} \quad \hat{\Pi}=\left(
        \begin{array}{cc}
            \widehat{\sigma}_{\epsilon}^2 & \widehat{\sigma}_{\epsilon\delta} \\
            \widehat{\sigma}_{\epsilon\delta} & \widehat{\sigma}_{\delta}^2
        \end{array}
    \right).
\end{align}
If complex algorithms, which could be prone to overfitting, are employed in estimating $f$ and $g$, one attractive possibility is to compute $\widehat\Pi$ on the $\Ical$ sample:
\begin{align*}
    \hat{\sigma}_\epsilon^2 &= \frac{1}{n}\sum_{i\in\Ical}\{Y_i - \hat{g}(X_i)\}^2, \quad \hat{\sigma}_\epsilon^2 = \frac{1}{n}\sum_{i\in\Ical}\{D_i - \hat{f}(X_i)\}^2, \\
    \hat{\sigma}_{\epsilon\delta} &= \frac{1}{n}\sum_{i\in\Ical}\{Y_i - \hat{g}(X_i)\}\{D_i - \hat{f}(X_i)\}.
\end{align*}
Although \eqref{eq: resample noise} uses Gaussian perturbations, the construction extends to non-Gaussian distributions. The essential requirement is that the generating distribution is sufficiently dispersed such that with high probability, at least one perturbed nuisance estimate lies close to the truth. Supplementary Section \ref{sec:heavy_tailed_resamp} explores heavy-tailed distributions to generate perturbation noises, including Laplace, symmetric lognormal and Student-t distributions. Across these choices, our proposal continues to generate a perturbed estimate $\hat{\beta}^\m$ close enough to $\hat{\beta}^{\rm ora}$ and achieve the desired coverage, though heavier tails tend to yield longer intervals because extreme perturbations occur more frequently. Supplementary Section \ref{sec:heavy_tailed_dgp} instead introduces heavy-tailed noise in the data-generating process while retaining Gaussian perturbations. All settings achieve the nominal coverage with a moderate $M$, with intervals based on heavier-tailed data being wider.

Given the collection of perturbed estimated nuisance models, we propose computing $M$ estimates of $\beta$ on sample $\Ical$ as
\begin{align}
    \widehat\beta^\m = \frac{\sum_{i \in \Ical} \{Y_i - \widehat{g}^\m(X_i)\}\{D_i - \widehat{f}^\m(X_i)\}}{\sum_{i \in \Ical}\{D_i - \widehat{f}^\m(X_i)\}^2}. \label{eq: dml estimator m general}
\end{align}
For each $m$, we then construct the Wald interval $\text{CI}^{[m]}$ centered at $\widehat\beta^\m$  as in \eqref{eq: dml ci m} {with $\hat{\rm SE}(\hat{\beta})$ defined in \eqref{eq:se for beta hat}}. Our proposed confidence interval consists of a filtered union of these Wald intervals. We propose using the same filtering approach discussed in Section \ref{sec: method section}. Suppose that the perturbation step successfully ensures that there exists $m^\star$ such that $\widehat\beta^\mstar$ is sufficiently close to $\widehat\beta^{\rm ora}$. Then, following the reasoning of Section \ref{sec: method section}, we have that $|\widehat\beta^\mstar - \widehat\beta|$ should be within $\widehat\sigma_\beta / \sqrt{n} + 1.01 \rho_n$ with high probability, where $\rho_n$ is an upper bound on the nuisance bias $T_n$ (with general formula given in \eqref{eq:Tn}), i.e., $|T_n| \leq \rho_n$. For example, if $f$ and $g$ are $\alpha$- and $\beta$-H\"{o}lder smooth (and estimators tailored to these classes are employed), then $\rho_n$ can be a constant multiple of $n^{-\frac{\alpha}{2\alpha + p}} \cdot n^{-\frac{\beta}{2\beta + p}}$, which is the product of the optimal root-mean-square-errors for estimating H\"{o}lder-smooth, $p$-dim regression functions (see, e.g., Chapter 1 in \cite{tsybakov2008nonparametric}). Thus, the filtering radius would be $\widehat\sigma_\beta / \sqrt{n} + 1.01 \rho_n$ \footnote{We remark that $n^{-\frac{\alpha}{2\alpha + p}} \cdot n^{-\frac{\beta}{2\beta + p}}$ is \textit{not} the optimal rate for estimating functionals like $\psi$ in H\"{o}lder smoothness models \citep{robins2009semiparametric}. We conjecture that a better filtering radius, tailored to the smoothness model, can be obtained by using higher-order estimators instead of the (first-order) DML estimator as done here.}.
As in the high-dimensional linear case, one can either directly specify the filtering radius above or filter out the Wald intervals corresponding to the $100 \cdot \pi^* \% $ largest differences $|\widehat\beta^\m - \widehat\beta|$, for some cutoff $\pi^*$, e.g., $\pi^* = 0.95$. We summarize the general version of our proposed perturbation and  filtering approach in Algorithm \ref{alg:resampled DML}. 

\begin{algorithm}[ht]
\caption{Perturbed DML with general nonlinear nuisance models}
\label{alg:resampled DML}
\begin{algorithmic}[1]
\Require Observed data $\{Y_i, D_i, X_i\}_{1\leq i\leq 2n}$; Number of perturbations $M$; Filtering proportion $\pi^*$; Confidence level $\alpha$.
\Ensure Confidence interval $\text{CI}$.
\State Split the data into two non-overlapping samples, $\mathcal{I}$ and $\mathcal{I}^c$, each of size $n$;
\State Fit $\widehat{g}$ and $\widehat{f}$ using machine learning methods on fold $\Ical^c$; 
\State Compute DML estimator $\widehat{\beta}$ using fold $\Ical$ as in \eqref{eq:dml estimator cross-fitted general}; \Comment{\textbf{Steps 1-3: DML}}
\For{$m=1, 2, \cdots, M$}
\State Generate the simulated noises $\{\epsilon_i^{[m]}, \delta_i^{[m]}\}_{i\in\Ical^c}$ as in \eqref{eq: resample noise};
\State Fit $\hat{g}^\m,\hat{f}^\m$ using machine learning methods on fold $\Ical^c$;
\State Compute the perturbed DML estimator $\hat{\beta}^\m$ using fold $\Ical$ as in \eqref{eq: dml estimator m general};
\State Construct the confidence interval $\text{CI}^{[m]}$ as in \eqref{eq: dml ci m}; 
\EndFor \Comment{\textbf{Steps 4-10: Perturbation}}
\State Construct the filtered perturbation set $\mathcal{M}$ as in \eqref{eq: filtering 2}; \Comment{\textbf{Filtering}}
\State Return the CI defined in \eqref{eq:filtered union CI}. 
\end{algorithmic}
\end{algorithm}

Similarly to the high-dimensional linear models, the perturbed DML with general machine learning will also require hyperparameter tuning. In line with Section \ref{sec: method section}, one approach is to perform cross-validation while restricting the candidate tuning parameters' values to a small set anchored at the values obtained by cross-validation for the unperturbed optimizations. For more general ML algorithms, the precise relationship between noise reduction and optimal tuning is less clear. Nevertheless, in {our simulations}, we have observed that, fixing the tuning parameters in all perturbations to the values selected in the original DML performs well for ML methods such as XGBoost.

\section{Empirical Investigation of Interval Length} 
\label{sec: compare ci with bias bound}
The CI in \eqref{eq:filtered union CI} is formed by taking the union of Wald intervals across a filtered set of perturbations, with the filtered set theoretically defined in \eqref{eq: filtering 1} and modified in \eqref{eq: filtering 2} for practical implementation.
Although such aggregation provides the coverage guarantee in a conceptually straightforward way, a primary concern is that this comes at the expense of an overly conservative confidence set, especially as the perturbation size $M$ increases.
This section aims to mitigate this concern through three empirical investigations, using simulated data from Example 1, with $n=1000$ and $p=500$. 
In Section \ref{sec:behavior_length_to_M}, we examine how the interval length behaves for a range of the perturbation number $M$. 
In Section \ref{sec:comp_to_OBA_and_CI_B}, we show that Perturbed DML is comparable to an oracle benchmark and avoids the severe conservatism of the bias-bound interval obtained by inflating the Wald interval by the bias bound $\rho_n$ to account for nuisance bias. We run 1000 simulations in each investigation.

\subsection{Finite-Perturbation Behavior of Interval Length}
\label{sec:behavior_length_to_M}

Increasing $M$ in our proposed procedure directly corresponds to adding more candidate intervals in the aggregation step, raising the concern that the resulting union CI is overly wide. To evaluate the dependence on $M$, we vary $M$ over various orders of magnitude and track the resulting interval length using simulated data generated as in Example 1.
Intuitively, since the union CI aggregates intervals across perturbations, the only channel through which $M$ can increase length is by generating more extreme perturbed estimates. We therefore track the maximal deviation $\max_{1\leq m\leq M}|\hat{\beta}^\m - \hat{\beta}|$ as $M$ grows. 

\begin{figure}[ht]
    \centering
    \includegraphics[width=0.95\linewidth]{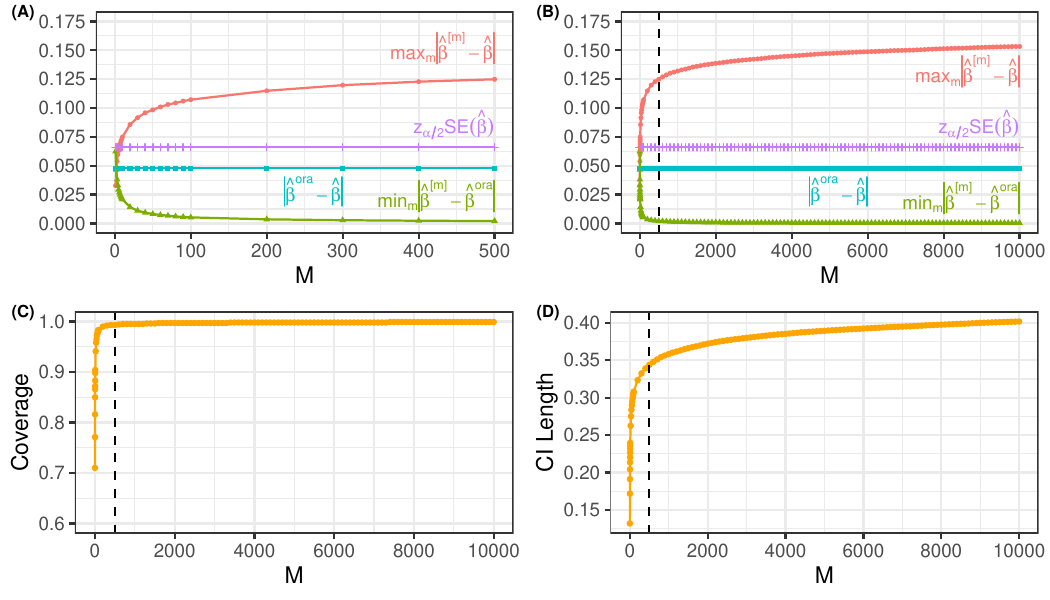}
    \caption{\small Empirical analysis of proposed CI in Example 1 with $n=1000, p=500, s=120, \pi^*=100\%$ (no filtering) and $M$ ranging from 1 to $10^4$. (A): Maximum of deviations $\{|\hat{\beta}^\m - \hat{\beta}|\}_{1\leq m\leq M}$, minimum of distances $\{|\hat{\beta}^\m - \hat{\beta}^{\rm ora}|\}_{1\leq m\leq M}$ , the deviation to the oracle estimator $|\hat{\beta}^{\text{ora}} - \hat{\beta}|$ and the half of Wald CI width $z_{\alpha/2}\hat{\rm SE}(\hat{\beta})$ on $M\leq500$. (B): Zoom out (A) on $M\leq10^4$. (C): Empirical coverages of proposed CIs. (D): Average of CI lengths. The black dashed line marks $M=500$. Results are averaged across 1000 simulations. }
    \label{fig:deviation}
\end{figure}

In Figure \ref{fig:deviation}, as comparison benchmark, we report the half of Wald CI width, $z_{\alpha/2}{\rm SE}(\widehat{\beta})$, of the DML estimator and the nuisance bias component $|\widehat{\beta}^{\rm ora}-\widehat{\beta}|$. We range $M$ from $1$ to $10^4$ and track the minimal perturbation error $\min_{m}|\widehat{\beta}^{\m}-\hat{\beta}^{\rm ora}|$ and the maximal deviation $\max_{ m}|\widehat{\beta}^\m - \hat{\beta}|$ as $M$ grows. In Panel (A), we consider $M\leq 500$ and observe that the minimal distance between $\hat{\beta}^\m$ and $\hat{\beta}^{\rm ora}$ quickly shrinks toward zero, suggesting {$M\geq 200$}  suffices to produce a $\hat{\beta}^\m$ close to $\hat{\beta}^{\rm ora}$. Meanwhile, the maximal deviation exhibits attenuated growth as $M$ increases, so the interval length does not increase markedly as $M$ becomes large.
Panel (B) extends the range of $M$ to $10^4$, where the maximal deviation continues to increase only gradually, reaching about 0.15 at $M=10^4$. Panel (C) and (D) show that our proposal reaches the 95\% coverage for $M>30$, and the average CI length only increases by 17\% from $M=500$ to $M=10^4$.

\subsection{Comparison to Oracle Bias-aware CI and Bias-bound Interval}
\label{sec:comp_to_OBA_and_CI_B}

Following the empirical demonstration that our proposed CI has length with mild growth as increasing $M$, we fix $M=500$ in the remaining of the paper and compare our proposal with other methods.
We first evaluate the relative efficiency of Perturbed DML by benchmarking it against the following oracle bias-aware (OBA) method.
For the standard DML estimator $\widehat{\beta}$ in \eqref{eq:dml estimator cross-fitted general}, we follow \citet{armstrong2020bias} (equation (6)) and  construct the oracle confidence interval using the oracle bias {$\E\hat{\beta} - \beta$} and the oracle standard error $\widehat{\text{SE}}_{\rm emp}(\widehat{\beta})$ as
\begin{equation}\label{eq: OBA ci}
	\left(\widehat\beta - \chi, \widehat\beta + \chi \right), \quad\text{ with } \quad \chi = \widehat{\text{SE}}_{\text{emp}}(\widehat\beta) \cdot \sqrt{\text{cv}_\alpha \left( |\E\widehat{\beta} - \beta|^2/[\widehat{\text{SE}}_{\text{emp}}(\widehat\beta)]^2 \right)},
\end{equation}
where $\text{cv}_\alpha(B^2)$ is the $1-\alpha$ quantile of the $\chi^2$ distribution with 1 degree of freedom and non-centrality parameter $B^2$. We approximate $\E\widehat{\beta} - \beta$ and $\hat{\rm SE}_{\rm emp}(\hat\beta)$ by sample average and sample standard deviation from 1000 Monte Carlo simulations.

\begin{figure}[ht!]
    \centering
    \includegraphics[width=0.85\linewidth]{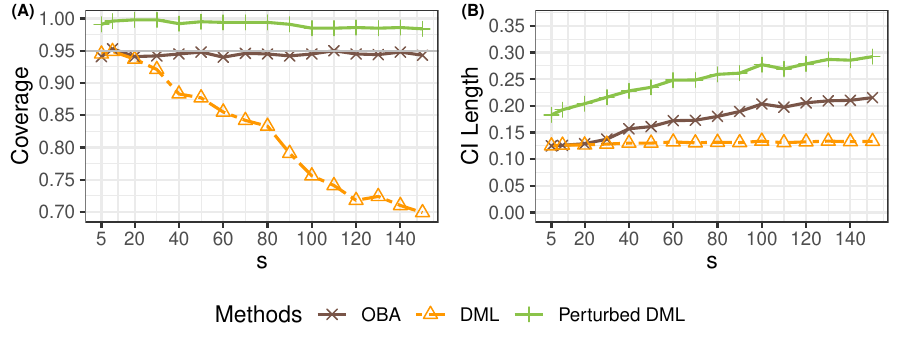}
    \caption{\small  
    Comparison of the oracle bias-aware procedure, the standard DML and our proposal with varying sparsity levels in Example 1 with $n=1000$, $p=500$, $\pi^*=0.95$ and $M=500$. (A): Empirical coverages. (B): Average interval lengths. Results are averaged across 1000 simulations.
    }
    \label{fig: F1 summary}
\end{figure}

In Panel (A) of Figure \ref{fig: F1 summary}, although the coverage of the standard DML procedure deteriorates as $s$ increases, our procedure maintains coverage (albeit conservatively) thanks to the existence of oracle-like perturbations such as those resulting in $\widehat\beta^\mstar$.
Crucially, as shown in Panel (B), the cost of this robustness is a CI length approximately 43\% greater than the infeasible OBA benchmark.

The OBA method introduced above is not practically feasible since it relies on oracle bias information. An empirical way
to address the undercoverage of the Wald interval is to increase the interval length to include a bound on the nuisance bias. 
Usually, such a confidence interval has length of order larger than root-$n$ and requires extra information for construction, such as the sparsity level. In the sparse linear setting,  assuming $\eta$ and $\gamma$ are $s_\eta$- and $s_\gamma$-sparse, respectively, this would mean enlarging the Wald interval to
\begin{equation}\label{eq: ci bound}
    \textrm{CI}_{\rm B} = \l[ \hat{\beta} - z_{\alpha/2}\hat{\rm SE}(\hat{\beta}) - \rho_n, \hat{\beta} + z_{\alpha/2}\hat{\rm SE}(\hat{\beta}) + \rho_n\r] \ \text{with } \rho_n = c^*\l(s_\gamma + \sqrt{s_\eta s_\gamma}\r) \frac{\log p}{n}
\end{equation}
where $\rho_n$ defined in \eqref{eq: Tn bound} serves as an upper bound for the absolute value of the nuisance bias $T_n$ in \eqref{eq:Tn}. Such a construction has been adopted in \cite{cai2017confidence} to justify the existence of a CI achieving the minimax optimal expected length. 
Importantly, even if the sparsity $s_\eta$ and $s_\gamma$ were known,
the specification of an inaccurate constant $c^*$ in \eqref{eq: ci bound} may lead to a dramatically wide confidence interval.

We now compare our proposal with ${\rm CI}_{\rm B}$ in \eqref{eq: ci bound} under  a practical scenario where we only have access to an overly conservative bias bound $\rho_n$. 
Our proposed CI and ${\rm CI}_{\rm B}$ incorporate the bias bound $\rho_n$ in fundamentally different ways. In \eqref{eq: ci bound}, the bound is used in a worst-case fashion directly widening the Wald interval by $\pm\rho_n$, thereby assuming that the maximum bias may be attained by some extremely poor estimates. In contrast, our procedure employs $\rho_n$ as a threshold to screen perturbations. 
By virtue of this difference, even when $\rho_n$ is overly conservative, the length of ${\rm CI}_{\rm B}$ increases by $2\rho_n$. In contrast, as highlighted in Figure \ref{fig:deviation}, the maximum deviation $\max_{1\le m\le M}|\widehat\beta^{[m]}-\widehat\beta|$ stabilizes for a sufficiently large $M$ that ensures valid coverage. This means that no perturbed DML estimator is filtered out and our proposal retains all Wald intervals, returning a confidence interval that is valid and much shorter than adding $2\rho_n$.

\begin{figure}[ht!]
    \centering
    \includegraphics[width=\linewidth]{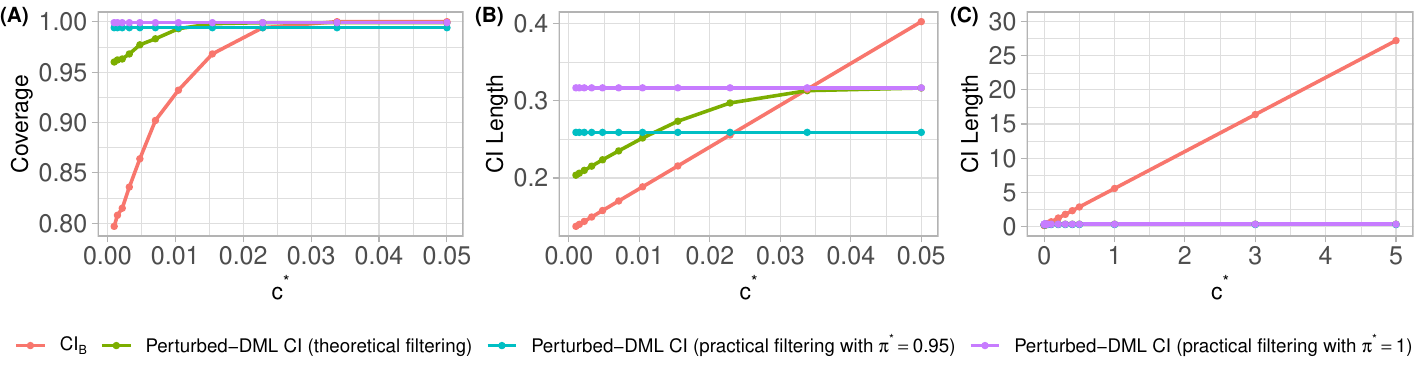}
    \caption{\small Comparison of ${\rm CI}_{\rm B}$ and proposed CIs with different filtering criterion in Example 1 with $n=1000,p=500$, $s=80$ and $M=500$. (A): Empirical coverages of CIs when $c^*\leq0.05$. (B): Average of CI lengths when $c^*\leq0.05$. (C): Same CI length as in Panel (B) but evaluated for $c^*\leq 5$. Results are averaged across 1000 simulations. }
    \label{fig:efficiency comparison}
\end{figure}

Figure \ref{fig:efficiency comparison} empirically compares our proposal and ${\rm CI}_{\rm B}$ in terms of coverage and length.
On Panel (A), ${\rm CI}_{\rm B}$ attains the desired coverage when $c^*\geq 0.015$, indicating that such a $c^*$ ensures $\rho_n$ serves as a valid upper bound for the nuisance bias $T_n$. In practice, however, oracle knowledge of {such} $c^*$ would most likely be unavailable as it would depend on the data generating process. Panel (B) and (C) report the interval lengths: (1) when $\rho_n$ is small (with $c^* \leq 0.05$), both our CI and ${\rm CI}_{\rm B}$ have comparable lengths; (2) as $\rho_n$ increases, the length of ${\rm CI}_{\rm B}$ can become over 80 times longer than that of our CI. Specifically, the length of ${\rm CI}_{\rm B}$ grows linearly with $\rho_n$, whereas in our method, a larger $\rho_n$ than needed simply relaxes the filtering threshold, admitting more perturbations. Once $\rho_n$ is sufficiently large so that all perturbations are retained, that is, the filtered set $\mathcal{M}$ includes all {$M =500$} perturbations, the length of our CI stabilizes even if a more conservative $\rho_n$ is adopted.

In all three panels, the theoretical and practical filtering with $\pi^*=1$ yield CIs with the same coverage and length when $c^*$ is sufficiently large (e.g., beyond $0.05$ in this setting). At this point, all perturbations have deviation $|\hat{\beta}^\m - \hat{\beta}|$ below the theoretical threshold in \eqref{eq: filtering 1}, which coincides with how $\mathcal{M}$ is constructed in \eqref{eq: filtering 2} with $\pi^*=1$. Hence, the proposed CIs from theoretical and practical filtering with $\pi^*=1$ are identical for large $\rho_n$. Notably, when we use the proportion $\pi^*=0.95$ to filter, our CI can be shortened by around 20\% compared to that with $\pi^*=1$ with little loss in coverage.

\section{Theoretical Justification: High-Dimensional Linear Models}
\label{sec: theory}
We provide theoretical justification for our proposal when the nuisances are high-dimensional sparse linear models. We introduce the following assumptions for analysis.

\begin{assumption} \textcolor{white}{Placeholder} \label{assumption:main_lasso}
\begin{enumerate}[(\textup{A}1)]
\item \ The outcome model satisfies $Y_i = X_i^\intercal \eta + \epsilon_i$ with $\E(\epsilon_i \mid X_i) = 0$;  the treatment model satisfies $D_i = X_i^\intercal \gamma + \delta_i$ with $\E(\delta_i \mid X_i) = 0$. The vectors $\eta$ and $\gamma$ are $s_\eta$- and $s_\gamma$-sparse, respectively. The sparsity parameters $s_\eta$ and $s_\gamma$ satisfy $s_{\eta} \log p \log (np)/n \rightarrow 0$ and $s_{\gamma} \log p \log (np)/n \rightarrow 0$.
\item \ The covariate vector $X_i\in\RR^p$ is sub-Gaussian  with $\Sigma_X=\E(X_iX_i^\T)$ satisfying $c_0 \leq \lambda_{\rm min}(\Sigma_X) \leq \lambda_{\rm max}(\Sigma_X) \leq C_0$, where {$C_0 \geq c_0 > 0$} are positive constants. The noise random variables $\epsilon_i$ and $\delta_i$ are sub-Gaussian. 
{Conditioning on the covariates $X_i$,} the covariance matrix $\Pi$ of the noise vector $\begin{bmatrix} \epsilon_i & \delta_i \end{bmatrix}^\T$ satisfies $c_1 \leq \lambda_{\rm min}(\Pi) \leq \lambda_{\rm max}(\Pi) \leq C_1$ for some positive constants {$C_1\geq c_1>0$}.
\end{enumerate}
\end{assumption}

Condition (A1) imposes a sparsity condition ensuring the nuisance models to be consistently estimated Modulo the $\log(np)$ factor \citep{bickel2009simultaneous,buhlmann2011statistics}. This condition can be weakened to $s_\eta\log p/n \to 0$ and $s_\gamma \log p/n \to 0$ if we use homoscedastic-type estimators of ${\Sigma}$ and ${\Lambda}$, for example, $\hat{\Sigma} = \hat{\sigma}_\epsilon^2 \cdot \frac{1}{n}\sum_{i\in\Ical^c} X_i X_i^\T$ with $\hat{\sigma}_\epsilon^2 = \frac{1}{n}\sum_{i\in\Ical^c}(Y_i-X_i^\T\hat{\eta})^2$. In Condition (A2), we assume that both the covariance matrix of the covariates $X_i$ and the conditional covariance matrix of the noise variables are well conditioned. This requirement would be satisfied as long as the covariates are not highly collinear and the noise components $\epsilon$ and $\delta$ are not perfectly correlated.  

To facilitate the discussion, we introduce the rate $\err$ that characterizes $\min_{1 \leq m \leq M} \|\xi - \xi^\m\|_\infty$ and $\min_{1 \leq m \leq M} \|\kappa - \kappa^\m\|_\infty$, where $\xi$ and $\kappa$ are random vectors specified in \eqref{eq: lasso optimization problem unperturbed}, and $\xi^\m$ and $\kappa^\m$ are generated perturbed copies. Let $\alpha_0\in(0,0.01]$ denotes the probability that $\|\xi\|_2$ and $\|\kappa\|_2$ falls in the $\alpha_0$-tail of their distributions. We define
\begin{equation}\label{eq: supp def err}
    \err = c_1 \cdot [c_*(\alpha_0)]^{-\frac{1}{\sqrt{p}}} \cdot  \l( \frac{4\log n}{M}\r)^{\frac{1}{2p}},
\end{equation}
where $c_1>0$ and $c_*(\alpha_0)>0$ are positive constants specified as in \eqref{eq:supp constants in err} in the supplementary material. Notice that, for fixed $n$ and $p$, $\err$ vanishes to zero as $M \rightarrow \infty$. 

The following Theorem shows that $\min_{1\leq m\leq M}\left|\hat{\beta}^\m - \hat{\beta}^{\rm ora}\right|$ is scaled to $\err.$ 

\begin{theorem}\label{thm: mstar lasso}
     Suppose Assumption \ref{assumption:main_lasso} holds and the penalty parameters $\lambda_\eta^\m$ and $\lambda_\gamma^\m$ in \eqref{eq: lasso optimization problem eta m} satisfy {$\lambda_\eta^\m = C  n^{-1/2}\err$ and $\lambda_\gamma^\m = C n^{-1/2}\err$} for some constant $C>1$. 
   There exists some other constant $C'>0${ independent of $n$ and $p$ such that}
   \begin{equation}
    \begin{aligned}
        \liminf_{n,p\to\infty}\liminf_{M\to\infty}&\P\bigg(\exists m\in \{1,\dots,M\}: |\hat{\beta}^\m - \hat{\beta}^{\rm ora}| \\
        & \leq C'\bigg(\frac{\sqrt{s_\eta}+\sqrt{s_\gamma}}{n}\err + \frac{\sqrt{s_\eta s_\gamma}+s_\gamma}{n}\err^2\bigg)\bigg) \geq 1-\alpha_0,
    \end{aligned}
    \label{eq: ora convergence}
    \end{equation}
with the oracle DML estimator $\hat{\beta}^{\rm ora}$ defined in \eqref{eq: betaHat ora} and $\err$ defined in \eqref{eq: supp def err}.
\end{theorem}

This theorem formally states that our procedure yields, with high probability, the minimal distance between a perturbed estimator $\hat{\beta}^\mstar$ and $\hat{\beta}^{\rm ora}$ is at most a constant multiple of a rate which goes to zero as $M$ grows. In the proof, this closeness is derived from the fast convergence rate of the nuisance estimators in the $m^*$-th perturbation; see more explanations in Supplementary Section \ref{sec: notation in Thm1}.

We shall remark that it is impossible to locate the exact perturbation $m^*$ and we are only able to justify that such an $m^*$ exists with high probability. In Section \ref{sec:simulation}, extensive simulations show that such $\hat{\beta}^\mstar$ indeed exists and its empirical distribution closely matches that of the theoretical distribution of $\hat{\beta}^{\rm ora}$ across different data generating processes; {see Figures \ref{fig: F1 summary} and \ref{fig: F2F3 summary}}.

Building upon the core properties established in Theorem \ref{thm: mstar lasso}, we establish coverage and length of the filtered union confidence interval ${\rm CI}$ in \eqref{eq:filtered union CI}.
\begin{theorem}\label{thm: coverage lasso}
    Suppose that conditions of Theorem \ref{thm: mstar lasso} hold.
    The confidence interval ${\rm CI}$ defined in \eqref{eq:filtered union CI} satisfies
    \begin{equation*}
        \liminf_{n,p\to\infty}\liminf_{M\to\infty}\Prob(\beta \in {\rm CI}) \geq 1-\alpha,
    \end{equation*}
    where $\alpha \in (0,1/2)$ is the significance level used to construct the CI in \eqref{eq: dml ci m}. 
    Furthermore, the length of ${\rm CI}$ satisfies
    \begin{equation}
\liminf_{n,p\to\infty}\liminf_{M\to\infty} \Prob\l(\len({\rm CI}) \leq {2.02}\rho_n + \frac{(4+c) \sigma_\beta}{\sqrt{n}}\r) = 1,
\label{eq: length}
\end{equation}
where $\rho_n = c^*(s_\gamma+\sqrt{s_\eta s_\gamma})\frac{\log p}{n}$ as defined in \eqref{eq: ci bound}, {$\sigma_\beta = \sqrt{\Var\{\varphi(O_i; \beta)\}}$} and $c>0$ is an arbitrarily small positive constant.
\end{theorem}

Theorem \ref{thm: mstar lasso} shows that, with high probability, there exists $m^*$ such that $\widehat\beta^\mstar$ is sufficiently close to $\widehat\beta^{\rm ora}$ so that its associated Wald  interval ${\rm CI}^\mstar$ retains asymptotic nominal coverage. Theorem \ref{thm: coverage lasso} crucially establishes that, with high probability, such special $m^*$ is retained in the filtered set $\mathcal{M}$ defined in \eqref{eq: filtering 1}. The inclusion of ${\rm CI}^\mstar$ in the union ensures the coverage of the proposed CI. The length of the final confidence interval is of order $\rho_n + n^{-1/2}$.

We also note that Theorems \ref{thm: mstar lasso} and \ref{thm: coverage lasso} require the perturbation size $M$ to diverge with the sample size $n$ and dimension $p$. Our proofs make the scale explicit: it suffices to take $\log M \gtrsim \log\log n + p^{2}$. 
The price is computational: even for moderate $p$, the implied $M$ can be large. However, we emphasize that this large $M$ requirement appears to be a proof artifact. In practice, modest choices (e.g., $M=500$) produce reliable confidence intervals; see Figure \ref{fig:deviation} and the additional sensitivity analysis in Supplementary Section \ref{sec: sensitivity to tuning}.

\noindent {\bf Optimality and Adaptivity.}
We evaluate the optimality of our proposal by considering the parameter space
\begin{equation}
\Theta(s) 
= \Bigl\{ \theta = (\beta,\eta,\gamma,\Psi,\sigma_\epsilon):\ 
s_\eta = s_\gamma \le s,\ 
c\le\lambda_{\rm min}(\Psi)\le\lambda_{\rm max}(\Psi)\le C,\ 
\sigma_\epsilon \le C_1 \Bigr\},
\label{eq: space}
\end{equation}
where $\Psi = \E[W_iW_i^\T]$ denotes the second-order moment of $W_i = (D_i\;\;X_i^\T)^\T \in \RR^{p+1}$, {and $\sigma_\epsilon$ stands for the standard deviation of the noise $\epsilon_i$,} and $c,C,C_1$ are positive constants independent of $n$ and $p$. By Theorem 2 in \citet{cai2017confidence}, we obtain that the minimax expected length of a 
confidence interval with correct coverage over $\Theta(s)$ is
\begin{equation}
\frac{1}{\sqrt{n}}+\frac{s \log p}{n}.
\label{eq: opt length}
\end{equation}
By taking $s_{\eta}=s_{\gamma}=s$, the length result in \eqref{eq: length} implies that our proposed $\text{CI}$ in \eqref{eq:filtered union CI} attains the optimal length in \eqref{eq: opt length} over $\Theta(s)$ up to constants.

We discuss adaptivity in confidence interval construction, focusing on the regime $s_{\eta}=s_{\gamma}=s$. A crucial step for attaining the optimal length in \eqref{eq: opt length} is the filtering step in \eqref{eq: filtering 1}, whose theoretical threshold requires knowledge of $s$. Without filtering, taking the union of all perturbed Wald intervals guarantees coverage but cannot ensure the minimax expected length. Importantly, \citet{cai2017confidence} show that, when the Wald interval does not provide valid coverage, constructing confidence intervals of optimal length requires knowledge of the sparsity level. In particular, when $\sqrt{n}/\log p\lesssim s\lesssim n/\log p$, their Theorem 3 establishes the impossibility of adaptation to $s$: one cannot attain the optimal length in \eqref{eq: opt length} without knowing $s$. For the regime with known $s$, \citet{cai2017confidence} construct a confidence interval as in \eqref{eq: ci bound} using a bias bound; our detailed comparison in Section \ref{sec: compare ci with bias bound} shows that the proposed perturbed DML interval is significantly shorter than the bias-bound interval in \eqref{eq: ci bound}. Related results on the (im)possibility of adaptive confidence intervals include \citet{robins2006adaptive} and \citet{nickl2013confidence}.

\section{Simulation Studies}
\label{sec:simulation}

In this section, we compare our proposal with the standard inference procedure based on the Wald interval centered at the influence-function-based estimator. In Supplementary Section \ref{sec: sensitivity to tuning}, we demonstrate that our proposal is insensitive to both the number of perturbations $M$ and the filtering proportion $\pi^*$ when $M\geq100$ and $\pi^*\geq0.95$.  In all our simulation studies, we implement our proposal using Algorithm \ref{alg:resampled DML} (with two-fold cross-fitting). We implement the standard DML procedure using the R / Python package {\tt DoubleML} with two-fold cross-fitting based on five splits \citep{bach2022doubleml}. All results are summarized based on 1000 simulations. {The code for replicating the simulation is available at \url{https://github.com/makaylazheng/DML-nonregular-inference}.}

We start with introducing the data generating processes. We generate the outcome $Y_i$ and the treatment $D_i$ following the correctly specified partially linear model {, $Y_i = D_i\psi + h(X_i) + e_i$, as specified in \eqref{eq:PLR}} and $D_i = f(X_i)+\delta_i$ with $f(X_i) = \E[D_i \mid X_i]$. Under this correct model, the coefficient $\psi$ equals our target parameter $\beta$. 
Across all settings, we vary only the functional forms of $f$ and $h$, keeping all the other components in data generation fixed. We first generate $W_i\in\R^p$ following a multivariate normal distribution with mean zero and covariance matrix $A$ where $A_{k,l} = 0.5^{|k-l|}$ for $k,l=1,\dots,p$. Let $X_{i,j} = \Psi(W_{i,j})$ for $i=1,\dots,n$ and $j=1,\dots,p$ where $\Psi(\cdot)$ is the cumulative distribution function of the standard normal. After the transformation, $X_{i,j}$ follows correlated uniform distributions on $(0,1)$ for $j=1,\dots,p$. {The noise terms $e_i$ and $\delta_i$} are independently drawn from the standard normal distributions. {Supplementary Section \ref{sec:hetero} considers heteroscedastic noises and finds similar performance.} The true treatment effect is set to $\psi = \beta=0.5$, and the sample size is fixed at $n=1000$. 

In the following, we present the comparison of our proposal to other benchmark methods under various nuisance models. Specifically, we consider {two} generating models for the nuisance functions: (F1) generalized additive model (GAM); (F2) nonlinear model with interaction terms. 
For both settings, we apply estimation methods suited to the model structure when implementing  the standard DML and our proposed Perturbed DML procedures. 
In particular, we employ the R package {\tt mgcv} \citep{wood2017generalized} to fit the generalized additive nuisance models in (F1), and the Python package {\tt xgboost} \citep{chen2016xgboost} to fit nonlinear models in (F2) with interaction terms. {We also consider an additional simulation setting in Supplementary Section \ref{sec:additional_simulation}, where the nuisance functions are linear and are estimated using ordinary least squares.}

{We now introduce the specific generating models for nuisance functions and then present the corresponding results.}

\vspace{2mm}
\noindent \textbf{F1 (Generalized Additive Nuisance Models):} $f(X_i) = \sum_{j=1}^p f_j(X_{i,j})$ and $h(X_i) = \sum_{j=1}^p h_{j}(X_{i,j})$ where the {univariate} functions $f_j$ and $h_j$ are both cyclically assigned from a predefined set of nonlinear functions: $s_1(z)=3\sin(z)/2,  s_2(z)=2e^{-z/2}, s_3(z)=(z-1)^2-25/12, s_4(z)=z-1/3, s_5(z)=3z/4, s_6(z)=z/2$. We assign $f_j=s_{j \text{ mod } 6}$ and $h_j = s_{(j+2) \text{ mod } 6}$, {where for any integer $k$, $k \textrm{ mod } 6$ equals the remainder of $k$ division upon 6, except that a zero remainder is recorded as 6.}

\vspace{2mm}
\noindent \textbf{F2 (Nonlinear Nuisance Models with Interactions):} $f(X_i) = \sum_{j=1}^p f_j(X_{i,j}) + \sum_{j=1}^{p-1}X_{i,j}X_{i,j+1}$ and $h(X_i) = \sum_{j=1}^p h_{j}(X_{i,j}) + \sum_{j=1}^{p-2}X_{i,j}X_{i,j+1}X_{i,j+2}$ with $f_j$ and $h_j$ assigned using the same rule as in setting F3. 
\vspace{2mm}

{In both F1 and F2, we vary the dimension $p$ from 2 to 20. }We implement both standard DML and our proposal with specified nuisance estimation methods. For computational efficiency, we adopt the penalty parameter selection method from Section \ref{sec: method section} in F1, while in F2 we set all the XGBoost-related parameters in the perturbed optimizations equal to those obtained by cross-validation from the unperturbed optimization.

\begin{figure}[ht!]
    \centering
    \includegraphics[width=0.95\linewidth]{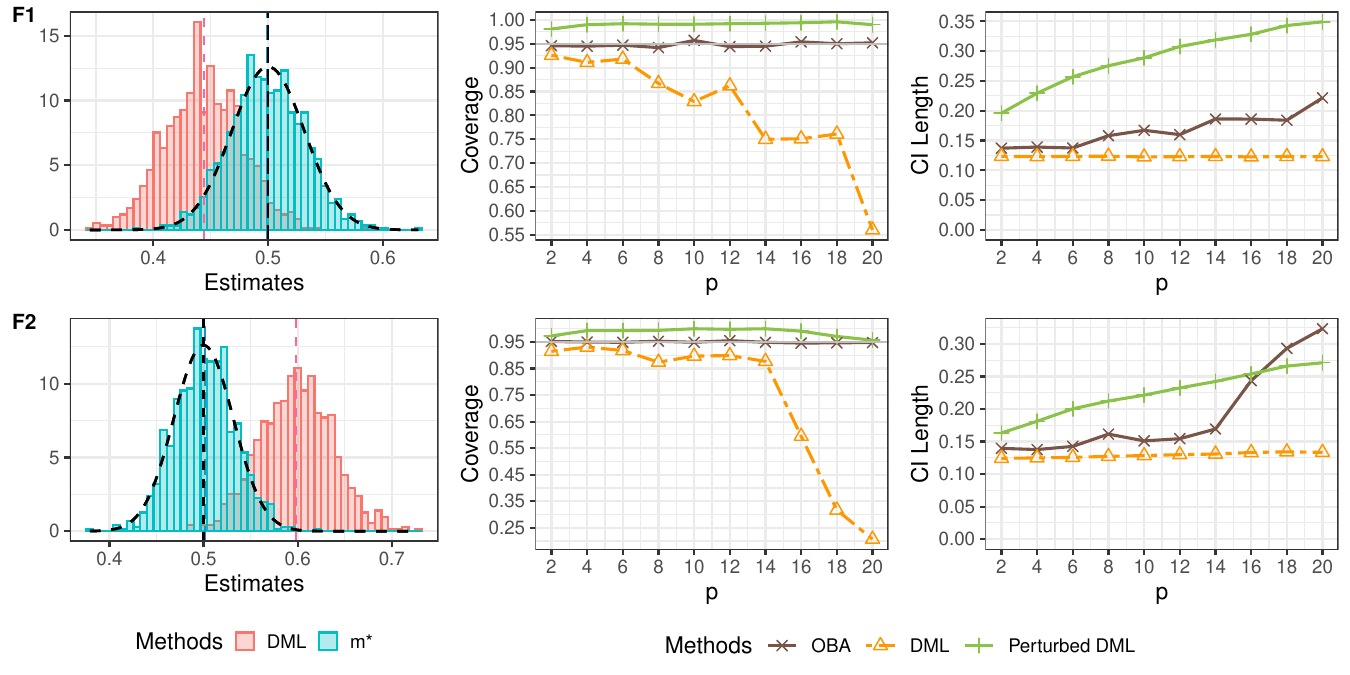}
    \caption{\small 
    Settings F1 and F2 with $n=1000$ and $p$ ranging from 2 to 20. The leftmost column compares the empirical distributions of $\hat{\beta}^\mstar$ and $\hat{\beta}$ when $p=20$, {where the black dashed curve represents the distribution $N(\beta,n^{-1}\Var\{\varphi(O_i;\beta)\})$.}The middle and rightmost columns report the empirical coverage and average CI lengths based on {OBA}, {DML} and Perturbed DML.
    }
    \label{fig: F2F3 summary}
\end{figure}

Figure \ref{fig: F2F3 summary} reports the results under settings F1 and F2. They present similar performance to that in Figure \ref{fig: F1 summary} under Example 1 where the nuisance functions are sparse linear. In the leftmost column, the original DML estimator $\hat{\beta}$ exhibits both large bias and slightly inflated variance while $\hat{\beta}^\mstar$ remains unbiased and has variance close to $n^{-1}\Var\{\varphi(O_i;\beta)\}$. As $p$ increases, the Wald interval centered at $\widehat\beta$ fails to achieve nominal coverage, whereas our inference method remains valid (albeit conservative) across all $p$. Notably, in Setting F2, the proposed CI becomes even shorter than oracle-bias-aware one when $p\geq 18$.

\section{Real Data Analysis}
\label{sec:real_data}

We revisit the estimation of gun ownership's effect on the homicide rates, as presented in Chapter 9.2 of \cite{chernozhukov2024applied}. Following \cite{chernozhukov2024applied}, let $Y_{i,t}$ denote the log homicide rate in county $i$ at time $t$, and let $D_{i, t-1}$ be the log fraction of suicides committed with a firearm in county $i$ at time $t-1$, which serves as a proxy for gun ownership at time $t-1$ as designed in \cite{cook2006social}. We also denote by $Z_{i,t}$ a vector of 195 demographic and economic characteristics of county $i$ at time $t$, such as the income distribution, home ownership rates, etc. Consistent with \cite{chernozhukov2024applied}, we account for the fixed heterogeneity across counties, common time factors, and deterministic time trends by augmenting the control variables. We include within-county and within-year averages of $Z_{i,t}$ (denoted by $\bar{Z}_i$ and $\bar{Z}_t$), initial conditions in county $i$ ($Z_{i,0}$ and $Y_{i,0}$), and the time index ($t$). As shown in \cite{wooldridge2025two}, including $\bar{Z}_i$ and $\bar{Z}_t$ in a linear regression is equivalent to including two-way fixed effects for time and county. Consequently, the full set of observed covariates is defined as $X_{i,t} = \{Z_{i,t}, \bar{Z}_i, \bar{Z}_t, Z_{i,0}, Y_{i,0}, t\}$. The dataset comprises 3,900 observations from 195 large United States counties spanning 1980 through 1999. 

We apply both the standard DML \citep{Chernozhukov2018} and our proposal to estimate the effect. We implement standard DML with five-fold cross-fitting based on one sample split.
For Perturbed DML, we specify a perturbation size of $M=500$ and a filtering proportion of $\pi^*=0.95$. More details are provided in Supplementary Section \ref{section:data_analysis_implementation}.

\begin{figure}[htp!]
    \centering
    \includegraphics[width=0.9\linewidth]{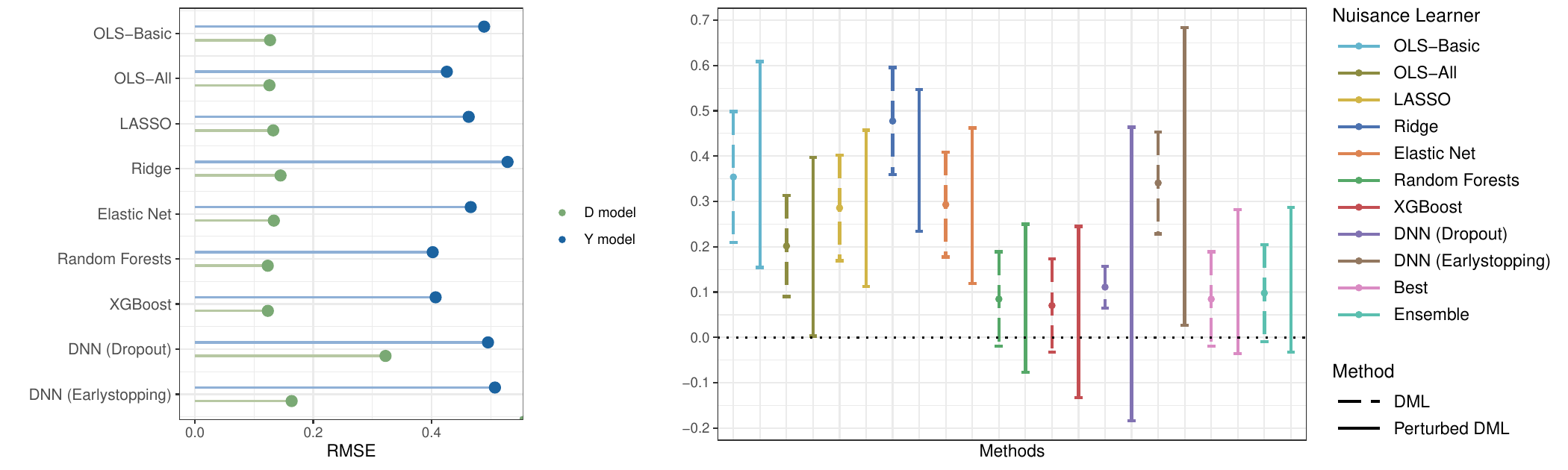}
    \caption{Root-mean-square-errors (RMSEs) and CIs from standard DML and Perturbed DML. The left subplot presents the cross-fitted RMSEs for predicting the outcome $Y_{i,t}$ and  treatment $D_{i,t-1}$ using various nuisance learners. The right subplot displays the corresponding CIs from standard DML and Perturbed DML. The ``Best'' learner selects the learner with the lowest RMSEs for each of $Y_{i,t}$ and $D_{i,t-1}$. The nuisance learner ``Ensemble'' uses the linear combination of all learners that achieves the lowest RMSEs for each of $Y_{i,t}$ and $D_{i,t-1}$. In Perturbed DML, the selection of ``Best'' and the weights of ``Ensemble'' are reidentified and recomputed in every perturbation. }
    \label{fig:data}
\end{figure}

Figure \ref{fig:data} presents the analysis results from the standard DML and Perturbed DML. The left subplot demonstrates the root-mean-square-errors (RMSEs) when estimating $\E[Y_{i,t} \mid X_{i,t}]$ and $\E[D_{i,t-1} \mid X_{i,t}]$ using unperturbed data. Among all single nuisance learners considered, the tree-based methods achieve lower RMSEs than other methods. In particular, Random Forests achieves the lowest RMSEs 
for both conditional mean models, indicating the best predictive performance. 

In the right subplot, the CIs for standard DML are broadly comparable in length, ranging from 0.2 to 0.29. A notable exception is DNN with dropout, where the poor nuisance prediction may lead to small estimated variance. In comparison, Perturbed DML produces wider CIs, with lengths between 0.3 and 0.45 except for the two DNNs. Disregarding these two exceptions, the Perturbed DML CIs are, on average, approximately 50\% longer than those from standard DML. This demonstrates that the proposed Perturbed DML obtains the robustness without being overly conservative, and delivers approximately the same conclusions as standard DML. The widening effect is most pronounced for the DNN learners, where the Perturbed DML CIs are roughly two to six times the width of those from standard DML. This observation aligns with literature suggesting that DNNs are more sensitive to small data perturbations and hyperparameter variability compared to tree-based methods \citep{borisov2023deeptlf, shavitt2018regularization}. We also provide the CI comparison with $M=5000$ in Supplementary Section \ref{sec:data_additional}: on average, Perturbed DML CIs are approximately 68\% longer than standard DML CIs excluding two DNNs; and increasing $M$ from 500 to 5000 enlarges our CIs by only 7.8\% on average across all learners. These observations further supports the claim that our proposal is not excessively conservative.

\section{Conclusion and Discussions}

We study inference on a low-dimensional functional in the presence of infinite-dimensional nuisance parameters. We move beyond the regular regime where Wald intervals have coverage and construct confidence intervals that remain valid even when the nuisance estimators converge at rates slower than $n^{-1/4}$. 
Our proposed perturbation-based DML offers a simple, implementable safeguard that preserves efficiency in favorable cases and maintains validity well beyond the classical \(n^{-1/4}\) regime, while opening a path toward adaptive, learner-agnostic semiparametric inference. We highlight two directions for further study. The first question is on the \emph{minimal} number of perturbations needed for our proposed confidence interval to have coverage. {In Theorems \ref{thm: mstar lasso} and \ref{thm:mstar ml beta} in the supplementary material, our proofs ensure validity whenever at least one of the \(M\) perturbations produces nuisance fits within the bias envelope required for selection.} This yields a sufficient (and potentially conservative) lower bound on \(M\). Empirically, we observe that a much smaller \(M\) already suffices to deliver valid coverage beyond the regime where the Wald interval is reliable. It would be desirable to capture the minimum perturbation budget in theory and develop data-dependent rules of choosing $M$ that is adaptive to the problem difficulty. Secondly, although the present paper focuses on inference for $\beta$, semiparametric theory encompasses a much broader class of summary functionals. Our perturbation idea promises valid inference for other functionals studied in the semiparametric efficiency literature, such as effects defined by continuous treatments \citep{kennedy2017non,takatsu2025debiased} or instrumental variables \citep{Chernozhukov2018,emmenegger2021regularizing}, as well as covariance measures \citep{lundborg2024projected, ShahPeters2020HardnessCI}; extending our guarantees to such functionals is an important next step.

\section*{Acknowledgment}

{The authors are grateful to Prof. Yuansi Chen for providing the first draft of the isoperimetric proof along with insightful discussions.}


\bibliographystyle{plainnat}
\bibliography{refs}

\newpage
\appendix

\section{Related Literature}\label{section:lit_review}

\subsection{Root-\texorpdfstring{$n$}{} inference under weaker dependence on the nuisance error}
One potential solution to the inference problem when the nuisance functions are not estimated accurately enough is to look for new estimators or modifications of the DML procedure that enjoy more favorable dependence on the nuisance estimation error. In this section, we review several promising avenues when the nuisances are H\"{o}lder smooth or follow high-dimensional generalized linear models (GLMs). We emphasize that these methods are conceptually quite distant from the approach we take in this work, since they mostly aim to weaken the requirements for $\sqrt{n}$-CAN (with a few exceptions deriving central limit theorems, and thus inference, in slower-than-root-$n$ regimes, e.g., \citeapp{robins2016asymptotic} and \citeapp{mcclean2024double}). Our goal, on the other hand, is to design a procedure that yields valid inference regardless of whether the parameter is root-$n$ estimable or the estimator asymptotically Gaussian. In particular, our approach is not based on further debiasing the DML estimator.

We start by reviewing well-established improvements over the DML estimator when the nuisances are H\"{o}lder-smooth. Under this general model, the theory of higher-order influence functions\footnote{Strictly speaking, the theory is not tied to H\"{o}lder smoothness; see \citeapp{10.1214/20-STS804}.} (HOIFs) has proven instrumental in obtaining new estimators that are minimax optimal under certain conditions \citepapp{robins2008higher, robins2009quadratic, robins2009semiparametric, robins2017minimax}. It has also been used to derive falsification tests of coverage of the Wald interval centered at the DML estimator \citepapp{liu2024assumption}. Estimators based on the (approximate) $m^{\text{th}}$-order influence function, henceforth referred to as $m^{\text{th}}$-order estimators, are U-statistics with kernel of order $m$, and, under certain conditions, exhibit a dependence on the nuisance error of order $m+1$. In the context of parameters having a first-order influence function, the first-order estimator corresponds to what we refer to as the DML estimator; see, e.g., \citeapp{robins2009quadratic} for a comprehensive discussion of first and second-order estimators. 

Despite the substantial theoretical gains, higher-order estimators are still rarely used in practice. To our knowledge, one reason is that these estimators crucially depend on an additional tuning parameter, the size of the dictionary of basis functions, in order to further de-bias the first-order / DML estimator. Such parameter is challenging to tune data-adaptively; see \citeapp{liu2021adaptive} for an application of Lepski's method in this context when the estimator is a second-order one (i.e., a U-statistic with kernel of order two). In addition to computational challenges, inference based on second-order estimators, whose asymptotic normality is established in \citeapp{robins2016asymptotic},  however, still requires certain higher-order nuisance error terms (of the form of products of three nuisance errors) to vanish sufficiently fast. 

In addition, higher-order corrections in low-smoothness regimes, for which the convergence rate is slower than $n^{-1/2}$, require estimating inverses of Gram matrices whose dimensions exceed the sample size. This effectively prevents the use of the corresponding empirical counterparts as they would not be invertible, thus considerably complicating the implementation. We refer the readers to \citeapp{robins2017minimax}, \citeapp{liu2017semiparametric}, \citeapp{liu2023new}, \citeapp{chen2024method}, and \citeapp{chen2025computing} for the state-of-the-art regarding the implementation of these methods\footnote{See also the GitHub repository: \\ \url{https://github.com/cxy0714/Falsification-using-higher-order-influence-functions}}. Finally, there is also a line of work, with promising examples by \citeapp{newey2018cross}, \citeapp{kennedy2024minimax} and \citeapp{mcclean2024double}, aiming at obtaining more practical estimators with similar guarantees as those enjoyed by HOIFs-based ones. They are based on particular forms of sample splitting coupled with undersmoothed estimation of the nuisance parameters. In particular, \citeapp{mcclean2024double} derives new estimators of the expected conditional covariance functional (the numerator in our leading example discussed in Section \ref{sec: model}) and establishes a slower-than-root-$n$ CLT when the density of the covariates is known. 
We remark that these methods are tailored to H\"{o}lder smooth nuisances or particular estimators. In contrast, our work strives to derive an inferential procedure that is agnostic with respect to the analyst's modeling choices and such that its validity does not rest on the assumption that the estimator is converging at $n^{-1/2}$-rates nor that it is asymptotically normally distributed.

Another stream of literature has discovered new estimators in settings where the nuisances are assumed to belong to sparse, or approximately sparse\footnote{Approximate sparsity was introduced by \citeapp{bradic2019minimax} to describe nuisance models by sparse linear combinations of functions taken from a set where the elements are not naturally ordered. This is a generalization of approximating a function via a dictionary of (ordered) basis functions.}, high-dimensional generalized linear models. For a class of functionals with the so-called mixed-bias property \citepapp{rotnitzky2021characterization}\footnote{Informally, these are parameters that depend on two nuisance components such that the DML estimator has an overall nuisance error depending only on the product of the individual nuisance errors.}, \citeapp{liu2023root} show that $n^{-1/2}$-consistency is attainable as long as at least one of the high-dimensional GLMs is sparse enough to be estimable at $n^{-1/4}$-rates; see also \citeapp{bradic2019minimax}. A separate line of work has derived similar doubly-robust inferential results in different nonparametric models \citepapp{van2014targeted,benkeser2017doubly,bonvini2024doubly,van2024doubly}. Although these refinements enlarge the validity regime of the Wald interval, inference still requires at least one nuisance estimator to converge faster than $n^{-1/4}$. Our method, which can, in principle, be combined with these new developments, addresses the fundamentally different problem of valid inference when Wald intervals are invalid.  

Finally, recent work on structure-agnostic functional estimation by \citeapp{balakrishnan2023fundamental} has highlighted how, for many functionals of interest, there is a strong sense in which improvements over the DML estimator are possible only under additional structural conditions that are not fully exploited by it. In particular, the DML estimator's dependence on the nuisance error is minimax optimal over the so-called \textit{structure-agnostic} function class\footnote{A structure-agnostic class of data generating distributions can be loosely defined as consisting of all distributions over which the nuisance estimators can attain certain convergence rates.}. See also follow-up work by \citeapp{jin2024structure} and  \citeapp{jin2025s}. The optimality of the DML estimator over this space certainly does not mean that one should not pursue the goal of designing estimators that improve upon it under certain conditions (and possibly perform as well if the conditions are not met). However, it does point to the fundamental difficulty of carrying out estimation and inference in a structure-agnostic way, i.e., without assuming smoothness or sparsity. Our procedure strives to seamlessly accommodate the analyst's choice for the nuisance estimators even when they are complex, black-box algorithms. It thus aims to be as structure-agnostic as possible in the sense that it incorporates the analyst's structural assumptions only in the specification of the filtering radius.

To summarize, exciting progress has been made towards weakening the requirements for $\sqrt{n}$-CAN of estimators of many functionals of interest, yielding efficient inference at the parametric rate. However, even for structured function classes, $\sqrt{n}$-consistency may only be attainable for a special subclass of all data generative distributions, as established by \citeapp{robins2009semiparametric} (H\"{o}lder classes) and \citeapp{bradic2019minimax} (approximately sparse classes). In this work, we derive a novel inferential strategy that does not rely on further debiasing the DML estimator nor does it aim to weaken the requirement for $\sqrt{n}$-CAN; rather, it augments the DML strategy with two additional steps, perturbation and filtering, to yield a confidence set that is valid outside the parametric regime of convergence and that is meant to be agnostic with respect to the analyst's choice for the nuisance models. In the next section, we review existing methods for conducting nonstandard inference in the challenging regime where the estimator is not $\sqrt{n}$-CAN, and we contrast these with our proposed approach.

\subsection{Inference outside the root-\texorpdfstring{$n$}{} regime}

When the estimator is not $\sqrt{n}$-CAN, the Wald interval is generally invalid. Several authors have considered inference in the challenging regime where the asymptotic normality fails. For instance, \citeapp{wasserman2020universal} develop a universal approach based on the likelihood-ratio principle and cross-fitting. It accommodates nuisance parameters via likelihood profiling; however, in the semiparametric settings considered here, it is unclear how to profile the nuisance parameters effectively. 
More recently, \citeapp{kuchibhotla2024hulc} derive sample-splitting confidence intervals that depend on a bound for the estimator’s median bias. In our settings, translating nuisance-estimation error into a nontrivial bound on median bias outside the parametric convergence regime appears difficult. Closer in spirit to our work, \citeapp{xie2024repro} propose injecting artificially sampled noise to construct confidence sets that, like \citeapp{wasserman2020universal}, do not rely on asymptotics or regularity conditions; however, their focus is on discrete parameters (e.g., the number of mixtures), and their handling of nuisances again relies on likelihood profiling, which is hard to generalize to our context. \citeapp{guo2024statistical} and \citeapp{guo2025robust} employ sampling/perturbation techniques to address nonregular inference arising from boundary conditions and model selection. By contrast, the present work targets a distinct challenge within semiparametric efficiency theory, namely how to conduct inference when the nuisance estimators converge more slowly than $n^{-1/4}$.

\section{Algorithm, Optimality and Adaptivity with High-Dimensional Linear Nuisance Models}
\label{section:algorithm_optimality_adaptivity}

\subsection{Perturbed DML Algorithm with High-Dimensional Linear Nuisance Models}
\label{section:algorithm_sparse_linear}

We include an algorithm for Perturbed DML algorithm with high-dimensional linear nuisance models.
\begin{algorithm}[ht]
\caption{Perturbed DML with high-dimensional linear nuisance models}
\label{alg:resampled DML lasso}
\begin{algorithmic}[1]
\Require Observed data $\{Y_i, D_i, X_i\}_{1\leq i\leq 2n}$; Number of perturbations $M$; Filtering proportion $\pi^*$; Significance level $\alpha$. 
\Ensure Confidence interval $\text{CI}$.
\State Split the data into two non-overlapping samples, $\mathcal{I}$ and $\mathcal{I}^c$, each of size $n$;
\State Compute $\hat{\eta}$ and $\widehat{\gamma}$ using fold $\Ical^c$ as in \eqref{eq: lasso optimization problems}; 
\State Compute DML estimator $\widehat{\beta}$ with $\hat{g}(X_i)=X_i^\T\hat{\eta}$ and $\hat{f}(X_i)=X_i^\T\hat{\gamma}$ using fold $\Ical$ as in \eqref{eq:dml estimator cross-fitted general};
\For{$m=1, 2, \cdots, M$}
\State Generate the simulated terms $\xi^{[m]}, \kappa^{[m]}$ as in \eqref{eq: resample xi kappa};
\State Fit perturbed nuisance estimators $\hat{\eta}^\m,\hat{\gamma}^\m$ using fold $\Ical^c$ as in \eqref{eq: lasso optimization problem eta m};
\State Compute the perturbed DML estimator $\hat{\beta}^\m$ using fold $\Ical$ as in \eqref{eq: dml estimator m};
\State Construct the confidence interval $\textrm{CI}^\m$ as in \eqref{eq: dml ci m}; 
\EndFor
\State Construct the filtered perturbation set $\mathcal{M}$ as in \eqref{eq: filtering 2};
\State Return the CI defined in \eqref{eq:filtered union CI}.
\end{algorithmic}
\end{algorithm}

\subsection{Optimality and Adaptivity with High-Dimensional Linear Nuisance Models}
\label{section:optimality_adaptivity}

We discuss the difference between the parameter space $\Theta(s)$ defined in \eqref{eq: space}
and that used in \citeapp{cai2017confidence} and comment how to leverage the results in \citeapp{cai2017confidence} to establish the optimality result \eqref{eq: opt length} over the parameter space $\Theta(s).$
The boundedness condition $\lambda_{\rm max}(\Psi)\le C$ in the parameter space $\Theta(s)$  in \eqref{eq: space} implies that the variance of $D_i$ and that of $\delta_i$ are bounded. The parameter space $\Theta(s)$ is a subspace of the parameter space considered in \citeapp{cai2017confidence}, where we additionally require a sparsity condition on the parameter $\eta$ associated with the outcome model. However, the lower bound results in \citeapp{cai2017confidence} are essentially established over the subspace $\Theta(s)$ in \eqref{eq: space} by setting $s_{\eta}=s_{\gamma}$; hence we directly apply Theorem 2 in \citeapp{cai2017confidence} to establish \eqref{eq: opt length}. 
When there is prior information that one of the sparsity levels $s_{\eta}$ and $s_{\gamma}$ is much smaller than the other, the minimax expected length can be better than \eqref{eq: opt length} since the prior information defines a smaller parameter space; see 
\citeapp{JavanmardMontanari2018AoS} for an example. Our proposal can be extended to this setting by adopting their estimator and using the corresponding convergence rate as the filtering radius. We expect the resulting interval to achieve the corresponding minimax expected length.

\section{Justification for Perturbed DML with General Machine Learning}
\label{sec:general theory}

In this subsection, we provide an informal justification of our approach in the case where the nuisances are fitted by more general machine learning models. By ``informal” we mean that the argument rests on strong high-level conditions that are difficult to verify in practice. We include it because it clarifies the mechanism by which the perturbation idea aids valid inference in this more general context. In close analogy to the Lasso case, our analysis suggests that, with high probability, there exists a perturbation yielding an estimator $\hat{\beta}^\mstar$ that approximates the following oracle estimator $\hat{\beta}^{\rm ora}$ at a rate faster than $n^{-1/2}$, 
\begin{align}
    \hat{\beta}^{\rm ora} = \frac{\sum_{i\in\Ical} (D_i-f(X_i))(Y_i - g(X_i))}{\sum_{i\in\Ical}(D_i-f(X_i))^2}. \label{eq: betaHat ora ML}
\end{align}

For theoretical purposes, we assume that the estimated noise covariance matrix $\hat\Pi$ defined in \eqref{eq: resample noise} is computed using the sample $\Ical_0$ that is independent of $\Ical$ and $\Ical^c$. We generate new noise realizations with this $\hat{\Pi}$, after which the remaining steps proceed as before: fitting perturbed nuisance models on the sample $\Ical^c$, and conducting inference on the sample $\Ical$. In practice, such an independent sample $\Ical_0$ is not required and the procedure implemented as described in Section \ref{sec: method general setting} performs well; see Figure \ref{fig: F2F3 summary} for details. 

To facilitate the discussion, we introduce the following notations to highlight the dependence of the fitted nuisance functions on the training data. For the unperturbed ML models  $\hat{g}$ and $\hat{f}$ in \eqref{eq:general optimization}, we emphasize that they are fitted using observations $\{Y_i,D_i,X_i\}_{i\in\Ical^c}$ by writing
$$\hat{g}(\cdot)=\hat{g}(\cdot;\{Y_i,X_i\}_{i\in\Ical^c}), \quad \hat{f}(\cdot)=\hat{f}(\cdot;\{D_{i},X_{i}\}_{i\in\Ical^c}),$$
where $\cdot$ indicates the covariate vector in $\RR^{p}$ that we shall apply the constructed ML to. For the perturbed case, define the perturbed outcome and treatment variables as $Y_i^\m = Y_i - \epsilon_i^\m$ and $D_i^\m = D_i - \delta_i^\m$ with $\epsilon_i^\m$ and $\delta_i^\m$ generated in \eqref{eq: resample noise} for $i\in\Ical^c$. The corresponding perturbed nuisance models $\hat{g}^\m$ and $\hat{f}^\m$ defined in \eqref{eq:general optimization perturbed} are then denoted by $$\hat{g}^\m(\cdot)=\hat{g}(\cdot;\{Y_i^\m,X_i\}_{i\in\Ical^c}), \quad \hat{f}^\m(\cdot)=\hat{f}(\cdot;\{D_{i}^\m,X_{i}\}_{i\in\Ical^c}).$$ 
Note that the above notations explicitly highlights the dependence on the fitted data. 

Let $\P_X$ be the marginal distribution of covariates, and define the out-of-sample prediction error norms of $\widehat{g}$ and $\widehat{g}^{[m]}$:
\begin{align*}
        \|\hat{g} - g\|_{q,\P_X} &:= \l(\E_{X_k\sim \P_X}[(\hat{g}(X_k;\{Y_i,X_i\}_{i\in\Ical^c}) - {g}(X_k) )^q]\r)^{1/q},\\
        \|\hat{g}^{[m]} - g\|_{q,\P_X} &:= \l(\E_{X_k\sim \P_X}[(\hat{g}(X_k;\{Y^{[m]}_i,X_i\}_{i\in\Ical^c}) - {g}(X_k) )^q]\r)^{1/q},
\end{align*}
where $q\geq 1$ is a positive integer and $\E_{X_k\sim \P_X}$ means that we take an expectation over an independent copy $X_k$ generated from the distribution $\P_X$. Analogously, we define $\|\hat{f} - f\|_{q,\P_X}$ and $\|\hat{f}^\m - f\|_{q,\P_X}$ using corresponding fitted nuisance models. 

We now introduce the first assumption on the convergence rate of the ML algorithms.

\begin{assumption}\label{assump:nuisance convergence} (Convergence {and boundedness} of the nuisance learners)
\textcolor{white}{Placeholder} 
\begin{enumerate}[(\textup{B}1)] 
    \item \ There exist positive sequences $\tau_n \to 0$ and $R_{2, g}, \ R_{2, f}, \ R_{4, g}, \ R_{4, f} \to 0$ such that, with probability at least $1-\tau_n$:
    \begin{align*}
        \|\hat{g} - g\|_{2,\P_X} &\lesssim R_{2, f}, \quad \|\hat{f} - f\|_{2,\P_X} \lesssim R_{2, f}; &\|\hat{g} - g\|_{4,\P_X} &\lesssim R_{4, g}, \quad \|\hat{f} - f\|_{4,\P_X} \lesssim R_{4, f}.
    \end{align*}

    \item \ The function classes $\mathcal{G}$ and $\mathcal{F}$ in \eqref{eq:general optimization} are $C$-uniformly bounded. That is, there exists some constant $C>0$ such that for all $\tilde{g}\in\mathcal{G}$ and $\tilde{f}\in\mathcal{F}$,
$     \|\tilde{g}\|_\infty := \sup_{x\in\RR^p} |\tilde{g}(x)| \leq C$ and $\|\tilde{f}\|_\infty := \sup_{x\in\RR^p} |\tilde{f}(x)| \leq C.  
$

    \item \ There exists constant $C>0$ such that for all $x\in\RR^p$,
        $|g(x)| \leq C$ and $|f(x)| \leq C.$
\end{enumerate}
\end{assumption}

Conditions (B1) in Assumption \ref{assump:nuisance convergence} requires that for unperturbed nuisance estimators obtained from the optimization procedures in \eqref{eq:general optimization}, the out-of-sample prediction errors converge in $\ell_q(\P_X)$ norm, $q = 2, 4$, norm with high probability. Such conditions are closely related to Assumption 3.2 in \citeapp{Chernozhukov2018}, which places moment and rate restrictions on the score function. By formulating them directly in terms of the nuisance estimators, Condition (B1) is conceptually aligned with this standard requirement in the DML literature. However, unlike the requirement for building the Wald interval based on the DML estimator, the convergence rates $R_{2,g}$ and $R_{2,f}$ in Condition (B1) are allowed to be slower than $n^{-1/4}$. Convergence rates {are available} for several ML algorithms under certain conditions, including reproducing kernel Hilbert space regression \citepapp[e.g.]{caponnetto2007optimal}, deep neural networks \citepapp[e.g.]{johannes2020nonparametric} and Random Forests \citepapp[e.g.]{scornet2015consistency}. 

{Condition (B2) requires that all models obtained from the optimization problems in \eqref{eq:general optimization} are uniformly bounded.} {This is a condition we impose to derive our theoretical guarantees, and note that a similar boundedness assumption can be found, for example, in Chapter 14 of \citeapp{wainwright2019high}.} 
Condition (B3) assumes the true nuisance functions are uniformly bounded. Such a condition is standard in facilitating nonparametric analysis; see Section 7 in \citeapp{gyorfi2002distribution} and Section 2.5 in \citeapp{tsybakov2008nonparametric}.

We now make an important assumption to justify our perturbation procedure for the ML setting. We assume that with high probability, the out-of-sample prediction vectors of the ML algorithms are Lipschitz continuous with respect to the training response vector.  
\begin{assumption}\label{assump:ml lipschitz}(Lipschitz continuity of the nuisance learners)
    For any two outcome variables $Y_i,Y_i'\in\RR$ and treatment variables $D_i,D_i'\in\RR$ from the sample $\Ical^c$, there exist positive sequences $L_g,L_f>0$ and  $\tau_n\to0$ such that with probability at least $1-\tau_n$,
    \begin{align*}
        \sum_{k\in\Ical}\l(\hat{g}(X_k;\{Y_i,X_i\}_{i\in\Ical^c}) - \hat{g}(X_k;\{Y_i',X_i\}_{i\in\Ical^c})\r)^2 &\leq L_g \sum_{i\in\Ical^c}(Y_i-Y'_i)^2, \\
        \sum_{k\in\Ical}\l(\hat{f}(X_k;\{D_i,X_i\}_{i\in\Ical^c}) - \hat{f}(X_k;\{D_i',X_i\}_{i\in\Ical^c})\r)^2 &\leq L_f \sum_{i\in\Ical^c}(D_i-D'_i)^2.
    \end{align*}
\end{assumption}

Assumption \ref{assump:ml lipschitz} requires the nuisance learners to satisfy a Lipschitz condition with respect to the outcome. Although not verified for all machine learning models, such a Lipschitz condition ensures the model’s stability, in that the predicted values do not change dramatically in response to small outcome perturbations.

In the ordinary least squares regression, if $n>p$ and the design matrix $X_{\Ical^c}$ has full column rank, then the Lipschitz constants $L_g$ and $L_f$ can be taken as 
$\|X_\Ical(X_{\Ical^c}^\T X_{\Ical^c})^{-1}X_{\Ical^c}^\T\|_\op$. When the covariates are Subgaussian, the existing random matrix theory (e.g. Theorem 4.4.3 and 4.6.1 in \citeapp{vershynin2018high}) implies that, with high probability, $L_g$ and $L_f$ are of order $(\sqrt{n}+\sqrt{p})/(\sqrt{n}-\sqrt{p})$. {When $n>Cp$ for a positive integer $0<C<1$, we have $L_g$ and $L_f$ are of constant orders.}
In Lasso regression, Theorem 3.1 in \citeapp{meng2024lipschitz} establishes that for fixed training covariates $X_{\Ical^c}$, the Lasso estimator is Lipschitz in the response vector, but their proof relies on the geometry of the solution set and a closed-form expression for the Lipschitz constant is not provided. Consequently, while the existence of $L_g$ and $L_f$ is guaranteed for any fixed $X_{\Ical^c}$, it remains unclear how these constants depend on the random matrix $X_{\Ical^c}$ and whether we can obtain high-probability bounds for $L_g$ and $L_f$ as $n$ and $p$ grow. 

To state the theorem, we further need to quantify how likely the oracle estimator $\hat{\beta}^{\rm ora}$ lies within the conditional distribution of $\hat{\beta}^\m$ given observed data $\mathcal{O}$. Note that $\hat{\beta}^{\rm ora}$ is a random variable depending on the observed data $\mathcal{O}$. Since $\hat{\beta}^{\rm ora}$ is a sample analog of $\beta$ and $\hat{\beta}^{\rm ora}-\beta$ is asymptotically normal, we capture the  fluctuation of $\hat{\beta}^{\rm ora}$ around $\beta$ by considering the following interval, 
\begin{align}
    T_0 = \l[ \beta - z_{\alpha_0/2}\sqrt{{\Var(\varphi(O_i;\beta))}/{n}}, \beta + z_{\alpha_0/2}\sqrt{{\Var(\varphi(O_i;\beta))}/{n}} \r], \label{eq:interval betaHat ora}
\end{align}
where $\alpha_0\in(0,0.01]$ is the same small positive constant. The constant $\alpha_0$ is used throughout the paper to control the probability of rare events. In the Lasso case, introducing $\alpha_0$ ensures that the observed noise norms, $\|\xi\|_2$ and $\|\kappa\|_2$, do not fall in the tails of their distributions, so that there is a nonzero chance that the artificial noise generating step can nearly recover the true noise vectors. In the general ML case, $\alpha_0$ accounts for the probability of $\hat{\beta}^{\rm ora}$ not falling into $T_0$. Note that $T_0$ contains $\hat{\beta}^{\rm ora}$ with probability larger than $1-\alpha_0$, that is, $\liminf_{n\to\infty}\P(\hat{\beta}^{\rm ora} \in T_0) = 1-\alpha_0$. Given this fixed interval $T_0$, we make the following assumption that the conditional distribution of $\hat{\beta}^\m$ covers the interval $T_0$ with a positive probability. This means that, when $\hat{\beta}^{\rm ora}$ does not show up in its own tail (i.e., falling inside $T_0$), $\hat{\beta}^{\rm ora}$ falls into the support of the conditional distribution of $\hat{\beta}^\m$; see Figure \ref{fig:assumption illustration} for an illustration. 

\begin{assumption}\label{assump:resample distribution} The interval $T_0$ defined in \eqref{eq:interval betaHat ora} lies strictly within the conditional support of the perturbed target estimators $\hat{\beta}^\m$, that is, 
\({\alpha}_{T_0} = \min\{\P(\hat{\beta}^\m < v_1 \mid \mathcal{O}), \;\; \P(\hat{\beta}^\m \allowbreak > v_2 \mid \mathcal{O})\}>0,\)
where $v_1$ and $v_2$ denote the lower and upper ends of the interval $T_0$ and $\mathcal{O}$ denotes the observed data.
\end{assumption}
The quantity $\alpha_{T_0}$ defined in Assumption \ref{assump:resample distribution} captures the smallest conditional tail probability that $\hat{\beta}^\m$ assigns to points in $T_0$. By requiring this quantity to be positive, Assumption \ref{assump:resample distribution} rules out cases where $T_0$ lies partly or entirely out of the support of the conditional distribution of $\widehat{\beta}^{[m]}$. {Figure \ref{fig:assumption illustration} illustrates a situation where the assumption is satisfied. Starting from the generated noises $\{\epsilon_i^\m, \delta_i^\m\}_{i \in \Ical^c}$, we construct perturbed nuisance estimators, which in turn yield the perturbed target estimator $\hat{\beta}^\m$. The mapping from the high-dimensional noise space to the real-valued $\hat{\beta}^\m$ may be highly nonlinear and many-to-one (blue paths). The induced distribution of $\hat{\beta}^m$ is required to cover the entire interval $T_0$ (orange segment), ensuring that the tail probability $\alpha_{T_0}$ (shadowed gray area) is strictly positive.} {Assumption \ref{assump:resample distribution} is indeed a strong assumption imposed to facilitate our analysis and we acknowledge that the tail probability $\alpha_{T_0}$ may depend on $n$, $p$ and function classes of $g$ and $f$.}

\begin{figure}[ht]
\centering
\begin{tikzpicture}

\coordinate (E) at (-4.0,0.7); 
\draw[very thick] (E) ellipse [x radius=1.2cm, y radius=0.8cm];
\node[align=left] at (-4.0,2.0) {Space of $\{\epsilon_i^\m, \delta_i^\m\}_{i\in\Ical^c}$ };

\fill[blue]             (-4.9,0.7)  circle (2pt);
\fill[blue]             (-3.4,0.6)  circle (2pt);
\fill[brown!80!black]   (-4.3,0.3)  circle (2pt);

\begin{axis}[
  name=plot,
  at={(0,0)}, anchor=origin,
  domain=0:10, samples=300,
  axis lines=none,          
  xtick=\empty, ytick=\empty,
  width=8cm, height=3.5cm,   
  clip=false
]

\pgfmathdeclarefunction{dens}{1}{%
  \pgfmathparse{0.9*exp(-((#1-3)^2)/2.0) + 0.6*exp(-((#1-7)^2)/3.0)}%
}

\def\Tzero{6}
\def\alphaval{1.5}
\pgfmathsetmacro{\alphat}{\alphaval*\Tzero}
\pgfmathsetmacro{\densalpha}{dens(\alphat)}

\addplot[very thick,black,domain=0:10,samples=300] {dens(x)};

\draw[thick] (axis cs:0,0) -- (axis cs:10,0);

\draw[fill=orange!70,draw=none,opacity=0.8]
  (axis cs:\Tzero,-0.05) rectangle (axis cs:\alphat,0.05);

\addplot[domain=\alphat:10,fill=lightgray,opacity=0.5]
  {dens(x)} -- (axis cs:10,0) -- (axis cs:\alphat,0) -- cycle;

\addplot[densely dotted, thick] coordinates {(\alphat,0) (\alphat,\densalpha)};

\node[below] at (axis cs:{(\Tzero+\alphat)/2},0) {$T_0$};
\node[below] at (axis cs:{\alphat+0.5},0) {$\alpha_{T_0}$};

\node[align=center] at (rel axis cs:0.55,1.15)
  {Conditional distribution of $\hat{\beta}^{[m]}$};

\coordinate (blueTarget) at (axis cs:4.0,0);
\coordinate (redTarget)  at (axis cs:6.7,0);

\fill[blue]             (blueTarget) circle (2pt);
\fill[brown!80!black]   (redTarget)  circle (2pt);

\end{axis}

\draw[blue, dashed, very thick, -{Stealth[length=3mm]}]
  (-4.9,0.7) .. controls (-2.8,1.8) and (-0.6,1.0) .. (blueTarget);
\draw[blue, dashed, very thick, -{Stealth[length=3mm]}]
  (-3.4,0.6) .. controls (-1.8,1.0) and (0.2,0.6) .. (blueTarget);

\draw[brown!70!black, dashed, very thick, -{Stealth[length=3mm]}]
  (-4.3,0.3) .. controls (-2.3,-1.0) and (0.5,-0.8) .. (redTarget);

\end{tikzpicture}
\caption{Illustration of Assumption \ref{assump:resample distribution}. The orange segment represents the interval $T_0$ defined in \eqref{eq:interval betaHat ora}. The shadowed grey area refers to the tail probability $\alpha_{T_0}$ defined in Assumption \ref{assump:resample distribution}.}
\label{fig:assumption illustration}
\end{figure}

Under Assumptions \ref{assump:nuisance convergence} to \ref{assump:resample distribution}, we next show that, as the number of perturbation $M$ grows, at least one perturbed estimator $\hat{\beta}^\m$ will lie arbitrarily close to $\hat{\beta}^{\rm ora}$. The key technical ingredient is an isoperimetric inequality for the Gaussian distribution, combined with the Lipschitz continuity in Assumption \ref{assump:ml lipschitz}, which ensures that the probability of $\hat{\beta}^{\m}$ does not vanish on any arbitrarily small interval{inside $T_0$ with a sufficiently large $M$}. This rules out the case where the density of $\hat{\beta}^\m$ would drop to zero within the interior of its support. 

Similarly to \eqref{eq: supp def err}, we define
\begin{align*}
    \errml = \max\{L_g,L_f\} \cdot \frac{1}{\alpha_{T_0}^2\sqrt{n}} \l( \l\lfloor\frac{(1-2\alpha_{T_0})M}{\log(\sqrt{n}M)}\r\rfloor \r)^{-1}
\end{align*}
to measure the minimum distance $|\hat{\beta}^\m-\hat{\beta}^{\rm ora}|$ among all $1\leq m\leq M.$ {Here, we slightly abuse the notation $\err$ by replacing $\alpha_0$ with $\alpha_{T_0}$ to highlight the analogous roles of $\err$ and $\errml$ in characterizing the mininum distance $|\hat{\beta}^\m - \hat{\beta}^{\rm ora}|$. Nevertheless, the two constants $\alpha_0$ and $\alpha_{T_0}$ have distinct interpretations: $\alpha_0$ corresponds to the tail probability of the observed data, whereas $\alpha_{T_0}$ reflects the smallest tail probability of $T_0$ assigned by the conditional distribution of $\hat{\beta}^\m$.}
Note that the rate $\errml$ vanishes when $M\to\infty$ but its scale gets larger if $\alpha_{T_0}$ is close to zero. With this rate, the following theorem establishes that among all $M$ perturbed estimates, at least one of them nearly recovers $\hat{\beta}^{\rm ora}$ with high probability. 
\begin{theorem}\label{thm:mstar ml beta}
    Suppose Assumptions \ref{assump:nuisance convergence}, \ref{assump:ml lipschitz} and \ref{assump:resample distribution} hold. Then there exists some constant $\bar{C}>0$ such that
    \begin{align*}
        \liminf_{n\to\infty}\liminf_{M\to\infty}\P\l( \exists 1\leq m\leq M: |\hat{\beta}^\m - \hat{\beta}^{\rm ora}| \leq \bar{C}\cdot\errml \r) \geq  1-\alpha_0,
    \end{align*}
    where $\alpha_0\in(0,0.01]$ is a small positive constant specified in \eqref{eq:interval betaHat ora}, ${\alpha}_{T_0}$ is defined in Assumption \ref{assump:resample distribution}, and $\hat{\beta}^{\rm ora}$ is defined in \eqref{eq: betaHat ora}. {Consequently, the confidence interval CI defined in \eqref{eq:filtered union CI} satisfies}
    \begin{align*}
        \liminf_{n \to \infty} \liminf_{M \to \infty} \Prob\left(\beta \in \rm{CI}\right) \geq 1 - \alpha, \quad \liminf_{n\to\infty}\liminf_{M\to\infty} \P\l( \len(\rm CI) \leq 2.02 \rho_n + (4 + c) \sigma_\beta / \sqrt{n} \r)
    \end{align*}
{where $\rho_n \asymp R_{2,f}(R_{2,g} + R_{2,f})$ is a high-probability upper bound on the conditional bias term $|T_n|$ defined in \eqref{eq:Tn}, $\sigma_\beta = \sqrt{\Var\{\varphi(O_i; \beta)\}}$ and $c$ is an arbitrarily small positive constant.}
\end{theorem}

Compared with the convergence result \eqref{eq: ora convergence}, both Theorem \ref{thm: mstar lasso} and \ref{thm:mstar ml beta} require $M\to\infty$, but the dependence of the minimum distance between $\hat{\beta}^\m$ and $\hat{\beta}^{\rm ora}$ on $M$ differs due to different proof strategies. In Theorem \ref{thm: mstar lasso}, the minimum distance shrinks at the rate $M^{-1/(2p)}$ which deteriorates quickly as the dimension $p$ grows, while Theorem \ref{thm:mstar ml beta} converges at the rate of $\log M/(\alpha_{T_0}^2(1-2\alpha_{T_0})M)$. 
Although this new rate appears to have a better dependence on $M$, a small tail probability $\alpha_{T_0}$ may still lead to a large number of perturbations $M$. With these caveats, Theorem \ref{thm:mstar ml beta} nonetheless suggests how the perturbation step can facilitate inference for functionals when modern machine learning estimators are used.

\section{Additional Simulation Results}

\subsection{Linear Models}\label{sec:additional_simulation}
In the following, we present the simulation results in the setting where the nuisance functions are linear models.

\vspace{2mm}
\noindent \textbf{F3 (Linear Nuisance Models):} $f(X_i) = X_{i}^{\intercal}{\gamma}$ and $h(X_i) = X_{i}^{\intercal}{\mu}$ with ${\gamma} = (\gamma_1, \dots, \gamma_p)^{\intercal}$ and ${\mu} = (\mu_1, \dots, \mu_p)^{\intercal}$ where $p$ varies from 5 to 240. The coefficients $\gamma_j$ and $\mu_j$ for $j=1,\dots,p$ are independently sampled from the uniform distribution on $(0,1)$. 

\begin{figure}[ht!]
    \centering
    \includegraphics[width=0.95\linewidth]{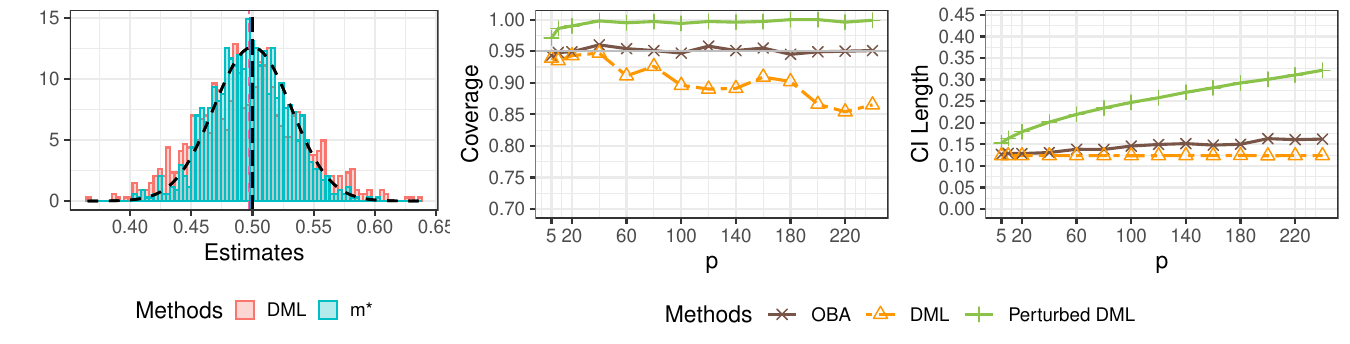}
    \caption{\small Setting F3 with $n=1000$ and $p$ from 5 to 240. The leftmost subfigure compares the empirical distributions of $\hat{\beta}^\mstar$ and $\hat{\beta}$ when $p=240$, {where the black dashed curve represents the reference distribution $N(\beta,n^{-1}\Var\{\varphi(O_i;\beta)\})$.} The middle and rightmost subfigures illustrate empirical coverages and {average lengths of confidence intervals based on {OBA}, {DML} and Perturbed DML.}
    }
    \label{fig: F3 summary}
\end{figure}

Figure \ref{fig: F3 summary} summarizes the performance of our proposed method under the setting F3. As shown in the leftmost subfigure, both $\hat{\beta}^\mstar$ and $\hat{\beta}$ are unbiased, but $\hat{\beta}$ has slightly inflated variance relative to the reference distribution $N(0, n^{-1}\Var\{\varphi(O_i;\beta)\})$, which is the theoretical limiting distribution of the central limit term $Z_n$. This is likely due to inaccurate nuisance estimation when $p=240$ is large relative to $n=1000$. In contrast, $\hat{\beta}^\mstar$ displays comparable variance to the reference variance  $n^{-1}\Var\{\varphi(O_i; \beta)\}$, demonstrating how a favorable injection of simulated noise can lead to much more accurate inference. The inflated variance of $\hat{\beta}$ explains the degrade in coverage as $p$ increases (middle subfigure). Conversely, our proposed filtered union confidence interval maintains the coverage above 95\% across increasing values of $p$. Finally, the rightmost subfigure indicates that the proposed CI is not overly conservative compared to the OBA CI. 

\subsection{Heteroscedastic settings}
\label{sec:hetero}

\begin{figure}
    \centering
    \includegraphics[width=0.95\linewidth]{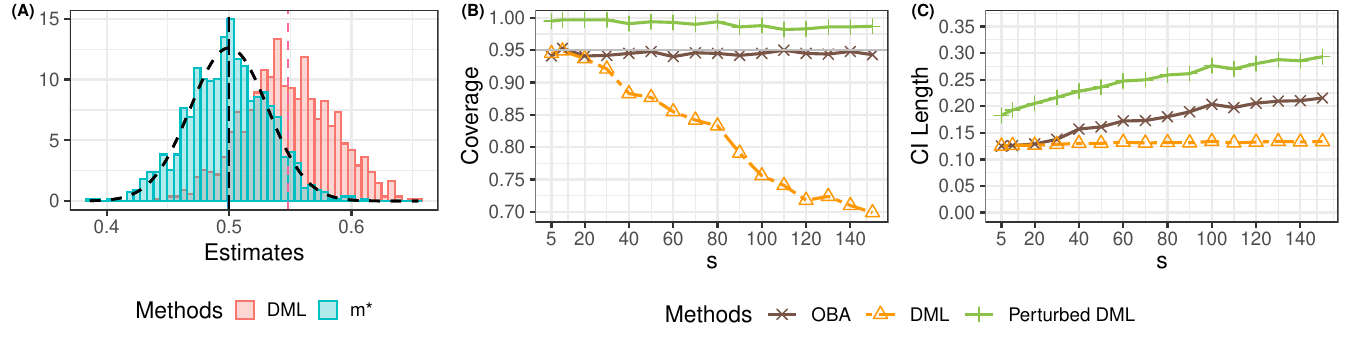}
    \caption{\small Comparison of the oracle bias-aware procedure, the standard DML and our proposal with varying sparsity levels in Example 1 with $n=1000$, $p=500$, $\pi^*=0.95$, $M=500$ and heteroscedastic noises. (A): Comparison of empirical distributions of $\hat{\beta}^\mstar$ and $\hat{\beta}$ when $s=150$, {where the black dashed curve represents the reference distribution $N(0,n^{-1}\Var\{\varphi(O_i;\beta)\})$.} (B): Empirical coverages of the CIs. (C): Average interval lengths. Results are averaged across 1000 simulations.}
    \label{fig:hetero}
\end{figure}

In this section, we show that our proposal is robust to heteroscedastic noises. 
We retain the data-generating process of Example 1 but replace the homoscedastic errors by heteroscedastic errors with covariate-dependent variances.
Specifically, for the $i$-th observation, let $\tilde{f}(X_i)$ and $\tilde{h}(X_i)$ denote the the standardised nuisance components $f(X_i)$ and $h(X_i)$. We generate 
\begin{align*}
    \delta_i = \sigma_{\delta_i}Z_{\delta,i}, \quad e_i = \sigma_{e,i} Z_{e,i} \quad \textrm{with} \ Z_{\delta,i}, Z_{e,i} \overset{i.i.d.}{\sim} N(0,1)
\end{align*}
with heteroscedastic standard deviations $\sigma_{\delta,i} \propto 0.5 + 0.8|\tilde{f}(X_i)|$ and $\sigma_{e,i} \propto 0.5 + 0.8|\tilde{h}(X_i)|$.
The proportionality constants are chosen so that $n^{-1} \sum_{i=1}^n \sigma_{\delta,i}^2 = n^{-1} \sum_{i=1}^n \sigma_{e,i}^2 = 1$, which matches the average error variance in the homoscedastic setting for better comparison. 
Figure \ref{fig:hetero} presents the empirical performance of our procedure implemented with $M=500,\pi^*=0.95$ and Lasso nuisance learners. Across all three panels, the performance closely mirrors the homoscedastic case in Figure \ref{fig: F1 summary}, suggesting the robustness of our proposal to heteroscedasticity.

\subsection{Perturbed noise generation with heavy-tailed distributions}
\label{sec:heavy_tailed_resamp}

We next illustrate that the perturbation step does not rely on Gaussian noise. 
The key requirement is that the perturbation distribution be sufficiently dispersed so that, with high probability, at least one perturbed nuisance estimate is close to the truth.
With the default hyperparameters $M=500$ and $\pi^*=0.95$, we implement our proposal using perturbations drawn from Gaussian, Laplace, symmetric lognormal, and Student's-t distributions with degrees of freedom (df) 3 and 10. 
In all cases, the perturbations are scaled to match the estimated covariance matrix $\hat{\Pi}$ defined in \eqref{eq: resample noise}, with the marginal law adjusted to the chosen family.

\begin{figure}[ht]
    \centering
    \includegraphics[width=0.95\linewidth]{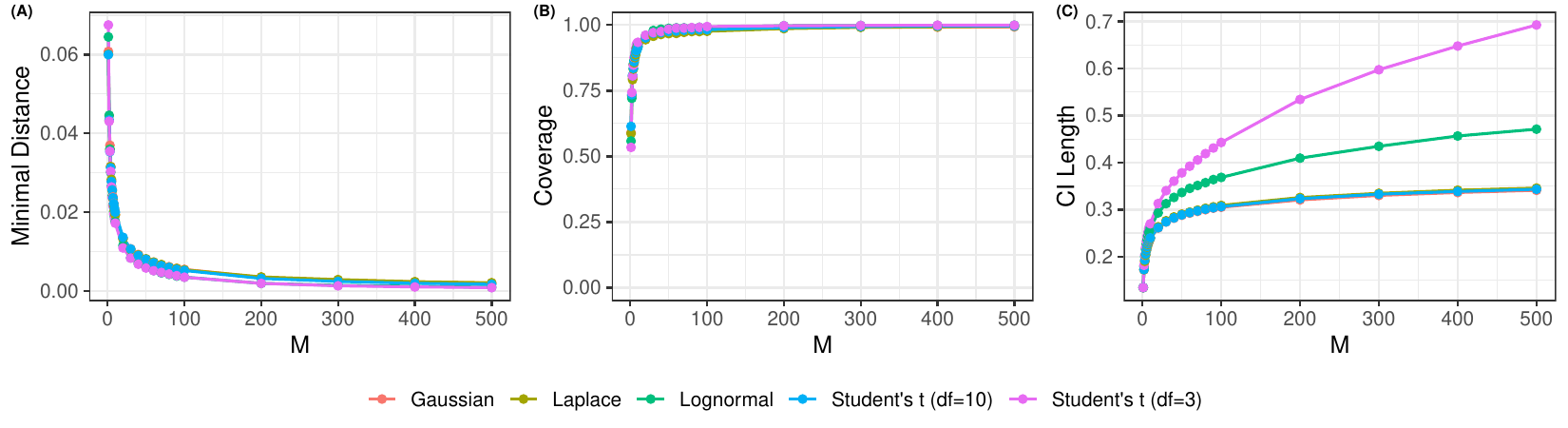}
    \caption{\small Empirical performance under Example 1 with $s=100$ using perturbation noises drawn from heavy-tailed families. (A): Minimal distance $\min_{1\leq m\leq M}|\hat{\beta}^\m - \hat{\beta}^{\rm ora}|$ with increasing $M$. (B): Empirical coverage. (C): Average length of Perturbed DML CIs. Results are averaged across 1000 simulations.}
    \label{fig:heavy-tailed}
\end{figure}

Figure \ref{fig:heavy-tailed} reports the resulting performance. In Panel (A) and (B), all distributions yield comparable values of the minimal distance $\min_{1\leq m\leq M}|\hat{\beta}^\m - \hat{\beta}^{\rm ora}|$ and similar empirical coverage, suggesting the feasibility of using non-Gaussian distributions to generate perturbation noises in our proposal. Panel (C) highlights how different distributions affect the efficiency: heavier-tailed perturbations tend to produce wider intervals. Particularly, the Student's-t distribution with df=10 behaves similarly to the Gaussian benchmark, whereas df=3 generates substantially longer intervals, consistent with its much heavier tails.

\subsection{Noise generation with heavy-tailed distributions in data generating process}
\label{sec:heavy_tailed_dgp}

In this section, we consider heavy-tailed errors in the data-generating process while retaining Gaussian perturbations. We normalise the error variance to one and draw the noise terms from Laplace, symmetric lognormal and Student-t distributions with degrees of freedom 3 and 10. In all cases, perturbed DML is implemented with Gaussian perturbations. 

\begin{figure}
    \centering
    \includegraphics[width=0.95\linewidth]{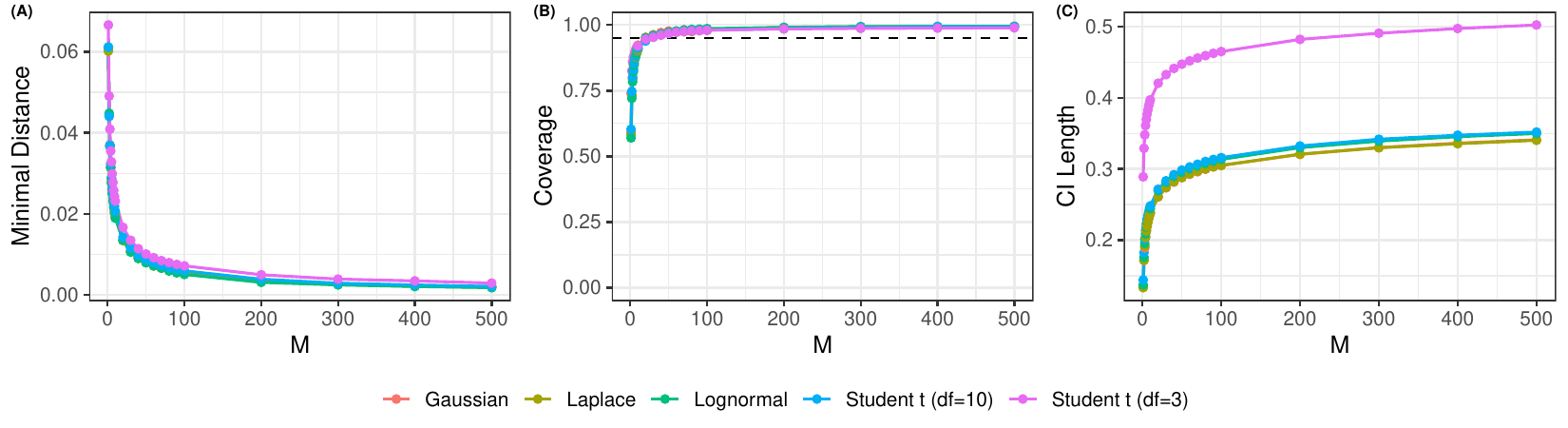}
    \caption{\small Empirical performance under Example 1 with $s=100$ and noises in data drawn from heavy-tailed families. (A): Minimal distance $\min_{1\leq m\leq M}|\hat{\beta}^\m - \hat{\beta}^{\rm ora}|$ with increasing $M$. (B): Empirical coverage. (C): Average length of Perturbed DML CIs. Results are averaged across 1000 simulations.}
    \label{fig:heavy-tailed-dgp}
\end{figure}

Figure \ref{fig:heavy-tailed-dgp} reports the minimal distance $\min_{1\leq m\leq M}|\hat{\beta}^\m - \hat{\beta}^{\rm ora}|$, empirical coverage and average CI length of our proposal under these settings. Panel (A) shows that the minimal distance decreases rapidly toward zero as $M$ increases across all settings, with the heaviest-tailed case (Student-t with df=3) converging slightly more slowly. Consistent with this, Panel (B) indicates that nominal coverage is achieved for moderate $M$ in every design. In Panel (C), Student-t distribution with df=3 yields substantially wider intervals, approximately 43\% longer than the other distributions. Although the noise variance is fixed at one by construction, heavy tails induce more extreme residuals and hence more variable and often larger finite-sample variance estimates. This inflates both the estimated standard error---reflected by different CI length at $M=1$---and the estimated scale used for Gaussian perturbations, leading to longer confidence intervals.

\subsection{Sensitivity to Choices of Tuning Parameters}
\label{sec: sensitivity to tuning}

In this section, we assess the sensitivity of the proposed method to the choices of perturbation size $M$ and filtering proportion $\pi^*$. {{As discussed in Section \ref{sec: method section}}, the tuning parameters $\lambda^\m_\eta$ and $\lambda_\gamma^\m$ in perturbed optimizations are chosen in data-driven ways (e.g., cross-validation).} When the perturbation size $M$ is small, the proposed procedure may fail to produce perturbed nuisance estimators close enough to true nuisance models. Similarly, when the filtering proportion $\pi^*$ is too small, for example $\pi^*\leq 0.9$, our procedure risks discarding perturbations that yield accurate estimates, thereby compromising coverage. 

We vary the perturbation size $M$ from 10 to 1300 and the filtering proportion $\pi^*$ from 85\% to 100\%. When evaluating the sensitivity to $M$, we set $\pi^*=95\%$, while when evaluating the sensitivity to $\pi^*$, we set $M = 500$. We consider Example 1 with $s=120$, and F1 and F2 with $p=20$. In these settings, as shown in Section \ref{sec: model} and \ref{sec:simulation}, the standard DML estimator exhibits large bias and inflated variance due to poor nuisance estimation. 

\begin{figure}[ht!]
    \centering
    \includegraphics[width=0.675\linewidth]{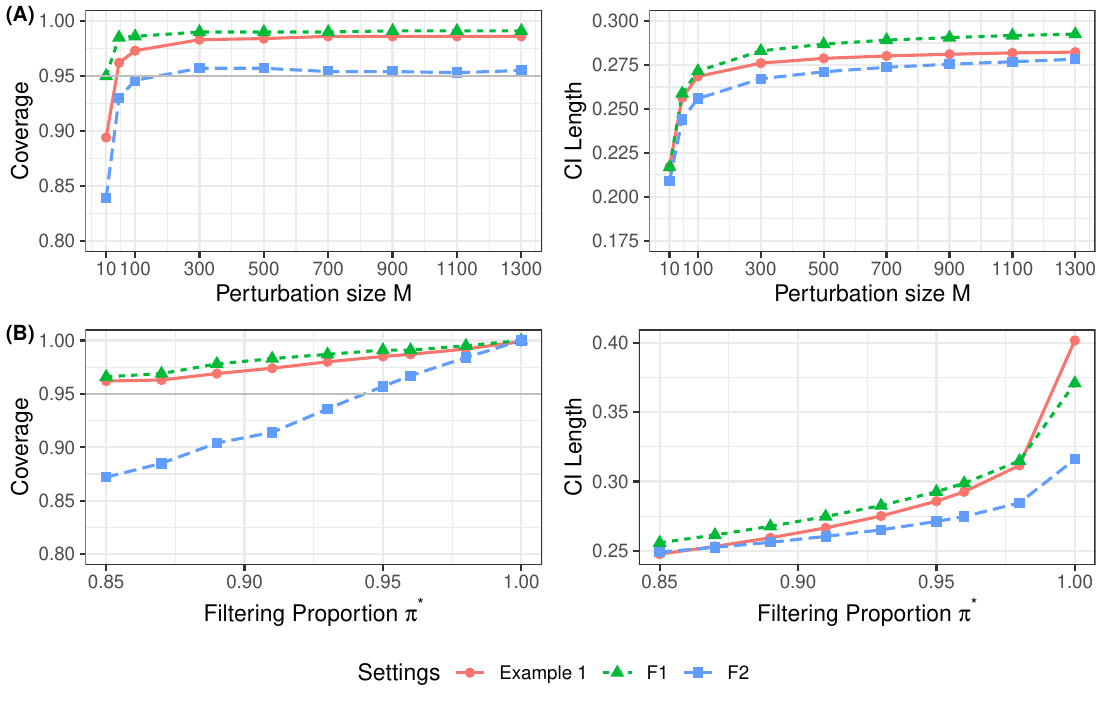}
    \caption{\small Sensitivity of Perturbed DML to the tuning parameters: (A) the perturbation size $M$; (B) the filtering proportion $\pi^*$. The left and the right columns show the empirical coverage and the average CI length of our proposal in Example 1 with $s=120$, F1 and F2 with $p=20$.}
    \label{fig: tuning primary}
\end{figure}

Figure \ref{fig: tuning primary} demonstrates that the proposed method exhibits robust performance when the perturbation size $M\geq100$ and the filtering proportion $\pi^*\geq0.95$. In panel (A), our proposal has coverage as soon as $M\geq 100$ across all settings. Notably, further enlarging $M$ beyond 100 results in only a marginal increase in CI length, suggesting that additional perturbations do not result in substantial efficiency loss. 
Panel (B) shows that our method has coverage as long as $\pi^* \geq 95\%$. As expected, the CI length increases with $\pi^*$ since more Wald intervals are retained in the filtered union $\mathcal{M}$. When $\pi^*=1$, the CI length becomes longer by 15\%-40\% compared to that based on $\pi^*=95\%$ across settings.

\section{Additional details on the data analysis}
\subsection{Implementation}\label{section:data_analysis_implementation}
We use nine out of ten nuisance learners as used in \citeapp{chernozhukov2024applied}, including linear regressions (baseline and full covariate sets), Lasso, Ridge, Elastic Net, Random Forests, XGBoost with decision trees, and deep neural networks (DNNs) with dropout and early stopping. In contrast to \citeapp{chernozhukov2024applied}, we explicitly tune the hyperparameters of all learners, except for the two linear regressions, 
 via cross-validation to optimize the nuisance predictions. Furthermore, we introduce a neural network architecture wider than the one used in \citeapp{chernozhukov2024applied}, featuring four hidden layers (with 512, 256, 128, and 64 nodes, respectively) and L2 regularization. 

We implement standard DML with five-fold cross-fitting based on one sample split.
For Perturbed DML, we specify a perturbation size of $M=500$ and a filtering proportion of $\pi^*=0.95$. The hyperparameter tuning strategy for Perturbed DML varies by learner: for penalized regressions (Lasso, Ridge and Elastic Net), we tune the perturbation-specific penalty parameters as discussed in Section \ref{sec: method section}; for tree-based methods and neural networks, we forgo additional tuning and fix hyperparameters to the values selected during the initial cross-validation to shorten computation time.
\subsection{Additional Results}
\label{sec:data_additional}

We report the CIs from standard and Perturbed DML using the perturbation size $M=5000$ and the filtering proportion $\pi^*=0.95$ for the real data analysis. 
The observed patterns are consistent with those found in the $M=500$ implementation.  
Excluding the two DNN variants, Perturbed DML CIs are, on average, approximately 68\% longer than standard DML CIs. 
For the DNN-based learners, the proposed CIs are 2.3 to 6.6 times wider than the standard Wald intervals.
Notably, increasing the perturbation size does not result in excessively wide CIs.
This demonstrates that our proposal maintains a certain degree of statistical efficiency while achieving robustness against slow-converging nuisance parameter estimation. 

\begin{figure}[htp]
    \centering
    \includegraphics[width=0.6\linewidth]{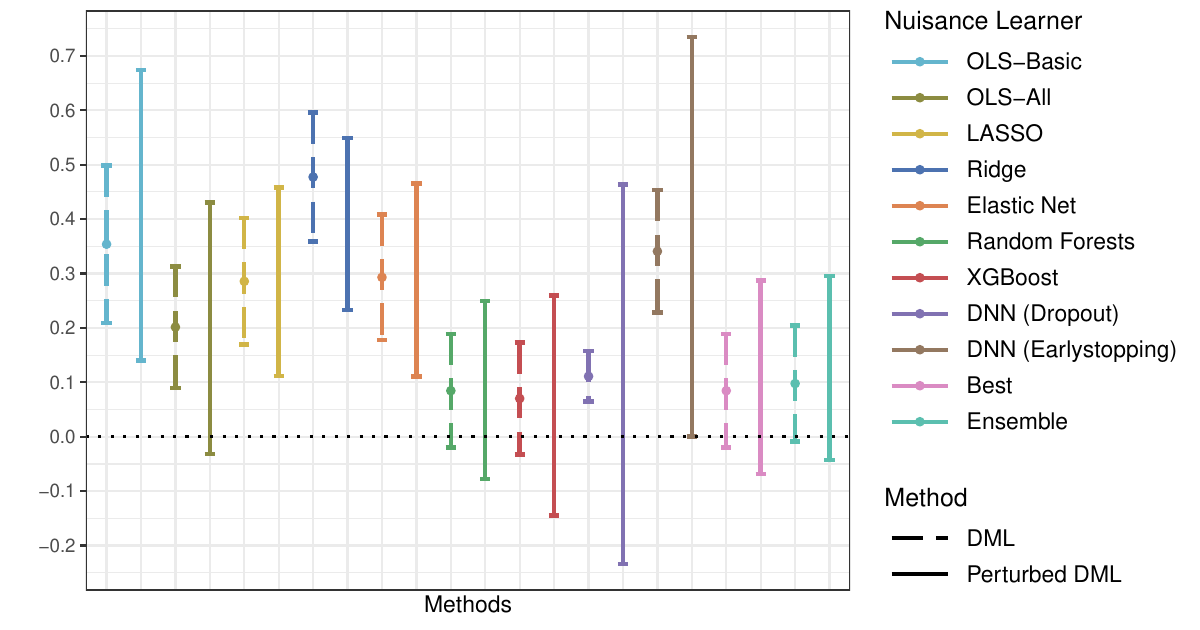}
    \caption{CIs from standard DML and Perturbed DML with $M=5000$. The ``Best'' learner selects the learner with the lowest RMSEs for each of $Y_{i,t}$ and $D_{i,t-1}$. The nuisance learner ``Ensemble'' uses the linear combination of all learners that achieves the lowest RMSEs for each of $Y_{i,t}$ and $D_{i,t-1}$. }
    \label{fig:data_M5000}
\end{figure}

\section{Proofs}

\subsection{Proof of Theorem \ref{thm: mstar lasso}}

\subsubsection{Preliminaries and notation}\label{sec: notation in Thm1}
Before starting the proof, we first restate Theorem \ref{thm: mstar lasso} by including an additional statement for the nuisance estimators. 

\begin{theorem}[Theorem 1 in the paper]
     Suppose Assumption \ref{assumption:main_lasso} holds and the penalty parameters $\lambda_\eta^\m$ and $\lambda_\gamma^\m$ in \eqref{eq: lasso optimization problem eta m} satisfy {$\lambda_\eta^\m = C  n^{-1/2}\err$ and $\lambda_\gamma^\m = C n^{-1/2}\err$} for some constant $C>1$. There exists some constant $C'>0${ independent of $n$ and $p$ such that}
    \begin{align*}
        \liminf_{n,p\to\infty}\liminf_{M\to\infty}\Prob \bigg(  
        \exists m \in \{1,\dots,M\}: 
        \|\widehat\eta^\m - \eta\|_2 & \leq C'\sqrt{\frac{s_\eta}{n}} \cdot \err,  \\
        \|\widehat\gamma^\m - \gamma\|_2 & \leq C'\sqrt{\frac{s_\gamma}{n}} \cdot \err 
        \bigg) \geq 1-\alpha_0,
    \end{align*}
    where $\err$ is defined in \eqref{eq: supp def err}. 
   Consequently, there exists some other constant $C'>0${ independent of $n$ and $p$ such that}
   \begin{equation*}
    \begin{aligned}
        \liminf_{n,p\to\infty}\liminf_{M\to\infty}&\P\bigg(\exists m\in \{1,\dots,M\}: |\hat{\beta}^\m - \hat{\beta}^{\rm ora}| \\
        & \leq C'\bigg(\frac{\sqrt{s_\eta}+\sqrt{s_\gamma}}{n}\err + \frac{\sqrt{s_\eta s_\gamma}+s_\gamma}{n}\err^2\bigg)\bigg) \geq 1-\alpha_0,
    \end{aligned}
    \end{equation*}
where the oracle DML estimator $\hat{\beta}^{\rm ora}$ is defined in \eqref{eq: betaHat ora}.
\end{theorem}

The additional statement for nuisance estimations in the above formally states that our procedure yields, with high probability, one pair of nuisance estimates $\hat{\eta}^\m$ and $\hat{\gamma}^\m$ such that their distances to the true nuisance parameters are at most a constant multiple of $ \sqrt{s_\eta / n} \cdot \err$ and $ \sqrt{s_\gamma / n} \cdot \err$, respectively. In contrast, the unperturbed Lasso estimator would satisfy, for example, a convergence rate for $\|\hat{\eta}-\eta\|_2$ of order $\sqrt{s_\eta / n} \cdot \|\xi\|_\infty \lesssim \sqrt{s_\eta \log p / n}$ with high probability (see \citeapp{bickel2009simultaneous} and \citeapp{zhou2009restricted}). In this light, for the $m^*$-th perturbation, the convergence rate is considerably faster than that achieved by the unperturbed Lasso optimization for a large $M$. {The fast convergence rate of nuisance estimations in the $m^*$-th perturbation translates to the closeness between the induced estimator $\hat{\beta}^\mstar$ and the oracle estimator $\hat{\beta}^{\rm ora}$, as established in \eqref{eq: ora convergence}.} 

We next start proving the results from the nuisance estimation and then derive the minimal distance between $\hat{\beta}^\m$ and $\hat{\beta}^{\rm ora}$ from that. Throughout this proof, we assume, for simplicity, that sample splitting is performed. That is, we suppose that all nuisance functions are estimated on fold $\Ical^c$ and fold $\Ical$ is used to compute $\widehat\beta$ and conduct inference. Both folds are assumed to be of size $n$. The arguments below go through if cross-fitting is performed as long as the number of folds is a constant independent of the sample size. 

Recall the notation $\xi = n^{-1/2}\sum_{i \in \Ical^c}X_i\epsilon_i$ and $\kappa = n^{-1/2} \sum_{i \in \Ical^c} X_i \delta_i$. We denote by $\widehat\eta$ and $\widehat\gamma$ the original, unperturbed Lasso estimates computed on sample $\Ical^c$. Further, conditioning on the observed data, let $\xi^\m \sim \mathcal{N}(\mathbf{0}, \hat{\Sigma}+\nu I)$ with $\hat{\Sigma} = n^{-1}\sum_{i\in\Ical^c}(Y_i - X_i^\T\hat{\eta})^2X_i X_i^\T$, and $\kappa^\m \sim \mathcal{N}(\mathbf{0}, \hat{\Lambda}+\nu I)$ with $\hat{\Lambda} = n^{-1}\sum_{i\in\Ical^c}(D_i - X_i^\T\hat{\gamma})^2X_i X_i^\T$. On fold $\Ical^c$, we solve the following Lasso optimizations:
\begin{align}
    & \hat\eta^\m = \argmin_{u\in\RR^p} u^{\intercal} \left(\frac{1}{2n} \sum_{i \in \Ical^c} X_i X_i^\T\right) u - u^{\intercal} \left(\frac{1}{n} \sum_{i \in \Ical^c} X_i Y_i - n^{-1/2} \xi^\m\right) +\lambda^\m_\eta \|\eta\|_1, \label{eq: supp thetaHat m def_eta}\\
    & \widehat\gamma^{[m]} = \argmin_{u\in\RR^p} u^\intercal \left(\frac{1}{2n} \sum_{i\in\Ical^c} X_i X_i^\T\right) u - u^\T\left(\frac{1}{n} \sum_{i\in\Ical^c} X_i D_i - n^{-1/2}\kappa^\m\right) + \lambda^{[m]}_\gamma \|\gamma\|_1. \label{eq: supp thetaHat m def_gamma}
\end{align}
Throughout the section, we let $X_{\rm tr}$ denote the $n \times p$ design matrix in sample $\Ical^c$. Recall that $X_{\rm tr}$ is assumed to be a random design matrix generated according to $X_{\rm tr}:= \Psi \Omega^{1/2}$, where $\Psi$ is a subgaussian $n\times p$ random matrix (Definition 1.3 and Theorem 1.6 in \citeapp{zhou2009restricted}) and $\Omega$ is a fixed $p \times p$-matrix that satisfies the restricted eigenvalue condition of Assumption 1.2 in \citeapp{zhou2009restricted}, which we restate below:
\begin{assumption}[Restricted Eigenvalue Condition (REC)] \label{assumption:rec}
Suppose $\Omega_{jj} = 1$, $\forall j = 1, \ldots, p$  and
for some integer $1 \leq s \leq p$ and a positive number $k_0$, the following condition holds,
\begin{align*}
    K\l(s, \kappa_0, \Omega\r):=\min_{\substack{J_0\subset\{1,\dots,p\}\\|J_0|\leq s}}\min_{\substack{v\neq0\\\|v_{J_0^c}\|_1\leq \kappa_0 \|v_{J_0}\|_1}} \frac{\|\Omega^{1/2} v\|_2}{\|v_{J_0}\|_2} > 0,
\end{align*}
\end{assumption}
The proof of Theorem 1 proceeds in four steps:
\begin{enumerate}
    \item Lemma \ref{zhou09:thm_1.6}. \citeapp{zhou2009restricted} shows that, under mild conditions, Assumption \ref{assumption:rec} holds with $\Omega^{1/2}$ replaced by $n^{-1/2}X_{\rm tr}$ with probability tending to 1 as $n, p \to \infty$.
    \item Lemma \ref{lemma:basic_ineqs}. Conditioning on the high probability event that Lemma \ref{zhou09:thm_1.6} holds for the sample design matrix $X_{\rm tr}$ with $s = \max(s_\eta, s_\gamma)$, with appropriately chosen tuning parameters $\lambda_\eta^\m$ and $\lambda_\gamma^\m$, there is a universal constant $C$ such that
    \begin{align*}
        \|\widehat\eta^\m - \eta\|_2 \leq C \cdot \sqrt{\frac{s_\eta}{n}} \cdot \|\xi - \xi^\m\|_\infty \quad \text{ and } \quad \|\widehat\gamma^\m - \gamma\|_2 \leq C \cdot \sqrt{\frac{s_\gamma}{n}} \cdot \|\kappa - \kappa^\m\|_\infty.
    \end{align*}
    \item Lemma \ref{lem: rate of xi(m) - xi}. Under Assumption \ref{assumption:main_lasso}, we establish that
    \begin{equation*}
\liminf_{n,p\to\infty}\liminf_{M\to\infty}\Prob\l( \min_{1\leq m\leq M} \max\left(\|\xi - \xi^\m\|_\infty, \|\kappa - \kappa^\m\|_\infty\right) \leq \err \r) \geq 1 - \alpha_0.
\end{equation*}
    \item Lemma \ref{lem: connect beta m with eta m gamma m}. Conditioning on the fold $\Ical^c$ such that $\|\hat{\eta}^\m-\eta\|_2$ and $\|\hat{\gamma}^\m-\gamma\|_2$ are fixed, we show that, with high probability as $n,p\to\infty$, $M\to\infty$, $\|\hat{\eta}^\m-\eta\|_2\to0$ and $\|\hat{\gamma}^\m-\gamma\|_2\to0$:
    $$|\hat{\beta}^\m - \hat{\beta}^{\rm ora}| \lesssim \frac{t_0(n)}{\sqrt{n}}\l(\|\hat{\eta}^\m-\eta\|_2+\|\hat{\gamma}^\m-\gamma\|_2\r) + \|\hat{\eta}^\m-\eta\|_2\|\hat{\gamma}^\m-\gamma\|_2 + \|\hat{\gamma}^\m - \gamma\|_2^2.$$
\end{enumerate}

\begin{lemma}\label{zhou09:thm_1.6}[Theorem 1.6 in \citeapp{zhou2009restricted}]
    Set $1 \leq n \leq p$, $s \leq p/2$, and $0 < \theta < 1$. Let $X_{\rm tr} = \Psi \Omega^{1/2}$, where each row in $\Psi$ is an independent $\psi_2$ isotropic random vector in $\R^p$, i.e., for every $u \in \RR^p$:
    \begin{align*}
    \E\left(\langle \Psi_{j, \cdot}, u \rangle^2\right) = \|u \|_2^2 \quad \text{and} \quad    \inf\left\{t: \E \exp\left(t^{-2}\langle \Psi_{j, \cdot}, u \rangle^2\right)\right\} \lesssim \|u\|_2.
    \end{align*} 
    Suppose $\Omega$ satisfies Assumption \ref{assumption:rec} and
    \begin{align*}
        \max_{\substack{\|t\|_2 = 1 \\ |\text{supp}(t)| \leq s}} \left\|\Omega^{1/2} t\right\|_2 < \infty.
    \end{align*}
    Then, for $n$ large enough and probability tending to 1,
    \begin{align*}
        & 1-\theta \leq \frac{\|X_j\|_2}{\sqrt{n}} \leq 1 + \theta \quad \text{ and } \quad (1-\theta) \leq \frac{\|X_{\rm tr} v\|_2}{\sqrt{n}} \leq (1+\theta).
    \end{align*}
    where $X_j$ is the $j^{\text{th}}$column of $X_{\rm tr}$, and $v\in \l\{ v: \|\Omega^{1/2}v\|_2=1 \; s.t. \; \|v_{T_0^c}\|_1 \leq \kappa_0 \|v_{T_0}\|_1 \r\}$, where $v_{T_0}$ denotes the sub-vector of $v$ confined to the locations of its $s$ largest coefficients.
\end{lemma}
\begin{lemma}\label{lemma:basic_ineqs}
{Let $\tau$ be a small constant in $(0, 1/2]$ and suppose that the events of Lemma \ref{zhou09:thm_1.6} hold, with $s = \max(s_\eta, s_\gamma)$ and $\kappa_0 = (2-\tau)/\tau$. For any fixed $m$, suppose the tuning parameters satisfy $(1-\tau)\lambda_\eta^\m = n^{-1/2}\|\xi^\m-\xi\|_\infty$ and $(1-\tau) \lambda_\gamma^\m = n^{-1/2}\|\kappa^\m - \kappa\|_\infty$. Then, there exists a constant $C>0$ such that}
\begin{align*}
& \|\widehat\eta^{[m]} - \eta\|_2 \leq C \sqrt{\frac{s_\eta}{n}} \cdot \|\xi - \xi^{[m]}\|_\infty \quad \text{and} \quad \|\widehat\gamma^{[m]} - \gamma\|_2 \leq C \sqrt{\frac{s_\gamma}{n}} \cdot \|\kappa - \kappa^{[m]}\|_\infty. 
\end{align*}
Consequently, it holds that
\begin{align*}
    \P\l(\|\widehat\eta^{[m]} - \eta\|_2 \gtrsim\sqrt{\frac{s_\eta \log p}{n}} \r) \lesssim p^{-c} \quad \text{and} \quad \P\left(\|\widehat\gamma^{[m]} - \gamma\|_2 \gtrsim  \sqrt{\frac{s_\gamma \log p}{n}}\right) \lesssim p^{-c}. 
\end{align*}
\end{lemma}
\begin{lemma}\label{lem: rate of xi(m) - xi}
Under Assumption \ref{assumption:main_lasso},
it holds that
\begin{equation*}
\liminf_{n,p\to\infty}\liminf_{M\to\infty}\Prob\l( \min_{1\leq m\leq M} \max(\|\xi - \xi^\m\|_\infty, \ \|\kappa - \kappa^\m\|_\infty) \leq \err \r) \geq 1 - \alpha_0.
\end{equation*}
\end{lemma} 
\begin{lemma}\label{lem: connect beta m with eta m gamma m}
    Let $t_0(n)$ be some slowly increasing sequence in $n$ (e.g. in the proof it may be set as $\log\log n$). Under Assumption \ref{assumption:main_lasso}, and for any fixed $m$, it holds that
    \begin{align*}
        \Prob\bigg(\left|\widehat\beta^\m - \widehat\beta^{\rm ora} \right| & \gtrsim \frac{\|\widehat\eta^\m - \eta\|_2 + \|\widehat\gamma^\m - \gamma\|_2}{\sqrt{n} / t_0(n)} + \|\widehat\gamma^\m - \gamma\|_2\cdot(\|\widehat\eta^\m - \eta\|_2 \\
        & + \|\widehat\gamma^\m - \gamma\|_2) \mid \Ical^c\bigg) \lesssim \frac{1}{t_0(n)}.
    \end{align*}
\end{lemma}

\subsubsection{Proof of Lemma \ref{lemma:basic_ineqs}}

We prove the statement for $\|\widehat\eta^\m - \eta\|_2$ as the one for $\|\widehat\gamma^\m - \gamma\|_2$ follows analogously. By definition \eqref{eq: supp thetaHat m def_eta}, $\widehat\eta^\m$ satisfies the basic inequality in terms of the true parameter $\eta$:
\begin{align*}
    & \frac{1}{2n} \|X_{\rm tr}\widehat\eta^\m\|_2^2 - \frac{1}{n}(\widehat\eta^
    \m)^{\intercal} (X_{\rm tr}^\T Y-\sqrt{n}\xi^\m) +\lambda^\m_\eta \|\hat\eta^\m\|_1 \\
    & \leq \frac{1}{2n} \|X_{\rm tr}\eta\|_2^2 - \frac{1}{n}\eta^{\intercal} (X_{\rm tr}^\T Y-\sqrt{n}\xi^\m) +\lambda^\m_\eta \|\eta\|_1,
\end{align*}
which can be rearranged as
\begin{gather*}
    \frac{1}{2n} \|X_{\rm tr}(\widehat{\eta}^\m - \eta)\|_2^2 + \lambda^\m_\eta \|\hat\eta^\m\|_1 \leq \l|\frac{1}{\sqrt{n}} \langle \widehat\eta^\m - \eta, \xi^\m - \xi \rangle \r| + \lambda^\m_\eta \|\eta\|_1. 
\end{gather*}
Denote the set of nonzero coordinates for $\eta$ as $S_\eta$, i.e. $S_\eta = \{1\leq j\leq p: \eta_j \neq 0\}$. Setting $(1-\tau)\lambda_\eta^\m = n^{-1/2}\|\xi^\m - \xi\|_\infty$ for some small $\tau\in(0,1/2]$, we have
\begin{gather}\label{eq: inequality containing cone condition}
    \frac{1}{2n} \|X_{\rm tr}(\widehat{\eta}^\m - \eta)\|_2^2 + \tau \lambda_\eta^\m \sum_{j \in S_\eta^c} \l|\widehat\eta^\m_j\r| \leq (2 - \tau)\lambda_\eta^\m \sum_{j \in S_\eta} \left|\eta_{j} - \widehat\eta^\m_j\right|.
\end{gather}
We proceed following the proof of Theorem 3.1 in \citeapp{zhou2009restricted}. Adding $\tau \lambda_\eta^\m \sum_{j \in S_\eta} \left|\widehat\eta_j^\m - \eta_j\right|$ to both sides of \eqref{eq: inequality containing cone condition} and multiply 2 to both sides, we have
\begin{align}\label{eq:basic_ineq_zhou}
    \frac{1}{n} \|X_{\rm tr}(\widehat{\eta}^\m - \eta)\|_2^2 + 2\tau \lambda_\eta^\m \|\widehat\eta^\m - \eta\|_1 \leq 4\lambda_\eta^\m \sum_{j \in S_\eta} \left|\eta_{j} - \widehat\eta^\m_j\right|.
\end{align}
By the inequality \eqref{eq: inequality containing cone condition}, we know that $\hat{\eta}^\m - \eta$ satisfies the cone condition, {i.e., that
\begin{align}\label{eq:cone condition of etaHat-eta}
    \sum_{j \in S_\eta^c} \left|\eta_j - \widehat\eta_j^\m \right| \leq \frac{2-\tau}{\tau} \sum_{j \in S_\eta} \left|\eta_j - \widehat\eta_j^\m\right|.
\end{align}}
By Proposition 1.4 in \citeapp{zhou2009restricted}, the cone condition in \eqref{eq:cone condition of etaHat-eta} implies that
\begin{align*}
    \|\widehat\eta_{T_0^c}^\m - \eta_{T_0^c}\|_1 \leq \frac{2-\tau}{\tau} \cdot \|\widehat\eta_{T_0}^\m - \eta_{T_0}\|_1,
\end{align*}
where $T_0$ denote the indices of the $s$ largest (in absolute values) coordinates of $\widehat\eta^\m - \eta$. In this light, on the events from Lemma \ref{zhou09:thm_1.6}, we have
\begin{align*}
    \frac{\left\|X_{\rm tr} (\widehat\eta^\m - \eta)\right\|_2}{\sqrt{n}} & \geq (1-\theta) \| \Omega^{1/2} (\widehat\eta^\m  - \eta)\|_2 \\
    & \geq (1-\theta) \cdot K(s_\eta, (2-\tau) / \tau, \Omega) \cdot \|(\widehat\eta^\m_{T_0} - \eta_{T_0})\|_2 \\ 
    & \geq (1-\theta) \cdot K(s_\eta, (2-\tau) / \tau, \Omega) \cdot \|(\widehat\eta^\m_{S_\eta} - \eta_{S_\eta})\|_2.
\end{align*}
We thus have that, given the events in Lemma \ref{zhou09:thm_1.6}, and setting $K_\eta:= (1-\theta) \cdot K(s_\eta, (2-\tau) / \tau, \Omega)$:
\begin{align}\label{eq:rec_Xtr}
    \|(\widehat\eta^\m_{S_\eta} - \eta_{S_\eta})\|_2 \leq \frac{1}{K_\eta} \cdot \frac{\left\|X_{\rm tr} (\widehat\eta^\m - \eta)\right\|_2}{\sqrt{n}}.
\end{align}
Together with \eqref{eq:basic_ineq_zhou} and \eqref{eq:rec_Xtr}, we have
\begin{align*}
    \frac{1}{n}\|X_{\rm tr}(\widehat\eta^\m - \eta)\|_2^2 + 2 \tau \lambda_\eta^\m \|\widehat\eta^\m - \eta\|_1 & \leq 4 \lambda_\eta^\m \sqrt{s_\eta} \|\widehat\eta_{S_\eta}^\m - \eta_{S_\eta}\|_2 \\
    & \leq 4 \lambda_\eta^\m \sqrt{s_\eta} \cdot \frac{1}{K_\eta}
    \cdot \frac{\|X_{\rm tr}(\widehat\eta^\m - \eta)\|_2}{\sqrt{n}} \\
    & \leq  4 (\lambda_\eta^\m)^2 s_\eta \cdot \frac{1}{K_\eta^2}
    +  \frac{\|X_{\rm tr}(\widehat\eta^\m - \eta)\|^2_2}{n},
\end{align*}
where the last inequality follows as $4ab \leq 4a^2 + b^2$.
This implies that
\begin{align}\label{eq:l1_ls_kappa_square}
  \|\widehat\eta^\m_{S_\eta} - \eta_{S_\eta}\|_1 \leq  \|\widehat\eta^\m - \eta\|_1 \leq \frac{2}{\tau} \cdot \lambda_\eta^\m s_\eta \cdot \frac{1}{K_\eta^2}.
\end{align}
Next we bound $\|\hat{\eta}^\m - \eta\|_2$. Let $T_0$ denote the $s_\eta$ largest (in absolute value) coordinates of $\widehat\eta^\m - \eta$. Reasoning as in Section A.2 in \citeapp{zhou2009restricted}, we have
\begin{align}\label{eq: supp zhou A.2}
\|\widehat\eta^\m- \eta\|_2 \leq \|\widehat\eta_{T_0}^\m - \eta_{T_0}\|_2 + s_\eta^{-1/2}\|\widehat\eta^\m - \eta\|_1. 
\end{align}
We now bound the two terms in \eqref{eq: supp zhou A.2}. For the first term $\|\widehat\eta_{T_0}^\m - \eta_{T_0}\|_2$, since the coordinates set $T_0$ of $\hat{\eta}^\m-\eta$ satisfies the cone condition, we can apply the universality of the RE condition and get 
\begin{align*}
    \|\hat{\eta}^\m_{T_0} - \eta_{T_0}\|_2 \leq \frac{1}{K_\eta} \cdot \frac{\|X_{\rm tr}(\hat{\eta}^\m -\eta)\|_2}{\sqrt{n}}.
\end{align*}
By \eqref{eq:basic_ineq_zhou} and \eqref{eq:l1_ls_kappa_square}, we further bound $\frac{1}{\sqrt{n}}\|X_{\rm tr}(\hat{\eta}^\m -\eta)\|_2$ and get
\begin{align}\label{eq: supp bound v in T0}
    \|\hat{\eta}^\m_{T_0} - \eta_{T_0}\|_2 \leq \frac{1}{K_\eta} \cdot 2 \sqrt{\lambda_\eta^\m \|\widehat\eta^\m_{S_\eta} - \eta_{S_\eta}\|_1} \leq \frac{2}{K_\eta^2} \sqrt{\frac{2}{\tau}} \cdot \sqrt{s_\eta} \cdot \lambda_\eta^\m.
\end{align}
For the second term in \eqref{eq: supp zhou A.2}, that is $s_\eta^{-1/2}\|\widehat\eta^\m - \eta\|_1$, we obtain the bound by \eqref{eq:l1_ls_kappa_square}:
\begin{align}\label{eq: supp bound for second term in 2 norm}
    s_\eta^{-1/2}\|\widehat\eta^\m - \eta\|_1 \leq \frac{2}{\tau K_\eta^2} \cdot \sqrt{s_\eta} \cdot \lambda_\eta^\m. 
\end{align}
Adding the bounds for two terms in \eqref{eq: supp bound v in T0} and \eqref{eq: supp bound for second term in 2 norm}, and recalling $(1-\tau)\lambda_\eta^\m = n^{-1/2}\|\xi - \xi^\m\|_\infty$, we finally obtain 
\begin{align*}
    \|\widehat\eta^\m- \eta\|_2 \leq \frac{2}{K_\eta^2\cdot(1-\tau)}\l(\frac{1}{\tau} + \sqrt{\frac{2}{\tau}}\r) \sqrt{\frac{s_\eta}{n}} \cdot \|\xi - \xi^\m\|_\infty. 
\end{align*}
Since $\xi_j = n^{-1/2}\sum_{i\in\Ical^c} X_{i,j}\epsilon_i$ is a normalized sum of independent, mean-zero sub-Exponential variables, by Corollary 5.17 of \citeapp{vershynin2010introduction} with $\varepsilon = \sqrt{\log p/n}$, we have
\begin{align*}
    \P(|\xi_j| \geq C\sqrt{\log p}) \leq 2p^{-c}. 
\end{align*}
Since $\xi^\m_j$ follows a mean-zero normal distribution given the data, we have
\begin{align*}
    \P\l(|\xi^\m_j| \geq C\sqrt{\log p} \mid \mathcal{O}\r) \lesssim p^{-c}.
\end{align*}
Taking the union bound and the expectation over $\mathcal{O}$, we get
\begin{align*}
    \P(\|\xi - \xi^\m\|_\infty \gtrsim \sqrt{\log p}) \leq \P(\|\xi\|_\infty \gtrsim \sqrt{\log p}) + \P(\|\xi^\m\|_\infty \gtrsim \sqrt{\log p})  \lesssim p^{-c}.
\end{align*}

\subsubsection{Proof of Lemma \ref{lem: rate of xi(m) - xi}}

Let $\zeta^\m = \max\left(\|\xi - \xi^\m\|_\infty, \|\kappa - \kappa^\m\|_\infty\right)$. Let the observed data be denoted by $\mathcal{O}$. We use $\Prob(\cdot \mid \mathcal{O})$ to denote the conditional probability with respect to the observed data $\mathcal{O}$. By the tower property of conditional probabilities, we have
\begin{align*}
    \Prob\l( \min_{1\leq m\leq M} \zeta^\m \leq \err \r) = \E\l[ \Prob\l( \min_{1\leq m\leq M} \zeta^\m \leq \err \mid \mathcal{O}\r) \r].
\end{align*}    
As the vectors $\xi, \kappa$ are fixed conditioning on the observed data $\mathcal{O}$, the randomness in the right-hand-side conditional probability comes solely from the sampling vectors $\xi^\m, \kappa^\m$. Furthermore, by independence given $\mathcal{O}$, we have
\begin{align}\label{eq: supp bound between m and mstar}
    \Prob\l( \min_{1\leq m\leq M} \zeta^\m \leq \err \mid \mathcal{O}\r) 
    &= 1- \Prob\l( \min_{1\leq m\leq M} \zeta^\m > \err \mid \mathcal{O}\r) \nonumber \\ 
    &= 1- \prod_{1\leq m\leq M} \l[1 - \Prob\l( \zeta^\m \leq \err \mid \mathcal{O}\r) \r] \nonumber \\
    & \geq 1 - \exp\l\{ -M\cdot\Prob\l( \zeta^\m \leq \err \mid \mathcal{O}\r) \r\},
\end{align}
where the last inequality follows by $1-x \leq e^{-x}$.

Next, by independence of $\xi^\m$ and $\kappa^\m$ conditioning on data $\mathcal{O}$, we have
\begin{align}
    &\quad \Prob\l( \zeta^\m \leq \err \mid \mathcal{O}\r) \nonumber \\
    &= \Prob\l( \|\xi-\xi^\m\|_\infty \leq \err, \ \|\kappa-\kappa^\m\|_\infty \leq \err \mid \mathcal{O}\r) \nonumber \\
    &= \Prob\l( \|\xi-\xi^\m\|_\infty \leq \err \mid \mathcal{O}\r) \cdot \Prob\l( \|\kappa-\kappa^\m\|_\infty \leq \err \mid \mathcal{O}\r). \label{eq: supp product of four probs mstar}
\end{align}
In the following, we bound the first term $\Prob\l( \|\xi-\xi^\m\|_\infty \leq \err \mid \mathcal{O}\r)$, noting that similar arguments carry over to the other term.

By construction, the density of $\xi^\m$ given the data is 
\begin{equation*}
    f_{\xi^\m} (u \mid \mathcal{O}) = \frac{1}{(2\pi)^{p/2}|\hat{\Sigma}+\nu I|^{1/2}}\exp\l\{-\frac{1}{2}u^{\intercal} (\hat{\Sigma}+\nu I)^{-1}u \r\},
\end{equation*}
where $\hat{\Sigma} = n^{-1}\sum_{i \in \Ical^c} (Y_i - X_i^\T\hat{\eta})^2 X_iX_i^\T$ and $\nu = \min_{1\leq j\leq p} \hat{\Sigma}_{j,j}$.
We lower bound $f_{\xi^\m} (u \mid \mathcal{O})$ as follows. Define the events
\begin{align*}
\mathcal{E}_1
&=
\bigg\{
\max_{1\le j\le p} (\hat{\Sigma}+ \nu I)_{j,j}
\le 2\max_{1\le j\le p}\Sigma_{j,j} + 2B(n,p,s_\eta) \\
&\qquad\text{and}\quad
\min_{1\le j\le p} (\hat{\Sigma}+ \nu I)_{j,j}
\ge 2\min_{1\le j\le p}\Sigma_{j,j} - 2B(n,p,s_\eta)
\bigg\}, \\
\mathcal{E}_2
&=
\left\{ \|\xi\|_2 \le c_\xi \sqrt{p}\,\log(1/\alpha_0) \right\}.
\end{align*}

where $B(n,p,s_\eta)= C\l(\log (np)\frac{s_\eta \log p}{n} + \frac{(\log n)^{5/2}}{\sqrt{n}}+ \frac{1}{\sqrt{n}}\r)$, and $c_\xi$ is the same constant appearing in Lemma \ref{lem: supp prob of event 2}, and $\alpha_0$ is the pre-specified constant in the statement of the theorem. 

The following lemmas show that both $\mathcal{E}_1$ and $\mathcal{E}_2$ holds with high probability. 

\begin{lemma}
\label{lem: supp prob of event 1}
    Under the conditions of Theorem \ref{thm: mstar lasso}, let $\nu = \min_{1\leq j\leq p}\hat{\Sigma}_{jj}$, then
    \begin{align*}
        \P(\mathcal{E}_1) \geq 1 - (np)^{-c} - p^{-c}.
    \end{align*}
\end{lemma}

\begin{lemma}\label{lem: supp prob of event 2}
    Under the conditions of Theorem \ref{thm: mstar lasso}, there exists constants $c_\xi$ such that the following holds:
    \begin{align*}
        \P(\mathcal{E}_2) \geq 1 - \frac{\alpha_0}{2}.
    \end{align*}
 \end{lemma}

Define the surrogate density function $\tilde{g}$ serving as a lower bound on $f_{\xi^\m} (u \mid \mathcal{O})$ given the event $\mathcal{E}_1 \cap \mathcal{E}_2$:
\begin{equation*}
    \tilde{g}(u) = \frac{1}{\{2\pi (2\max_{1\leq j\leq p}\Sigma_{j,j} + 2B(n,p,s_\eta))\}^{p/2}}\exp\l(-\frac{u^\T u}{2 (2\min_{1\leq j\leq p}\Sigma_{j,j} - 2B(n,p,s_\eta))}\r).
\end{equation*}
Conditioning on $\mathcal{E}_1$ and writing $A \succeq B$ to denote that the matrix $A-B$ is positive semidefinite, we have $ \hat\Sigma + \nu I \succeq (2\min_{1\leq j\leq p}\Sigma_{j,j} - 2B(n,p,s_\eta)) I$ and thus $f_{\xi^\m} (u \mid \mathcal{O}) \geq \tilde{g}(u)$ for $u\in\RR^{p}$. Note that $B(n,p,s_\eta) \to 0$ as $n,p\to\infty$. There exist large enough $n_0$ and $p_0$ such that $B(n,p,s_\eta) < \frac{1}{2}\min_{1\leq j\leq p} \Sigma_{j,j}$ for $n\geq n_0, p\geq p_0$. Therefore, we lower bound $\tilde{g}(u)$ by 
\begin{align*}
    \tilde{g}(u) \geq g(u) :=  \frac{1}{(8\pi \max_{1\leq j\leq p}\Sigma_{j,j})^{p/2}}\exp\l(-\frac{u^\T u}{2 \min_{1\leq j\leq p}\Sigma_{j,j}}\r), \quad \textrm{with } n\geq n_0, p \geq p_0,
\end{align*}
and get $\hat\Sigma + \nu I \succeq \min_{1\leq j\leq p}\Sigma_{j,j} I$. Moreover, on the event $\mathcal{E}_2$, we have $\|\xi\|_2^2 \leq c_\xi \sqrt{p} \cdot \log(1/\alpha_0)$.
By plugging the above bound in $g(\xi)$, we have, with $n\geq n_0, p \geq p_0$,
\begin{align}\label{eq: supp lowerbound for g}
    g(\xi) \cdot \mathbf{1}_{\mathcal{O}\in\mathcal{E}_1\cap\mathcal{E}_2}
    &\geq \frac{1}{(8\pi \max_{1\leq j\leq p}\Sigma_{j,j})^{p/2}}
    \exp\l\{-\frac{c_\xi}{2 \min_{1\leq j\leq p}\Sigma_{j,j}} \sqrt{p} \log (1 /\alpha_0) \r\} \cdot \mathbf{1}_{\mathcal{O}\in\mathcal{E}_1\cap\mathcal{E}_2} \\
    &= C_1^p C_{\alpha_0}^{\sqrt{p}} \cdot \mathbf{1}_{\mathcal{O}\in\mathcal{E}_1\cap\mathcal{E}_2},
\end{align}
where 
$C_1=(8\pi \max_{1\leq j\leq p}\Sigma_{j,j})^{-1/2}$ and $C_{\alpha_0}=\exp\{-c _\xi\log (1 / \alpha_0) / (2\min_{1\leq j\leq p}\Sigma_{j,j})\}$.

Since, on $\mathcal{E}_1$, $f_{\xi^\m}(u) \geq \tilde{g}(u) \geq g(u)$ with $n\geq n_0, p \geq p_0$, we have
\begin{align*}
    & \Prob\l(\l\|\xi - \xi^\m\r\|_\infty \leq \err \mid \mathcal{O}\r) \cdot \mathbf{1}_{\mathcal{O}\in\mathcal{E}_1\cap\mathcal{E}_2}  \\
    & = \int \mathbb{1}_{\l\|\xi - u\r\|_\infty \leq \err} \cdot f_{\xi^\m}(u \mid \mathcal{O}) du \cdot \mathbf{1}_{\mathcal{O}\in\mathcal{E}_1\cap\mathcal{E}_2} \\
    & \geq \int \mathbb{1}_{\l\|\xi - u\r\|_\infty \leq \err} \cdot g(u) du \cdot \mathbf{1}_{\mathcal{O}\in\mathcal{E}_1\cap\mathcal{E}_2}.
\end{align*}
Adding and subtracting $g(\xi)$, we decompose the above integral into two parts as
\begin{align}
    & \Prob\l(\l\|\xi - \xi^\m\r\|_\infty \leq \err \mid \mathcal{O}\r) \cdot \mathbf{1}_{\mathcal{O}\in\mathcal{E}_1\cap\mathcal{E}_2} \nonumber \\
    & \geq \left[\int \mathbb{1}_{\l\|\xi - u\r\|_\infty \leq \err} \cdot g(\xi) du + \int \mathbb{1}_{\l\|\xi - u\r\|_\infty \leq \err} \cdot \l\{g(u)-g(\xi)\r\} du\right] \cdot \mathbf{1}_{\mathcal{O}\in\mathcal{E}_1\cap\mathcal{E}_2}. \label{eq: supp decompose prob into two integrals}
\end{align}
To bound the first term in \eqref{eq: supp decompose prob into two integrals}, we apply the lower bound of $g(\xi)$ in \eqref{eq: supp lowerbound for g} and get, with $n\geq n_0, p \geq p_0$,
\begin{align}\label{eq: supp bound integral 1}
    \int \mathbb{1}_{\l\|\xi - u\r\|_\infty \leq \err} \cdot g(\xi) du \cdot \mathbf{1}_{\mathcal{O}\in\mathcal{E}_1\cap\mathcal{E}_2} \geq [2\err]^p \cdot C_1^p C_{\alpha_0}^{\sqrt{p}} \cdot \mathbf{1}_{\mathcal{O}\in\mathcal{E}_1\cap\mathcal{E}_2}. 
\end{align}
To bound the second term in \eqref{eq: supp decompose prob into two integrals}, note that for a value $\overline\xi_u$ between $u$ and $\xi$, we have
\begin{align*}
\left|g(u) - g(\xi)\right| = \l| \nabla g(\overline\xi_u)^\T (u-\xi) \r| \leq \| \nabla g(\overline\xi_u)\|_1\|u - \xi\|_\infty \leq \sqrt{p}\|\nabla g(\overline{\xi}_u)\|_2 \|u - \xi\|_\infty,
\end{align*}
with $\nabla g(\overline\xi_u) = -(\min_{1\leq j\leq p}\Sigma_{j,j})^{-1}g(\overline\xi_u) \cdot {\overline\xi_u}$. By the definition of $g(u)$, notice that
\begin{align*}
\|\nabla g(\overline\xi_u)\|_2 & = (\min_{1\leq j\leq p}\Sigma_{j,j})^{-1}g(\overline\xi_u) \cdot \|\overline\xi_u\|_2 \\
& = \frac{1}{(8\pi\max_{1\leq j\leq p}\Sigma_{j,j})^{p/2} \min_{1\leq j\leq p}\Sigma_{j,j}}\exp\l\{-\frac{\|\bar{\xi}_u\|_2^2}{2\min_{1\leq j\leq p}\Sigma_{j,j}} \r\} \cdot \|\overline\xi_u\|_2 \\ 
& \leq \l(8\pi\max_{1\leq j\leq p}\Sigma_{j,j}\r)^{-p/2} \l(e\min_{1\leq j\leq p}\Sigma_{j,j}\r)^{-1/2},
\end{align*}
where the last inequality follows because the function $x \mapsto x\exp\{-\frac{1}{2a}x^2\}$ achieves its maximum at $x=\sqrt{a}$. Therefore, with $n\geq n_0, p \geq p_0$, the second term in \eqref{eq: supp decompose prob into two integrals} is bounded as
\begin{align}
    & \quad \l| \int \mathbb{1}_{\l\|\xi - u\r\|_\infty \leq \err} \cdot \l\{g(u)-g(\xi)\r\} du \cdot \mathbf{1}_{\mathcal{O}\in\mathcal{E}_1\cap\mathcal{E}_2} \r| \nonumber \\
    & \leq \int \mathbb{1}_{\l\|\xi - u\r\|_\infty \leq \err} \cdot \sqrt{p}\|\nabla g(\overline{\xi}_u)\|_2 \|u - \xi\|_\infty du \cdot \mathbf{1}_{\mathcal{O}\in\mathcal{E}_1\cap\mathcal{E}_2} \nonumber \\
    & \leq [2\err]^p \cdot \frac{\sqrt{p} \cdot  \err}{\l(8\pi\max_{1\leq j\leq p}\Sigma_{j,j}\r)^{p/2} \l(e\min_{1\leq j\leq p}\Sigma_{j,j}\r)^{1/2}} \cdot \mathbf{1}_{\mathcal{O}\in\mathcal{E}_1\cap\mathcal{E}_2}. \label{eq: supp bound integral 2}
\end{align}
Putting together the first term bound in \eqref{eq: supp bound integral 1} and the second term bound in \eqref{eq: supp bound integral 2}, we get, with $n\geq n_0, p \geq p_0$,
\begin{align*}
    & \quad \Prob\l(\l\|\xi - \xi^\m\r\|_\infty \leq \err \mid \mathcal{O}\r) \cdot \mathbf{1}_{\mathcal{O}\in\mathcal{E}_1\cap\mathcal{E}_2} \\
    & \geq [2\err]^p \cdot \l(C_1^p C_{\alpha_0}^{\sqrt{p}} - \frac{\sqrt{p} \cdot  \err}{\l(8\pi\max_{1\leq j\leq p}\Sigma_{j,j}\r)^{p/2} \l(e\min_{1\leq j\leq p}\Sigma_{j,j}\r)^{1/2}}\r) \cdot \mathbf{1}_{\mathcal{O}\in\mathcal{E}_1\cap\mathcal{E}_2}.
\end{align*}
For any given $n$ and $p$, we have $\err$ tends to zero as $M\to\infty$, so that there exists a positive $M_0$ {satisfying $\log M_0 \gtrsim \log\log n +p^2$} such that for $M>M_0$, we have $\frac{\sqrt{p} \cdot  \err}{\l(8\pi\max_{1\leq j\leq p}\Sigma_{j,j}\r)^{p/2} \l(e\min_{1\leq j\leq p}\Sigma_{j,j}\r)^{1/2}} < \frac{1}{2}C_1^p C_{\alpha_0}^{\sqrt{p}}$. In this light, assuming $M>M_0$, we have
\begin{align*}
    \Prob\l(\l\|\xi - \xi^\m\r\|_\infty \leq \err \mid \mathcal{O}\r) \cdot \mathbf{1}_{\mathcal{O}\in\mathcal{E}_1\cap\mathcal{E}_2} \geq 2^{p-1} C_1^p C_{\alpha_0}^{\sqrt{p}} \cdot [\err]^p\cdot \mathbf{1}_{\mathcal{O}\in\mathcal{E}_1\cap\mathcal{E}_2}.
\end{align*}

Let $\mathcal{E}_1'$ and $\mathcal{E}_2'$ denote the events $\mathcal{E}_1$ and $\mathcal{E}_2$ written in terms of $\widehat\Lambda$ and $\kappa$. That is,
\begin{align*}
\mathcal{E}_1'
&=
\bigg\{
\max_{1\le j\le p} (\hat{\Lambda}+ \nu' I)_{j,j}
\le 2\max_{1\le j\le p}\Lambda_{j,j} + 2B(n,p,s_\gamma) \\
&\qquad\text{and}\quad
\min_{1\le j\le p} (\hat{\Lambda}+ \nu' I)_{j,j}
\ge 2\min_{1\le j\le p}\Lambda_{j,j} - 2B(n,p,s_\gamma)
\bigg\}, \\
\mathcal{E}_2'
&=
\left\{ \|\kappa\|_2 \le c_\kappa \sqrt{p}\,\log(1/\alpha_0) \right\}.
\end{align*}

with $\nu' = \min_{1\leq j\leq p} \hat{\Lambda}_{j,j}$. Following the similar reasoning in Lemma \ref{lem: supp prob of event 1} and \ref{lem: supp prob of event 2}, we have 
\begin{align}
    \P(\mathcal{E}_1') \geq 1 - (np)^{-c}-p^{-c}, \quad \textrm{and} \quad \P(\mathcal{E}_2) \geq 1- \frac{\alpha_0}{2}. \label{eq:supp prob E1' E2'}
\end{align}
We can similarly derive
    \begin{align*}
    &\quad \Prob\l(\l\|\kappa - \kappa^\m\r\|_\infty \leq \err \mid \mathcal{O}\r) \cdot \mathbf{1}_{\mathcal{O}\in\mathcal{E}'_1\cap\mathcal{E}'_2} \\ & \geq 2^{p-1} {(C'_1)}^p (C'_{\alpha_0})^{\sqrt{p}} \cdot [\err]^p\cdot \mathbf{1}_{\mathcal{O}\in\mathcal{E}'_1\cap\mathcal{E}'_2}.
\end{align*}
Thus, letting $\mathcal{E} = \cap_{l = 1}^2 \mathcal{E}_l \cap \mathcal{E}_l'$, we arrive at
\begin{align*}
     \Prob\l( \zeta^\m \leq \err \mid \mathcal{O}\r) \geq 2^{2p - 2} [\err]^{2p} (C_1C_1')^{p} (C_{\alpha_0}C_{\alpha_0}')^{\sqrt{p}} \cdot \mathbf{1}_{\mathcal{O}\in \mathcal{E}}.
\end{align*}
Finally, under the condition $M\geq M_0$, we plug this into \eqref{eq: supp bound between m and mstar} and have
\begin{align*}
    & \Prob\l( \min_{1\leq m\leq M} \zeta^\m \leq \err \r) \\
    & \geq  \left[1- \exp\left\{-M \cdot 2^{2p - 2} [\err]^{2p} (C_1C_1')^{p} (C_{\alpha_0}C_{\alpha_0}')^{\sqrt{p}} \right\}\right] \cdot \Prob(\mathcal{E}).
\end{align*}
Recall that $\err$ is defined in \eqref{eq: supp def err} to be equal to
\begin{align}\label{eq:supp constants in err}
    \err = c_1 \cdot [c_*(\alpha_0)]^{-\frac{1}{\sqrt{p}}} \cdot  \l( \frac{4\log n}{M}\r)^{\frac{1}{2p}}.
\end{align}
With $c_1 = (2\sqrt{C_1C_1'})^{-1}$ and $c_*(\alpha_0) = (C_{\alpha_0} C'_{\alpha_0})^{1/2}$, we arrive at
\begin{align*}
    \Prob\l( \min_{1\leq m\leq M} \zeta^\m \leq \err \r) \geq (1-n^{-1})\cdot \P(\mathcal{E}).
\end{align*}
Notice that by Lemma \ref{lem: supp prob of event 1} and \ref{lem: supp prob of event 2} and \eqref{eq:supp prob E1' E2'}, $\liminf_{n, p \to \infty} \Prob(\mathcal{E}) \geq 1 - \alpha_0$. The result then follows by taking $M \to \infty$ for any fixed $n\geq n_0$ and $p\geq p_0$.

\subsubsection{Proof of Lemma \ref{lem: connect beta m with eta m gamma m}}

For shorthand notation, let us define 
\begin{align*}
    \varphi_1(O_i) = \l(Y_i - X_i^\T\eta\r)\l(D_i - X_i^\T\gamma\r), \quad \varphi_2(O_i) = (D_i - X_i^\T\gamma)^2,
\end{align*}
For $1\leq m\leq M$, we define the corresponding estimators
\begin{align*}
& \widehat\varphi^\m_{1}(O_i) = (Y_i - X_i^\intercal \widehat\eta^\m)(D_i - X_i^\intercal \widehat\gamma^\m), \quad \widehat\varphi^\m_{2}(O_i) = (D_i - X_i^\intercal \widehat\gamma^\m)^2.
\end{align*}
Write
\begin{align*}
    \beta = \frac{\E\{\varphi_1(O)\}]}{\E\{\varphi_2(O)\}]} \equiv \frac{\psi_1}{\psi_2}, \quad \hat{\beta}^\m = \frac{n^{-1} \sum_{i\in\Ical} \widehat\varphi^\m_{1}(O_i)}{n^{-1} \sum_{i\in\Ical} \widehat\varphi^\m_{2}(O_i)} \equiv \frac{\hat{\psi}_{1}^\m}{\hat{\psi}_{2}^\m}, \quad \hat{\beta}^{\rm ora} = \frac{n^{-1} \sum_{i\in\Ical} \varphi_{1}(O_i)}{n^{-1} \sum_{i\in\Ical} \varphi_{2}(O_i)} \equiv \frac{\hat{\psi}_{1}^{\rm ora}}{\hat{\psi}_{2}^{\rm ora}}.
\end{align*}
With these notations, the distance between $\hat{\beta}^\m$ and $\hat{\beta}^{\rm ora}$ can be decomposed as 
\begin{align}
    \hat{\beta}^\m - \hat{\beta}^{\rm ora} &= \frac{\hat{\psi}_1^\m - \hat{\psi}_1^{\rm ora}}{\psi_2} + \frac{\hat{\psi}_1^\m - \hat{\psi}_1^{\rm ora}}{\psi_2}\l(\frac{\psi_2}{\hat{\psi}_2^\m} - 1\r) + \frac{\hat{\psi}_1^{\rm ora}}{\hat{\psi}_2^\m} - \frac{\hat{\psi}_1^{\rm ora}}{\hat{\psi}_2^{\rm ora}} \nonumber \\
    & = \frac{\hat{\psi}_1^\m - \hat{\psi}_1^{\rm ora}}{\psi_2} + \frac{\hat{\psi}_1^\m - \hat{\psi}_1^{\rm ora}}{\psi_2}\l(\frac{\psi_2}{\hat{\psi}_2^\m} - 1\r) \nonumber \\
    & \quad + (\hat{\psi}_1^{\rm ora} - \psi_1)\l(\frac{1}{\hat{\psi}_2^\m} - \frac{1}{\hat{\psi}_2^{\rm ora}}\r) + \psi_1\l(\frac{1}{\hat{\psi}_2^\m} - \frac{1}{\hat{\psi}_2^{\rm ora}}\r), \label{eq:supp decomp 1 beta m - beta ora}
\end{align}
where the last term in the parenthesis can be further decomposed as
\begin{align}
    \frac{1}{\hat{\psi}_2^\m} - \frac{1}{\hat{\psi}_2^{\rm ora}} = \frac{\hat{\psi}_2^{\rm ora} - \hat{\psi}_2^\m}{\psi_2^2}\l\{1+\l(\frac{\psi_2}{\hat{\psi}_2^\m} - 1\r)\l(\frac{\psi_2}{\hat{\psi}_2^{\rm ora}} - 1\r) + \l(\frac{\psi_2}{\hat{\psi}_2^\m} - 1\r) + \l(\frac{\psi_2}{\hat{\psi}_2^{\rm ora}} - 1\r) \r\} \label{eq:supp decomp 2 beta m - beta ora}
\end{align}
by $ab-1 = (a-1)(b-1)+(a-1)+(b-1)$. 

Define the events 
\begin{align*}
    \mathcal{B}_1 &= \l\{ |\hat{\psi}_1^\m - \hat{\psi}_1^{\rm ora}| \leq t_1(\hat{\eta}^\m, \hat{\gamma}^\m,n) \r\}, \quad \mathcal{B}_2 = \l\{ \l|\hat{\psi}_1^{\rm ora} - \psi_1 \r| \leq t_2(n) \r\},\\
    \mathcal{B}_3 &= \l\{ \l|\frac{\hat{\psi}_2^\m}{\psi_2} - 1\r| \leq t_3(\hat{\eta}^\m, \hat{\gamma}^\m,n) \r\}, \quad \mathcal{B}_4 = \l\{ \l|\frac{\hat{\psi}_2^{\rm ora}}{\psi_2} - 1\r| \leq t_4(n) \r\}, \\
    \mathcal{B}_5 &= \l\{|\hat{\psi}_2^\m - \hat{\psi}_2^{\rm ora}| \leq t_5(\hat{\eta}^\m, \hat{\gamma}^\m,n)\r\},
\end{align*}
with 
\begin{align*}
    t_1(\hat{\eta}^\m, \hat{\gamma}^\m,n) &= c\l(\frac{t_0(n)}{\sqrt{n}}\|\hat{\eta}^\m - \eta\|_2 + \frac{t_0(n)}{\sqrt{n}}\|\hat{\gamma}^\m - \gamma\|_2 + \|\hat{\eta}^\m - \eta\|_2\|\hat{\gamma}^\m - \gamma\|_2\r), \\
    t_2(n) & = c\sqrt{\frac{t_0(n)}{n}},\\
    t_3(\hat{\eta}^\m, \hat{\gamma}^\m,n) &= c\l( \frac{t_0(n)}{\sqrt{n}}\|\hat{\gamma}^\m - \gamma\|_2 + \|\hat{\gamma}^\m - \gamma\|_2^2 + \sqrt{\frac{t_0(n)}{n}} \r), \\
    t_4(n) &= c\sqrt{\frac{t_0(n)}{n}}, \\
    t_5(\hat{\eta}^\m, \hat{\gamma}^\m,n) &= c\l( \frac{t_0(n)}{\sqrt{n}}\|\hat{\gamma}^\m - \gamma\|_2 + \|\hat{\gamma}^\m - \gamma\|_2^2\r),
\end{align*}
where $t_0(n)$ is a slowly increasing rate in $n$, for example, $t_0(n)=\log\log n$. 

On the event $\mathcal{B}_3\cap\mathcal{B}_4$, we have
\begin{align*}
    \l| \frac{\psi_2}{\hat{\psi}_2^\m} - 1 \r| = \l| \frac{1 - \hat{\psi}_2^\m/\psi_2}{\hat{\psi}_2^\m/\psi_2} \r| \leq \frac{t_3(\hat{\eta}^\m, \hat{\gamma}^\m,n)}{1-t_3(\hat{\eta}^\m, \hat{\gamma}^\m,n)}, \quad \l| \frac{\psi_2}{\hat{\psi}_2^{\rm ora}} - 1 \r| = \l| \frac{1 - \hat{\psi}_2^{\rm ora}/\psi_2}{\hat{\psi}_2^{\rm ora}/\psi_2} \r| \leq \frac{t_4(n)}{1-t_4(n)}.
\end{align*}

Then, on the event $\cap_{1\leq j\leq 5}\mathcal{B}_j$, we can bound $|\hat{\beta}^\m - \hat\beta^{\rm ora}|$ each term based on the decompositions in \eqref{eq:supp decomp 1 beta m - beta ora} and \eqref{eq:supp decomp 2 beta m - beta ora}. As $n\to\infty,\|\hat{\eta}^\m-\eta\|_2\to0$ and $\|\hat{\gamma}^\m-\gamma\|_2\to0$, note that $t_3/(1-t_3)$ has the same rate as $t_3$ and $t_4/(1-t_4)$ has the same rate as $t_4$. Simplifying the above inequality by removing higher-order terms of  $\|\hat{\eta}^\m - \eta\|_2, \|\hat{\gamma}^\m - \gamma\|_2$ and $n$, the bound is of the order
\begin{align*}
    \frac{t_0(n)}{\sqrt{n}}\|\hat{\eta}^\m - \eta\|_2 + \frac{t_0(n)}{\sqrt{n}}\|\hat{\gamma}^\m - \gamma\|_2 + \|\hat{\eta}^\m - \eta\|_2\|\hat{\gamma}^\m - \gamma\|_2 + \|\hat{\gamma}^\m - \gamma\|_2^2.
\end{align*}
Then we establish the bound shown in Lemma \ref{lem: connect beta m with eta m gamma m}. It remains to show 
\begin{align}
    \P\l(\bigcap_{j=1}^5 \mathcal{B}_j \mid \Ical^c\r) \geq 1-\frac{c}{t_0(n)}. \label{eq:supp event prob for betam - betaora}
\end{align}

For $\mathcal{B}_1$, note that 
\begin{align*}
    \hat\psi_1^\m - \hat\psi^{\rm ora} = \frac{1}{n}\sum_{i\in\Ical} \epsilon_iX_i^\T(\gamma - \hat{\gamma}^\m) + \frac{1}{n}\sum_{i\in\Ical} \delta_iX_i^\T(\eta - \hat{\eta}^\m) + (\gamma - \hat{\gamma}^\m)^\T \l(\frac{1}{n}\sum_{i\in\Ical} X_iX_i^\T\r) (\eta - \hat{\eta}^\m).
\end{align*}
Conditioning on the fold $\Ical^c$ such that $\hat{\eta}^\m$ and $\hat{\gamma}^\m$ are fixed, by Markov inequality, we have
\begin{align}
    \P\l( \l| \frac{1}{n}\sum_{i\in\Ical} \epsilon_iX_i^\T(\gamma - \hat{\gamma}^\m) \r| \lesssim \|\Sigma\|_\op^{1/2}\frac{t_0(n)}{\sqrt{n}}\|\hat{\gamma}^\m - \gamma\|_2\mid \Ical^c\r) & \geq 1 - \frac{c}{t_0(n)}, \label{eq:supp psi diff term 1} \\
    \P\l( \l| \frac{1}{n}\sum_{i\in\Ical} \delta_iX_i^\T(\eta - \hat{\eta}^\m) \r| \lesssim \|\Lambda\|_\op^{1/2}\frac{t_0(n)}{\sqrt{n}}\|\hat{\eta}^\m - \eta\|_2\mid \Ical^c\r) & \geq 1 - \frac{c}{t_0(n)}. \label{eq:supp psi diff term 2}
\end{align}
We introduce the following lemma to bound the terms like $(\gamma - \hat{\gamma}^\m)^\T \l(\frac{1}{n}\sum_{i\in\Ical} X_iX_i^\T\r) (\eta - \hat{\eta}^\m)$. 

\begin{lemma}[Partly from Lemma 11, \citetapp{tony2020semisupervised}]\label{lem:Lemma11_Zijian}
    Let $\Sigma_X = \E(XX^\T)$. For given $w, v \in \RR^p$ and $t > 0$, then
    \begin{align}
        \Prob\left(\left|w^\T\left(\frac{1}{n} \sum_{i=1}^n X_iX_i^\T\right) v - w^T\Sigma_X v\right| \gtrsim t \frac{\|\Sigma_X^{1/2} w\|_2 \|\Sigma_X^{1/2} v\|_2}{\sqrt{n}}\right) \leq 2 \exp(-c t^2). \label{eq:supp zijian cnetered}
    \end{align}
    Consequently, with $t = \sqrt{t_0(n)}$ for some slowly increasing sequence $t_0(n)$ in $n$ (e.g., $t_0(n) = \log\log n$),
    \begin{align}
        \Prob\left(\left|w^\T\left(\frac{1}{n} \sum_{i=1}^n X_iX_i^\T\right) v\right| \gtrsim  \|\Sigma_X\|_\op {\|w\|_2 \|v\|_2}\right) \leq 2 \exp(-c t_0(n)). \label{eq:supp zijian mean noncentered}
    \end{align}
\end{lemma}
Let $v=\hat{\gamma}^\m - \gamma$ and $w = \hat{\eta}^\m - \eta$ and they are fixed vectors conditioning on $\Ical^c$. Recall that $\Sigma_X = \E[X_iX_i^\T]$. By \eqref{eq:supp zijian mean noncentered} in Lemma \ref{lem:Lemma11_Zijian}, we have
\begin{align}
    &\quad\P\l( \l| (\gamma - \hat{\gamma}^\m)^\T \l(\frac{1}{n}\sum_{i\in\Ical} X_iX_i^\T\r) (\eta - \hat{\eta}^\m) \r| \gtrsim \|\Sigma_X\|_\op\|\gamma - \hat{\gamma}^\m\|_2\|\eta - \hat{\eta}^\m\|_2 \mid \Ical^c\r) \nonumber \\
    &\leq 2\exp(-ct_0(n)) \lesssim \frac{1}{t_0(n)}. \label{eq:supp psi diff term 3}
\end{align}

By inequalities \eqref{eq:supp psi diff term 1}, \eqref{eq:supp psi diff term 2} and \eqref{eq:supp psi diff term 3}, we have,
\begin{align*}
    \P(\mathcal{B}_1 \mid \Ical^c) \geq 1 - \frac{c}{t_0(n)}.
\end{align*}

For $\mathcal{B}_3$ and $\mathcal{B}_5$, we have the decompositions
\begin{align*}
    \hat{\psi}_2^\m - \psi_2 &= \frac{2}{n}\sum_{i\in\Ical} X_i^\T \delta_i(\gamma - \hat{\gamma}^\m) + (\gamma - \hat{\gamma}^\m)^\T \l(\frac{1}{n}\sum_{i\in\Ical} X_iX_i^\T\r) (\gamma - \hat{\gamma}^\m) + \l(\frac{1}{n}\sum_{i\in\Ical} \varphi_2(O_i) - \psi_2\r), \\
    \hat{\psi}_2^\m - \hat{\psi}_2^{\rm ora} &= \frac{2}{n}\sum_{i\in\Ical} X_i^\T \delta_i(\gamma - \hat{\gamma}^\m) + (\gamma - \hat{\gamma}^\m)^\T \l(\frac{1}{n}\sum_{i\in\Ical} X_iX_i^\T\r) (\gamma - \hat{\gamma}^\m).
\end{align*}
By the similar Markov arguments and Chebyshev inequality, we have 
\begin{align*}
    \P(\mathcal{B}_3 \mid \Ical^c) \geq 1- \frac{c}{t_0(n)}, \quad \P(\mathcal{B}_5 \mid \Ical^c) \geq 1- \frac{c}{t_0(n)}.
\end{align*}

We bound the probabilities of both events $\mathcal{B}_2$ and $\mathcal{B}_4$ by Chebyshev inequality and get 
\begin{align*}
    \P(\mathcal{B}_2 \mid \Ical^c) \geq 1 - \frac{c}{t_0(n)}, \quad \P(\mathcal{B}_4 \mid \Ical^c) \geq 1 - \frac{c}{t_0(n)}.
\end{align*}

Applying the union bound to the above high probability events establishes \eqref{eq:supp event prob for betam - betaora} and further establishes Lemma \ref{lem: connect beta m with eta m gamma m}.

\subsection{Proof of Theorem \ref{thm: coverage lasso}}

\subsubsection{Preliminaries and notation}
We prove Theorem \ref{thm: coverage lasso} under a sample splitting scheme whereby observations in fold $\Ical^c$ are used to construct all nuisance functions while those in fold $\Ical$ are used to compute $\widehat\beta$ and conduct inference. In particular, the estimators $\hat{\eta}^\m, \hat{\gamma}^\m, \hat{\eta}, \hat{\gamma}$ are fitted on fold $\Ical^c$. 

Theorem \ref{thm: coverage lasso} follows by establishing that
\begin{align*}
\limsup_{n, p \to \infty}\limsup_{M \to \infty} \Prob\left(\beta \not\in \text{CI}\right) \leq \alpha.
\end{align*}
{In this proof, we slightly abuse the notation $m^*$ and} let $m^*$ be the smallest index such that the following event holds for some constant $C>0$,
\begin{align}\label{eq: supp event G1}
    \mathcal{G}_1^\mstar = \l\{ \|\widehat\eta^\mstar - \eta\|_2 \leq C\sqrt{\frac{s_\eta}{n}} \err, \quad
        \|\widehat\gamma^\mstar - \gamma\|_2 \leq C\sqrt{\frac{s_\gamma}{n}}\err  \r\},
\end{align}
with $\err$ defined in \eqref{eq: supp def err}.
To establish the coverage property of our constructed CI, we also need to control the errors incurred by the original Lasso estimators, i.e., $\|\hat{\eta} -\eta\|_2$ and $\|\hat{\gamma} - \gamma\|_2$. Thus, similarly to $\mathcal{G}_1^\mstar$, we define, for some other constant $C$:
\begin{equation}\label{eq: supp event G2}
    \mathcal{G}_1 = \l\{ \|\hat{\eta} - \eta\|_2 \leq C\sqrt{\frac{s_\eta \log p}{n}}, \quad \|\hat{\gamma} - \gamma\|_2 \leq C\sqrt{\frac{s_\gamma \log p}{n}} \r\}.
\end{equation}
From the definition of $\textrm{CI}$ in \eqref{eq:filtered union CI}, the event $\{\beta\notin\textrm{CI}\}$ implies two disjoint cases: $m
^*\notin \mathcal{M}$ or $m
^*\in \mathcal{M}$ but $\beta\notin\textrm{CI}^\mstar$. Therefore,
\begin{align}\label{eq: ci cover decomp1}
    \Prob\left(\beta \not\in \text{CI}\right) & \leq \Prob\left(\l\{\beta \not\in \text{CI}^\mstar\r\} \cap \mathcal{G}_1^\mstar \cap \mathcal{G}_1 \right) + \Prob\left(\l\{m^* \not\in \mathcal{M} \r\}\cap \mathcal{G}_1^\mstar \cap \mathcal{G}_1 \right) \nonumber \\
    & \hphantom{=} + \Prob((\mathcal{G}_1^\mstar)^c \cup \mathcal{G}_1^c). 
\end{align}
By Theorem \ref{thm: mstar lasso}, $\limsup_{n, p \to \infty} \limsup_{M \to \infty} \Prob([\mathcal{G}_1^\mstar]^c) \leq \alpha_0$. 
To control the event $\mathcal{G}_1$, we can view the original estimators $\widehat\eta$ and $\widehat\gamma$ (defined in \eqref{eq: lasso optimization problem unperturbed}) as solving the Lasso optimizations (Eqs. \eqref{eq: supp thetaHat m def_eta} and \eqref{eq: supp thetaHat m def_gamma}) with $\xi^\m = \kappa^\m = 0$. Thus, by Lemma \ref{lemma:basic_ineqs}, and conditioning on the event from Lemma \ref{zhou09:thm_1.6}, the estimators satisfy:
\begin{align*}
    \|\widehat\eta - \eta\|_2 \leq C \sqrt{\frac{s_\eta}{n}} \cdot \left| \max_{1 \leq j \leq p} n^{-1/2}X_j^\intercal \epsilon\right|, \quad \text{ and } \quad \|\widehat\gamma - \gamma\|_2 \leq C \sqrt{\frac{s_\gamma}{n}} \cdot \left| \max_{1 \leq j \leq p} n^{-1/2} X_j^\intercal \delta\right|.
\end{align*}
We bound the value $\left|\max_{1 \leq j \leq p} n^{-1/2}X_j^\intercal \epsilon\right|$ by the following lemma. The value $\left| \max_{1 \leq j \leq p} n^{-1/2} X_j^\intercal \delta\right|$ can be bounded following the similar reasoning. 

\begin{lemma}\label{lemma:xi_subgaussian}
    Suppose that $X_{ji}$ and $\epsilon_i$ are sub-Gaussian random variables with parameters $\sigma_X$ and $\sigma_\epsilon$, respectively. Further suppose, that $(X_{j1}, \epsilon_1), \ldots, (X_{jn}, \epsilon_n)$ are independent. Then, there exist positive constants $c$, $C$ and $C'$, such that $\xi_j = n^{-1/2} X_j^\T \epsilon = n^{-1/2} \sum_{i=1}^n X_{ji} \epsilon_i$ satisfies the following:
\begin{align*}
\Prob\left(\max_{1 \leq j \leq p}\left| \xi_j \right| \geq C\sqrt{\log p}\right) \leq p^{-c}.
\end{align*}

\end{lemma}

By Lemma \ref{lemma:xi_subgaussian}, on the event from Lemma \ref{zhou09:thm_1.6}, with probability tending to 1 as $p \to \infty$, we have that $\|\widehat\eta - \eta\|_2 \lesssim \sqrt{(s_\eta \log p) / n}$ and $\|\widehat\gamma - \gamma\|_2 \lesssim \sqrt{(s_\gamma \log p) / n}$. Thus, $\limsup_{p \to \infty} 
\Prob(\mathcal{G}_1^c) = 0$ and 
\begin{align}\label{eq: supp prob G1star and G1 bound}
    \limsup_{n, p \to \infty}\limsup_{M \to \infty} \Prob([\mathcal{G}_1^\mstar]^c \cup \mathcal{G}_1^c) \leq \alpha_0.
\end{align}
The result follows after showing that 
\begin{align}
    & \limsup_{n, p \to \infty} \limsup_{M \to \infty} \Prob\left(\{\beta \notin \mathrm{CI}^{\mstar}\}\cap \mathcal{G}_1^{\mstar} \cap \mathcal{G}_1 \right) = \alpha' \label{eq:betastar_not_covered}\\
    & \limsup_{n, p \to \infty} \limsup_{M \to \infty} \Prob\left(\l\{m^* \not\in \mathcal{M} \r\}\cap \mathcal{G}_1^\mstar \cap \mathcal{G}_1 \right) = 0.\label{eq:mstar_not_included}
\end{align} 

\subsubsection{Proof of Equation \texorpdfstring{\eqref{eq:betastar_not_covered}}{}}
\label{sec:proof beta in ci*}

Recall the notation $\varphi_1(O_i) = \l(Y_i - X_i^\T\eta\r)\l(D_i - X_i^\T\gamma\r)$,
\begin{align*}
    \varphi_2(O_i) = (D_i - X_i^\T\gamma)^2 \quad \text{and} \quad \varphi_\beta(O) = \frac{\{\varphi_1(O) - \beta \varphi_2(O)\}}{\E\{\varphi_2(O)\}}.
\end{align*}
We have
\begin{align}
    \beta = \frac{\E\{\varphi_1(O)\}]}{\E\{\varphi_2(O)\}]} \equiv \frac{\psi_1}{\psi_2}, \quad \hat{\beta}^\m = \frac{n^{-1} \sum_{i\in\Ical} \widehat\varphi^\m_{1}(O_i)}{n^{-1} \sum_{i\in\Ical} \widehat\varphi^\m_{2}(O_i)} \equiv \frac{\hat{\psi}_{1}^\m}{\hat{\psi}_{2}^\m},
\end{align}
and, for $1\leq m\leq M$, we define:
\begin{align*}
& \widehat\varphi^\m_{1}(O_i) = (Y_i - X_i^\intercal \widehat\eta^\m)(D_i - X_i^\intercal \widehat\gamma^\m), \quad \widehat\varphi^\m_{2}(O_i) = (D_i - X_i^\intercal \widehat\gamma^\m)^2.
\end{align*}
Notice that
\begin{align*}
    \widehat\varphi^\m_{1}(O_i) 
    & = \varphi_1(O_i) + (\hat{\eta}^\m - \eta)^\T X_i X_i^\T (\hat{\gamma}^\m - \gamma) - X_i^\T \epsilon_i (\hat{\gamma}^\m - \gamma) - X_i^\T \delta_i (\hat{\eta}^\m - \eta), \\
    \widehat\varphi^\m_{2}(O_i) 
    & = \varphi_2(O_i) + (\hat{\gamma}^\m - \gamma)^\T X_i X_i^\T (\hat{\gamma}^\m - \gamma) - 2X_i^\T \delta_i (\hat{\gamma}^\m - \gamma).
\end{align*}
Let $\sigma_\beta = \sqrt{\Var\{\varphi_\beta(O_i)\}}$, $\widehat\sigma_\beta = \sqrt{\widehat\Var\{\widehat\varphi_\beta(O_i)\}}$ and $\widehat{\text{SE}}(\widehat\beta) = \widehat\sigma_\beta / \sqrt{n}$. We have:
\begin{align*}
    \frac{\widehat{\beta}^\m - \beta}{\widehat{\text{SE}}(\widehat\beta)} & = \frac{n^{-1/2}\sum_{i\in\Ical} \{\hat{\varphi}_1^\m(O_i) - \beta \hat{\varphi}_2^\m(O_i)\} }{\hat{\psi}_{2}^\m \widehat\sigma_\beta} = \left(\frac{1}{\sqrt{n}}\sum_{i\in\Ical} \frac{\varphi_\beta(O_i)}{\sigma_\beta} + \frac{\Delta^\m}{\psi_2 \sigma_\beta}\right)\cdot \frac{\psi_2 \sigma_\beta}{\hat{\psi}_{2}^\m \widehat\sigma_\beta},
\end{align*}
where $\Delta^\m = \sqrt{n}(R_1^\m + R_2^\m)$ and
\begin{align}
R_1^\m & = \frac{1}{n}\sum_{i \in \Ical} \left\{2\beta X_i^\intercal\delta_i(\widehat\gamma^\m - \gamma) - X_i^\intercal \epsilon_i (\widehat\gamma^\m-\gamma) - X_i^\intercal \delta_i (\widehat\eta^\m-\eta) \right\} \label{eq: supp def R_1^m},\\
R_2^\m & = \left\{(\widehat\eta^\m - \eta) - \beta\cdot (\widehat\gamma^\m - \gamma)\right\}^\intercal \cdot \left(\frac{1}{n}\sum_{i \in \Ical} X_iX_i^\intercal\right)\cdot (\widehat\gamma^\m - \gamma). \label{eq: supp def R_2^m}
\end{align}
Therefore, we have
\begin{align*}
    \Prob\left(\left|\frac{\widehat{\beta}^\m - \beta}{\widehat{\text{SE}}(\widehat\beta)} \right| > z_{\alpha'/2}\right) = \Prob\left(\left| \frac{1}{\sqrt{n}}\sum_{i \in \Ical} \frac{\varphi_\beta(O_i)}{\sigma_\beta} \right| > z_{\alpha'/2} + z_{\alpha'/2} \cdot \left(\frac{\widehat\psi_2^\m\widehat\sigma_\beta}{\psi_2 \sigma_\beta} - 1\right) - \frac{|\Delta^\m|}{\psi_2 \sigma_\beta}\right).
\end{align*}
Let $\tau_0(n, M, p)$ be a sequence of constants converging to zero at an arbitrarily slow rate as $n, p, M \to \infty$; for example, we can set $\tau_0(n, M, p) = (\log \log n)^{-1}$. Define the following events:
\begin{align*}
    \mathcal{G}_2^\mstar &= \l\{\sqrt{n}\l|R_1^\mstar\r| \leq \tau_2(n, M, p)\r\}, \quad \mathcal{G}_3^\mstar = \left\{\sqrt{n}\l|R_2^\mstar\r| \lesssim \tau_3(n, M, p) \right\}, \\
    \mathcal{G}_4^\mstar &= \l\{ \l| \widehat\psi^\mstar_2 / \psi_2 - 1 \r| \leq \tau_4(n,M, p)\r\}, \quad \mathcal{G}_5 = \l\{ \l|\widehat\sigma_\beta / \sigma_\beta - 1 \r| \leq \tau_5(n, M, p)\r\},
\end{align*}
where
\begin{align*}
    & \tau_2(n, M, p) = \frac{c \cdot \sqrt{3}}{\sqrt{\tau_0(n, M, p)}} \cdot \l\{ \l(2|\beta| \|\Lambda\|_{\text{op}}^{1/2} + \|\Sigma\|_{\text{op}}^{1/2}\r) \cdot \sqrt{\frac{s_\eta}{n}} +  \|\Lambda\|_{\text{op}}^{1/2} \cdot \sqrt{\frac{s_\gamma}{n}} \r\} \cdot \err, \\
    & \tau_3(n, M, p) = \{\sqrt{n} + \tau^{-1/2}_0(n, M, p)\} \cdot c \cdot\|\Sigma_X\|_{\text{op}} \cdot \frac{\sqrt{s_\gamma s_\eta} + |\beta| \cdot s_\gamma}{n} \cdot \err^2,  \\
    & \psi_2 \cdot \tau_4(n, M, p) = \sqrt{\frac{\Var\{\varphi_2(O_i)\}}{n \cdot \tau_0(n, M, p)}} + \frac{c \cdot \|\Lambda\|^{1/2}_{\text{op}} \cdot \sqrt{s_\gamma} \cdot \err }{n \cdot \sqrt{\tau_0(n, M, p)}} \\
    & \hphantom{\psi_2 \cdot \tau_4(n, M, p) =} + \left(c + \frac{c}{\sqrt{n \cdot \tau_0(n, M,p)}}\right) \cdot \|\Sigma_X\|_{\text{op}} \cdot \frac{s_\gamma \cdot \err^2}{n},
\end{align*}
and $\tau_5(n, M, p)$ is defined in Lemma \ref{lemma:tau_5} below. Let
\begin{align*}
    \tau(n, M, p) = \tau_4(n, M, p) + \tau_5(n, M, p) + \tau_4(n, M, p)\cdot\tau_5(n, M, p) + \tau_2(n, M, p) + \tau_3(n, M, p)
\end{align*}
so that 
\begin{align*}
\Prob\left(\left|\frac{\widehat{\beta}^\mstar - \beta}{\widehat{\text{SE}}(\widehat\beta)}\right| > z_{\alpha'/2} \cap \mathcal{G}^\mstar_1 \right) & \leq \Prob\left(\left|\frac{1}{\sqrt{n}}\sum_{i \in \Ical} \frac{\varphi_\beta(O_i)}{\sigma_\beta}\right| > z_{\alpha'/2} -c \cdot \tau(n, M, p)\right) \\
& \hphantom{=} + \sum_{j=2}^4 \Prob\left((\mathcal{G}^\mstar_j)^c \cap \mathcal{G}^\mstar_1\right) + \Prob\left(\mathcal{G}_5^c \cap \mathcal{G}_1\right).
\end{align*} 
for some constant $c$. 

By the central limit theorem and Slutsky's theorem, the first term converges to $\alpha'$, since $\tau(n, M, p) \to 0$ as $n, M, p \to \infty$. We will show that
\begin{align}\label{eq:G1-G5_bound}
\limsup_{n, p \to \infty} \limsup_{M \to \infty} \sum_{j=2}^4 \Prob\left([\mathcal{G}^\mstar_j]^c \cap \mathcal{G}^\mstar_1\right) + \Prob\left(\mathcal{G}_5^c \cap \mathcal{G}_1^\mstar\right) = 0,
\end{align}
thereby establishing Equation \eqref{eq:betastar_not_covered}.

Notice that $R_1^\mstar$ has mean-zero given $\Ical^c$ and 
\begin{align*}
\Var(\sqrt{n} R_1^\mstar) \lesssim \E\left\{\|\Lambda\|_{\text{op}}( \beta^2 \|\widehat\gamma^\mstar - \gamma\|_2^2 + \|\widehat\eta^\mstar - \eta\|_2^2) + \|\Sigma\|_{\text{op}}\|\widehat\gamma^\mstar- \gamma\|_2^2\right\},
\end{align*}
so that $\Prob\left((\mathcal{G}_2^\mstar)^c \cap \mathcal{G}_1^\mstar\right) \lesssim \tau_0(n, M, p)$.

Let $v = (\widehat \eta^\mstar - \eta) - \beta (\widehat\gamma^\mstar - \gamma)$ and $w = \widehat\gamma^\mstar - \gamma$. On $\mathcal{G}_1^\mstar$, we have
\begin{align*}
    \tau_3(n, M, p) & \geq \{\sqrt{n} + \tau^{-1/2}_0(n, M, p)\} \cdot \|\E(XX^\intercal)\|_{\text{op}} \cdot(\|\widehat\eta^\mstar - \eta\|_2 + |\beta|\|\widehat\gamma^\mstar - \gamma\|_2) \\
    & \hphantom{\geq} \cdot \|\widehat\gamma^\mstar - \gamma\|_2 \\ 
    & \geq \sqrt{n} \cdot \left| v ^\intercal \Sigma_X w\right| + \frac{\|\Sigma_X^{1/2}v\|_2\|\Sigma_X^{1/2}w\|_2}{\sqrt{\tau_0(n, M, p)}}.
\end{align*}
In this light, we have:
\begin{align*}
& \Prob\left(\{\mathcal{G}_3^\mstar\}^c \cap \mathcal{G}_1^\mstar \mid \Ical^c \right) \\
& = \Prob\left(\left\{\sqrt{n} \left|v^\T \left(\frac{1}{n}\sum_{i \in \Ical} X_i X_i^\T\right) w\right| \gtrsim \tau_3(n, M, p) \right\} \cap \mathcal{G}_1^*\right) \\
& \leq \Prob\left(\left| v^\intercal \left\{\frac{1}{n} \sum_{i \in \Ical}X_i X_i^\intercal - \E(XX^\intercal)\right\} w\right| \gtrsim  \frac{\|\E(XX^\intercal)^{1/2}  v\|_2 \|\E(XX^\intercal)^{1/2}  w\|_2}{\sqrt{n \cdot \tau_0(n, M, p)}} \mid \Ical^c \right) \\
& \leq 2\exp(-c_3 \tau^{-1}_0(n, M, p)) \lesssim \tau_0(n, M, p),
\end{align*}
for $\tau_0(n, M, p)$ sufficiently small, and by Lemma \ref{lem:Lemma11_Zijian} applied with $t = \tau^{-1/2}_0(n, M, p)$ and some constant $c_3$. 

Next, we have
\begin{align*}
    \widehat\psi_2^\mstar - \psi_2 & = \frac{1}{n}\sum_{i \in \Ical} \{\varphi_2(O_i) - \psi_2\} + (\widehat\gamma^\mstar - \gamma)^\intercal \cdot \left(\frac{1}{n}\sum_{i \in \Ical} X_iX_i^\intercal \right) \cdot (\widehat\gamma^\mstar - \gamma) \\
    & \hphantom{=} - \frac{2}{n}\sum_{i \in \Ical} X_i^\intercal \delta_i (\widehat\gamma^\mstar - \gamma).
\end{align*}
Therefore, by Chebyshev inequality and Lemma \ref{lem:Lemma11_Zijian}, we similarly have:
\begin{align*}
\Prob([\mathcal{G}_4^\mstar]^c \cap \mathcal{G}_1^\mstar) \leq 2\exp(-c_3 n) + 2\tau_0(n, M,p) \lesssim \tau_0(n, M, p).
\end{align*}

Finally, we introduce Lemma \ref{lemma:tau_5} to bound $\P(\mathcal{G}_5^c \cap \mathcal{G}_1^\mstar)$. 

\begin{lemma}\label{lemma:tau_5}
Under the conditions of Theorem \ref{thm: coverage lasso}, it holds that
\begin{align*}
    \Prob\left(\mathcal{G}_5^c \cap \mathcal{G}_1\right) = \Prob\left(\left\{\left| \widehat\sigma_\beta / \sigma_\beta - 1\right| \geq \tau_5(n, p)\right\} \cap \mathcal{G}_1 \right) \lesssim \tau_0(n, p),
\end{align*}
where $t_0(n, p)$ is a sequence of constant slowly converging to zero and
\begin{align*}
    \tau_5(n, p) = 1 - \sqrt{1 - \sigma^2_\beta \cdot \tau_5'(n, p)},
\end{align*}
with $\tau_5'(n,p)$ specified in \eqref{eq:supp tau5'}.
\end{lemma}

Lemma \ref{lemma:tau_5} establishes $\Prob(\mathcal{G}_5^c \cap \mathcal{G}_1) \lesssim t_0(n, p)$, where $t_0(n, p)$ is a sequence of constants slowly converging to zero as $n, p \to \infty$. This concludes our proof of Equation \eqref{eq:G1-G5_bound}, and thus of Equation \eqref{eq:betastar_not_covered}.

\subsubsection{Proof of Equation \texorpdfstring{\eqref{eq:mstar_not_included}}{}}
\label{sec:proof m* in M}

Let $\widehat\beta$ denote the original, unperturbed procedure to estimate $\beta$, so that
\begin{align*}
& \widehat\beta - \beta = \left(\frac{1}{n}\sum_{i \in \Ical} \varphi_\beta(O_i) + \Delta\right) \cdot \frac{\psi_2}{\widehat\psi_2},
\end{align*}
where $\Delta = (R_1 + R_2) / \psi_2 $, with $R_1$ and $R_2$ defined as in Equations \eqref{eq: supp def R_1^m} and \eqref{eq: supp def R_2^m} simply with $\widehat\gamma^\m$ replaced by $\widehat\gamma$ and $\widehat\eta^\m$ replaced by $\widehat\eta$. Also, let us redefine $\Delta^\m$ to be $(R_1^\m + R_2^\m) / \psi_2$, so that
\begin{align*}
\widehat\beta^\mstar - \widehat\beta = \left\{\frac{1}{n}\sum_{i \in \Ical} \varphi_\beta(O_i) + \Delta^\mstar \right\} \cdot \frac{\psi_2}{\widehat\psi_2^\mstar} - \left\{\frac{1}{n}\sum_{i \in \Ical} \varphi_\beta(O_i) + \Delta\right\} \cdot \frac{\psi_2}{\widehat\psi_2}.
\end{align*}
Recall that the filtering radius in \eqref{eq: filtering 1} is 
\begin{align*}
    r_n = \rho_n + \rho_{n, M} + \widehat{\rm SE}(\widehat\beta) = \{c^* \log p + \bar{c}\cdot \err^2\} \cdot \frac{\sqrt{s_\gamma s_\eta} + s_\gamma}{n} + \frac{\widehat\sigma_\beta}{\sqrt{n}}.
\end{align*}
In this light, we have
\begin{align}
    & \Prob\left(\l\{m^* \not\in \mathcal{M}\r\} \cap \mathcal{G}_1^\mstar \cap \mathcal{G}_1 \right) \nonumber \\
    & \leq \Prob\l( \l\{ \l| \Delta^\mstar \r| > \frac{\widehat\psi_2^\mstar}{\psi_2} \cdot \left(\rho_{n, M} + \frac{\widehat\sigma_\beta}{4 \sqrt{n}} \right) \r\} \cap \mathcal{G}_1^\mstar \cap \mathcal{G}_1\r) \\
    & \hphantom{\leq} + \Prob\l( \l\{ \l| \Delta \r| > \frac{\widehat\psi_2}{\psi_2} \cdot \left(\rho_n  + \frac{\widehat\sigma_\beta}{4 \sqrt{n}}\right) \r\} \cap \mathcal{G}_1\r) \\
    & \hphantom{\leq} + \Prob\l( \l\{ \l| \frac{1}{n}\sum_{i\in \Ical}\varphi_\beta(O_i) \r| \cdot \left| \frac{\psi_2}{\widehat\psi_2^\mstar} - 1 \right| > \frac{\widehat\sigma_\beta}{4 \sqrt{n}} \r\} \cap \mathcal{G}_1^\mstar\r) \\
    & \hphantom{\leq} + \Prob\l( \l\{ \l| \frac{1}{n}\sum_{i\in \Ical}\varphi_\beta(O_i) \r| \cdot \left| \frac{\psi_2}{\widehat\psi_2} - 1 \right| > \frac{\widehat\sigma_\beta}{4 \sqrt{n}}\r\} \cap \mathcal{G}_1^\mstar\r). \label{eq: supp coverage prob term2}
\end{align}
Notice that, for $n$ sufficiently large, there exists a constant $C$ such that
\begin{align*}
    \rho_{n, M} + \frac{\widehat\sigma_\beta}{4\sqrt{n}} & = \rho_{n, M} + \frac{\sigma_\beta}{4\sqrt{n}} + \left(\frac{\widehat\sigma_\beta}{\sigma_\beta} - 1\right) \cdot \frac{\sigma_\beta}{4\sqrt{n}} \geq C \cdot \frac{\tau_3(n, M, p)}{\sqrt{n}} + \frac{\sigma_\beta}{4\sqrt{n}} +\left(\frac{\widehat\sigma_\beta}{\sigma_\beta} - 1\right) \cdot \frac{\sigma_\beta}{4\sqrt{n}},
\end{align*}
and, on the event $\mathcal{G}_1^\mstar$, by Lemma \ref{lemma:tau_5}, $|\widehat\sigma_\beta / \sigma_\beta - 1| > \tau_5(n, p)$ with probability no larger than (a constant multiple of) $\tau_0(n, p) \to 0$. Similarly, under even $\mathcal{G}_1^\mstar$, $|\widehat\psi_2^\mstar / \psi_2 - 1| > \tau_4(n, M, p)$ with probability no larger than (a constant multiple of) $\tau_0(n, M, p) \to 0$. In this respect, for $n$ sufficiently large:
\begin{align*}
    & \Prob\l( \l\{ \l| \Delta^\mstar \r| > \frac{\widehat\psi_2^\mstar}{\psi_2} \cdot \left(\rho_{n, M} + \frac{\widehat\sigma_\beta}{4 \sqrt{n}} \right) \r\} \cap \mathcal{G}_1^\mstar \cap \mathcal{G}_1\r) \\
    & \lesssim \Prob\l( \l\{ \l| \Delta^\mstar \r| > C \cdot \frac{\tau_3(n, M, p)}{\sqrt{n}} + \frac{\sigma_\beta\{1- \tau_5(n, p)\}}{4\sqrt{n}} - \frac{\tau_4(n, M, p)}{\sqrt{n}}\r\} \cap \mathcal{G}_1^\mstar \cap \mathcal{G}_1\r) + \tau_0(n, M, p) \\
    & \leq \Prob\l( \l\{ \l| \Delta^\mstar \r| > C' \cdot \frac{\tau_3(n, M, p) + \tau_2(n, M, p)}{\sqrt{n}} \r\} \cap \mathcal{G}_1^\mstar \cap \mathcal{G}_1\r) + \tau_0(n, M, p),
\end{align*}
since $\tau_3(n, M, p)$ is of order greater than $\tau_4(n, M, p)$, $\tau_5(n, p) \to 0$ and, given an appropriate choice of $\tau_0(n, M, p)$, $n^{-1/2} \cdot \tau_2(n, M, p)$ is of order no larger than $n^{-1/2}$. Therefore, there exists $\bar{c}$ and $C$ such that we can bound the probability above as
\begin{align*}
    & \Prob\l( \l\{ \l| \Delta^\mstar \r| > C' \cdot \frac{\tau_3(n, M, p) + \tau_2(n, M, p)}{\sqrt{n}} \r\} \cap \mathcal{G}_1^\mstar \cap \mathcal{G}_1\r) + \tau_0(n, M, p) \\
    & \lesssim \Prob\l((\mathcal{G}_2^\mstar)^c \cap \mathcal{G}_1^\mstar\r) + \Prob\l((\mathcal{G}_3^\mstar)^c \cap \mathcal{G}_1^\mstar\r) + \tau_0(n, M, p) \\
    & \lesssim \tau_0(n, M, p).
\end{align*}
A similar argument, with $\err$ replaced by $\sqrt{\log p}$, yields that
\begin{align*}
    \Prob\l( \l\{ \l| \Delta \r| > \frac{\widehat\psi_2}{\psi_2} \cdot \left(\rho_n  + \frac{\widehat\sigma_\beta}{4 \sqrt{n}}\right) \r\} \cap \mathcal{G}_1\r) \lesssim \tau_0(n, M, p).
\end{align*}
Finally, on $\mathcal{G}_4^\mstar$, we have:
\begin{align*}
\left|\frac{\psi_2}{\widehat\psi_2^\mstar} - 1\right| & = \frac{|1-\widehat\psi_2^\mstar / \psi_2|}{\widehat\psi_2^\mstar / \psi_2} \\
& \leq \frac{\tau_4(n, M, p)}{1-\tau_4(n, M, p)} \\
& \lesssim \frac{1}{\sqrt{n \cdot \tau_0(n, M, p)}} \cdot \left(1 + \sqrt{\frac{s_\gamma}{n}} \cdot \err + \frac{s_\gamma \cdot \err^2}{\sqrt{n}}\right),
\end{align*}
so that:
\begin{align*}
    & \Prob\left(\left|\frac{1}{n}\sum_{i \in \Ical}\varphi_\beta(O_i)\right| \cdot \left|\frac{\psi_2}{\widehat\psi_2^\mstar} - 1\right| > \frac{\widehat\sigma_\beta}{4 \sqrt{n}} \cap \mathcal{G}_1^\mstar \right) \\
    & \lesssim \Prob\bigg(\left|\frac{1}{n}\sum_{i \in \Ical}\varphi_\beta(O_i)\right| \cdot \left(1 + \sqrt{\frac{s_\gamma}{n}} \cdot \err + \frac{s_\gamma \cdot \err^2}{\sqrt{n}}\right) \\
    &\quad\qquad\gtrsim \sqrt{\tau_0(n, M, p)} \cdot \{\sigma_\beta  - \tau_5(n, M, p)\}\bigg) + \tau_0(n, M, p) \\
    & \lesssim \frac{1}{\tau_0(n, M, p) \cdot n}\left(1 + \frac{s_\gamma \cdot \err^2 + s_\gamma^2 \cdot \err^4}{n}\right) + \tau_0(n, M, p) \\
    & \to 0 \quad \text{ as } n, p, M \to \infty.
\end{align*}
The same bound holds when $\widehat\psi_2^\mstar$ is replaced by $\widehat\psi_2$ by replacing $\err$ with $\sqrt{\log p}$, thus proving \eqref{eq:mstar_not_included}.

\subsubsection{Length of the confidence interval \label{appendix:length_CI_lasso}}

Regarding the length of the confidence interval ${\rm CI}$ defined in \eqref{eq:filtered union CI}, note that
\begin{equation*}
    \len({\rm CI}) \leq 2\left\{\max_{m\in\mathcal{M}}|\hat{\beta}^\m - \hat{\beta}| + z_{\alpha'/2} \cdot\frac{\sigma_\beta}{\sqrt{n}} \cdot \l(\frac{\widehat\sigma_\beta}{\sigma_\beta} - 1\r) + z_{\alpha'/2} \cdot \frac{\sigma_\beta}{\sqrt{n}}\right\}.
\end{equation*}
By the construction of $\mathcal{M}$ in \eqref{eq: filtering 1}, we have 
\begin{align*}
    \max_{m\in\mathcal{M}} |\hat{\beta}^\m - \hat\beta| \leq 1.01 \cdot \rho_n + \frac{\widehat\sigma_\beta}{\sqrt{n}}.
\end{align*}
By Lemma \ref{lemma:tau_5}, we have that $|\widehat\sigma_\beta / \sigma_\beta - 1| \leq \tau_5(n, p) \to 0$ with probability tending to 1. Therefore, with probability tending to 1, there exists an arbitrarily small positive constant $c$ such that 
$$\len({\rm CI}) \leq 2.02 \cdot \rho_n + (4 + c)\sigma_\beta \cdot n^{-1/2}.$$

\subsection{Proof of Theorem \ref{thm:mstar ml beta} (in the supplement)}

\subsubsection{Quantifying \texorpdfstring{$|\hat{\beta}^\m - \hat{\beta}^{\rm ora}|$}{}}

We first define a mapping from the simulated noise terms to the perturbed DML estimator in each perturbation. Let $e_i^\m$ denote the generated noise vector in the perturbation step and let $z_i^\m$ denote the standardized $e_i^\m$, i.e. 
\begin{align*}
    e_i^\m = \bigg(
    \begin{array}{c}
        \epsilon_i^\m \\
        \delta_i^\m
    \end{array}
    \bigg) \sim \mathcal{N}_2(0, \hat{\Pi}), \quad z_i^\m = \hat{\Pi}^{-1/2}e_i^\m.
\end{align*}
By this construction, it holds that $z_i^\m \sim \mathcal{N}_{2}(0, I)$ conditional on the sample $\Ical_0$ which is used to generate $\hat{\Pi}$. We also define the stacking vector across individuals as 
\begin{align*}
    e^\m = \left(
    \begin{array}{c}
        e_1^\m \\
        \vdots \\
        e_n^\m
    \end{array}
    \right) \sim \mathcal{N}_{2n}(0, I_n \otimes \hat{\Pi}), \quad z^\m = (I_n \otimes \hat{\Pi}^{-1/2})e^\m = \left(
    \begin{array}{c}
        z_1^\m \\
        \vdots \\
        z_n^\m
    \end{array}
    \right) \sim \mathcal{N}_{2n}(0,I_{2n}),
\end{align*}
where $I_n$ is the $n\times n$ identity matrix and $\otimes$ denotes the Kronecker product. Concretely, for a $2\times 2$ matrix $A$, $I_n \otimes A$ is the $2n\times 2n$ block diagonal matrix with $A$ repeated along the diagonal $n$ times. Conditioning on the observed data, define the mapping 
\begin{align*}
    \psi: \RR^{2n} \to \RR, \quad  \psi(z^\m) = \hat{\beta}^\m,
\end{align*}
where the mapping $\psi$ is the composition of the following mappings:
\begin{align*}
    z^\m \to 
    \bigg(
    \begin{array}{c}
         \hat g^\m \\
         \hat f^\m
    \end{array}
    \bigg) \to \hat{\beta}^\m \quad \textrm{with }  \hat{\beta}^\m = \frac{\sum_{i\in\Ical}(Y_i - \hat{g}^\m(X_i))(D_i - \hat{f}^\m(X_i))}{\sum_{i\in\Ical} (D_i - \hat{f}^\m(X_i))^2}. 
\end{align*}
Here in the first step, $z^\m$ is injected into the perturbed ML training step to produce the nuisance predictors $\hat g^\m$ and $\hat f^\m$. The second step constructs the perturbed DML estimator $\hat{\beta}^\m$ using perturbed nuisance estimators. 
 
Next, we shall derive the isoperimetric inequality for the space $\RR$ of $\hat{\beta}^\m$. Given the mapping $\psi$, we define two corresponding  partitions for the space $\RR^{2n}$ of $z^\m$ and the space $\RR$ of $\hat{\beta}^\m$. Let $\P_z$ denote the standard Gaussian measure on $\RR^{2n}$ conditioning on the sample $\Ical_0$ and $\P_\beta$ denote the push-forward measure on $\RR$ via the mapping $\psi$ conditioning on the observed data $\mathcal{O}$, i.e. the conditional distribution of $\hat{\beta}^\m$. 
Notice that $\hat{\beta}^{\rm ora}$, the target to recover, satisfies $\sqrt{n}(\hat{\beta}^{\rm ora} - \beta) \indist \mathcal{N}(0, \sigma_\beta^2)$ with $\sigma_\beta^2=\E[(\epsilon_i-\beta\delta_i)^2\delta_i^2]/(\E[\delta_i^2])^2$. Thus we can use the interval $T_0$ defined in \eqref{eq:interval betaHat ora} to account for the uncertainty of $\hat{\beta}^{\rm ora}$. We employ this interval and the measure $\P_\beta$ to construct the partition in $\RR$. Let $\alpha_{T_0}$ be the smallest tail quantile of this interval under measure $\P_\beta$ as defined in Assumption \ref{assump:resample distribution}. Note that we have $\tilde{\alpha} = 2\alpha_{T_0}$. Let $\{B_0, B_1, \dots, B_K, B_{K+1}\}$ be a partition of $\RR$ into $K+2$ intervals, arranged sequentially along the real line such that 
\begin{align}
    \P_\beta(B_0) = \P_\beta(B_{K+1}) = \frac{\tilde\alpha}{2}, \quad \P_\beta(B_k) = \frac{1-\tilde\alpha}{K} \quad \textrm{for }k=1,\dots,K. \label{eq:supp construction of partiton in R}
\end{align}
The construction of $B_k$ depends on the cumulative distribution function of $\hat{\beta}^\m$. Specifically, for some constants $c_k$ and $1\leq k\leq K$, we have $B_0 = (-\infty, c_0], B_k = (c_{k-1}, c_{k}], B_{K+1} = (c_{K}, +\infty)$. Based on the constructed intervals $\{B_k\}_k$, we define the corresponding subsets in $\RR^{2n}$ by
\begin{align*}
    A_k = \{z: \psi(z) \in B_k \} \quad \textrm{for } k=0,1,\dots,K+1.
\end{align*}
The subsets $\{A_k\}_k$ are the preimages of $\{B_k\}_k$ under the mapping $\psi$. By definition of the push-forward measure, we have 
\begin{align}
    \P_z(A_k) = \P_\beta(B_k), \quad \textrm{for } k=0,1, \dots, K+1. \label{eq:supp same measure Ak and Bk}
\end{align}
In addition, we shall note that $\{A_k\}_k$ forms a partition of $\RR^{2n}$: (i) since $\{B_k\}_k$ are disjoint, their preimages $\{A_k\}_k$ are also disjoint; (ii) the union of $\{A_k\}_k$ cover the whole space $\RR^{2n}$ because for any $z^\m\in\RR^{2n}$, $\psi(z^\m)\in B_k$ for exactly one $k$.

With the defined measures and partitions, we can derive the isoperimetric inequality for $\RR$ and $\P_\beta$ from that of the standard Gaussian in $\RR^{2n}$. In the space $\RR^{2n}$, the input $z^\m$ follows the $2n$-dimensional standard Gaussian distribution. The following lemma from \citeapp{cousins2018gaussian} states a 3-set isoperimetric inequality for Gaussian (in fact for all strongly log-concave measures).

\begin{lemma}\label{lem:isoperimetric inequality}
    (Theorem 5.4 in \citeapp{cousins2018gaussian})
    Let $\P_z$ be the standard Gaussian measure. Let $S_1, S_2, S_3$ be a partition of $\RR^{2n}$. Then,
    \begin{align*}
        \P_z(S_3) \geq \log(2) \cdot d(S_1, S_2) \cdot \P_z(S_1) \cdot \P_z(S_2),
    \end{align*}
  where $d(S_1, S_2) := \min \l\{\|{x-y}\|_{2}: x \in S_1, y \in S_2\r\}$.
\end{lemma}
Since Lemma \ref{lem:isoperimetric inequality} is applied on a 3-set partition, for each $A_k$, we consider 
\begin{align*}
    S_1 = \bigcup_{j=0}^{k-1} A_j, \quad S_2 = \bigcup_{j=k+1}^{K+1} A_j, \quad S_3 = A_k \quad \textrm{for }k=1,\dots,K.
\end{align*}
The triple $\{S_1, S_2, S_3\}$ forms a valid partition of $\RR^{2n}$. Then, for $k=1,\dots,K$, we have
\begin{align}
    \P_z(A_k) & \geq \log(2) \cdot d\l(\cup_{j=0}^{k-1} A_j, \cup_{j=k+1}^{K+1} A_j\r) \cdot \P_z\l(\cup_{j=0}^{k-1} A_j\r) \cdot \P_z\l(\cup_{j=k+1}^{K+1} A_j\r) \nonumber \\
    & \geq \log(2) \cdot d\l(\cup_{j=0}^{k-1} A_j, \cup_{j=k+1}^{K+1} A_j\r) \cdot \P_z(A_0) \cdot \P_z(A_{K+1}). \label{eq:supp iso ineq for R^2p}
\end{align}

To introduce a similar isoperimetric inequality in the space $\RR$, we need to connect the distance in $\RR^{2n}$ to that in $\RR$. We introduce the following lemma to show that the mapping $\psi$ is $L^*$-Lipschitz continuous for some positive constant $L^*>0$, then we rely on this continuity to transform the distance $d(S_1, S_2)$ in $\RR^{2n}$ to the distance in $\RR$.

\begin{lemma}\label{lem:lipschitz continuity}
    Let $t_0(n)$ be some slowly increasing sequence in $n$ (e.g., $t_0(n)=\log\log n$). Under the conditions of Theorem \ref{thm:mstar ml beta}, with the probability at least $1 - c(1/t_0(n) + \tau_n)$, the mapping $\psi$ is $L^*$-Lipschitz continuous. That is, for any $z_1,z_2\in\RR^{2n}$, it holds that
    $$\l|\psi(z_1) - \psi(z_2)\r| \leq L^* \|z_1 - z_2\|_2$$
    with $L^* = \frac{C\max\{L_g, L_f\}}{\sqrt{n}}$ for some constant $C>0$.
\end{lemma}

Based on Lemma \ref{lem:lipschitz continuity}, we define the high-probability event
\begin{align*}
    \mathcal{E}_1 = \l\{ \l|\psi(z_1) - \psi(z_2)\r| \leq L^* \|z_1 - z_2\|_2 \textrm{ for any two } z_1, z_2 \in \RR^{2n}\r\}.
\end{align*}
On the event $\mathcal{E}_1$, by the definition of $d(\cdot, \cdot)$ in Lemma \ref{lem:isoperimetric inequality}, we connect the distance $d\l(\cup_{j=0}^{k-1} A_j, \cup_{j=k+1}^{K+1} A_j\r)$ used in \eqref{eq:supp iso ineq for R^2p} to the corresponding distance in $\RR$:
\begin{align}\label{eq:supp distance connect}
    d\l(\cup_{j=0}^{k-1} A_j, \cup_{j=k+1}^{K+1} A_j\r) & = \min \l\{\|{z_1 - z_2}\|_{2}: z_1 \in \cup_{j=0}^{k-1} A_j, z_2 \in \cup_{j=k+1}^{K+1} A_j\r\} \nonumber \\
    & \geq \frac{1}{L^*} \cdot \min \l\{|{\psi(z_1) - \psi(z_2)}|: \psi(z_1) \in \cup_{j=0}^{k-1} B_j, \psi(z_2) \in \cup_{j=k+1}^{K+1} B_j\r\} \cdot \mathbf{1}_{\mathcal{O}\in\mathcal{E}_1}\nonumber \\
    & = \frac{1}{L^*} \cdot d\l(\cup_{j=0}^{k-1} B_j, \cup_{j=k+1}^{K+1} B_j\r) \cdot \mathbf{1}_{\mathcal{O}\in\mathcal{E}_1},
\end{align}
where the inequality is obtained by applying Lemma \ref{lem:lipschitz continuity}.
By \eqref{eq:supp distance connect} and the measure property in \eqref{eq:supp same measure Ak and Bk}, we derive the isoperimetric ineuqality in the space $\RR$:
\begin{align}
    \P_\beta(B_k) \geq \frac{\log(2)}{L^*} \cdot d\l(\cup_{j=0}^{k-1} B_j, \cup_{j=k+1}^{K+1} B_j\r) \cdot \P_\beta(B_0) \cdot \P_\beta(B_{K+1}) \cdot \mathbf{1}_{\mathcal{O}\in\mathcal{E}_1}. \label{eq:supp iso ineq for R}
\end{align}
Based on the construction of intervals $\{B_k\}_k$ in \eqref{eq:supp construction of partiton in R}, we then have, for $k=1,\dots,K$, 
\begin{align}\label{eq:supp upper bound for |B_k|}
    \frac{1 - \tilde\alpha}{K} \geq \frac{\log(2)}{L^*} \cdot |B_k| \cdot \frac{\tilde\alpha^2}{4} \quad \Rightarrow \quad |B_k| \leq \frac{4L^*(1-\tilde\alpha)}{\log(2)K\tilde\alpha^2} \leq \frac{4L^*}{\log(2)K\tilde\alpha^2}.
\end{align}
where, for notation simplicity, $|B_k|$ is used to denote the distance $d\l(\cup_{j=0}^{k-1} B_j, \cup_{j=k+1}^{K+1} B_j\r)$. This distance is essentially the length of the interval $B_k$ because both $\cup_{j=0}^{k-1} B_j$ and $\cup_{j=k+1}^{K+1} B_j$ are intervals and are separated by $B_k$ by construction. When the partition size $K$ increases, the interval length $|B_k|$ decreases.

We first control the probability where none of $\hat\beta^\m$ falls in a given $B_k$ for $k=1,\dots,K$:
\begin{align*}
    \P\l(\bigcap_{m=1}^M \{\hat{\beta}^\m \notin B_k\} \mid \mathcal{O}\r)= \l(1-\P_\beta(B_k)\r)^{M}\leq \exp(-M \cdot \P_\beta(B_k)).
\end{align*}
Here we use the independence of $\hat{\beta}^{[m]}$ conditioning on the observed data $\mathcal{O}$. This implies that the probability of the event in which there exists an empty interval $B_k$ containing no $\hat{\beta}^m$ is:
\begin{align*}
    \P\l(\bigcup_{k=1}^K \bigcap_{m=1}^M \{\hat{\beta}^\m \notin B_k\} \mid \mathcal{O}\r) \leq \sum_{k=1}^K\exp(-M \cdot \P_\beta(B_k)) = K\exp(-M \cdot \P_\beta(B_k)),
\end{align*}
where the last equality is because $\P_\beta(B_k)$ is the same for $k=1,\dots,K$ by construction. Consequently, with probability at least $1 - K\exp(-M \cdot \P_\beta(B_k))$, 
every interval $B_k$ with $k=1,\dots,K$ contains at least one sample $\hat{\beta}^{(m)}$, i.e.,
\begin{align*}
    \P\l(\bigcap_{k=1}^K \bigcup_{m=1}^M \{\hat{\beta}^\m \in B_k\} \mid \mathcal{O}\r) \geq 1- K\exp(-M \cdot \P_\beta(B_k)).
\end{align*}
When this occurs, the union of intervals centered at $\hat{\beta}^\m$ covers $\cup_{k=1}^K B_k$, provided the interval radius $r$ is at least the length of each interval 
$B_k$ for $k=1,\dots,K$. Built upon this intuition, we have 
\begin{align}\label{eq:supp cover space event}
    \P\l( \l\{\bigcup_{k=1}^K B_k\r\} \subseteq \l\{\bigcup_{m=1}^{M} B(\hat{\beta}^\m, r)\r\} \mid \mathcal{O} \r) \geq 1 - K\exp(-M \cdot \P_\beta(B_k)),
\end{align}
where $B(\hat{\beta}^\m,r):= \{v: |\hat{\beta}^\m-v|\leq r\}$ is the interval with radius $r$ centered at $\hat{\beta}^\m$, and the radius $r$ satisfies $r \geq \max_{1\leq k\leq K}|B_k|$. Note that for all intervals $B_k$ with $k=1,\dots,K$, we have derived a common upper bound for their lengths in \eqref{eq:supp upper bound for |B_k|}. We can simply use this upper bound as a valid radius, so we set 
\begin{align}\label{eq:supp radius of balls}
    r = \frac{4L^*}{\log(2)K\tilde\alpha^2}.
\end{align}
To determine the partition size $K$, a simple choice is to set the probability in \eqref{eq:supp cover space event} as $1-1/\sqrt{n}$, which requires
\begin{align*}
    K\exp(-M \cdot \P_\beta(B_k)) = K\exp\l(-M \cdot \frac{1-\tilde{\alpha}}{K}\r) \leq \frac{1}{\sqrt{n}},
\end{align*}
implying 
\begin{align*}
    \frac{(1-\tilde\alpha)M}{K} \exp\l\{ \frac{(1-\tilde\alpha)M}{K} \r\} \geq (1-\tilde\alpha)\sqrt{n}M. 
\end{align*}
If $ye^y = x$, then $y = W(x)$ where $W(\cdot)$ is the Lambert W function \citepapp{lehtonen2016lambert}. Hence we take $K$ as 
\begin{align}\label{eq:supp choice of K}
    K = \l\lfloor\frac{(1-\tilde\alpha)M}{W((1-\tilde\alpha)\sqrt{n}M)}\r\rfloor.
\end{align}

With the choice of $r$ in \eqref{eq:supp radius of balls} and the choice of $K$ in \eqref{eq:supp choice of K}, the inequality in \eqref{eq:supp cover space event} becomes
\begin{align}\label{eq:supp cover union of B_k}
    \P\l( \l\{\bigcup_{k=1}^K B_k\r\} \subseteq \l\{\bigcup_{m=1}^{M} B\l(\hat{\beta}^\m, r\r)\r\} \mid \mathcal{O} \r) \geq \l(1 - \frac{1}{\sqrt{n}}\r) \cdot \mathbf{1}_{\mathcal{O}\in\mathcal{E}_1},
\end{align}
where $r = \frac{4L^*}{\log(2)\tilde\alpha^2} \l( \l\lfloor\frac{(1-\tilde\alpha)M}{W((1-\tilde\alpha)\sqrt{n}M)}\r\rfloor \r)^{-1}$.
By the construction of the partition $\{B_k\}_k$, conditioning on data, the interval $T_0=[\beta - z_{\alpha_0/2}\sigma_\beta \cdot n^{-1/2}, \ \beta + z_{\alpha_0/2}\sigma_\beta \cdot n^{-1/2}]$ is a subset of $\cup_{k=1}^K B_k$. Define the event 
\begin{align*}
    \mathcal{E}_2 = \l\{\hat{\beta}^{\rm ora} \in T_0\r\}.
\end{align*}
Note that $\liminf_{n\to\infty}\P(\mathcal{E}_2) = 1-\alpha_0$. On the event $\mathcal{E}_1 \cap \mathcal{E}_2$, the probability in \eqref{eq:supp cover union of B_k} implies 
\begin{align*}
    \P\l( \hat{\beta}^{\rm ora} \in \l\{\bigcup_{m=1}^{M} B\l(\hat{\beta}^\m, r\r)\r\} \mid \mathcal{O} \r) \geq \l(1 - \frac{1}{\sqrt{n}}\r) \cdot \mathbf{1}_{\mathcal{O}\in\mathcal{E}_1\cap\mathcal{E}_2}.
\end{align*}
That is, with a high probability, $\hat{\beta}^{\rm ora}$ falls into the neighborhood of at least one $\hat{\beta}^\m$. In other words, there exists one $\hat{\beta}^\m$ whose distance to $\hat{\beta}^{\rm ora}$ is controlled within $r$:
\begin{align*}
    \P\l( \exists 1\leq m\leq M: |\hat{\beta}^\m - \hat{\beta}^{\rm ora}| \leq r\mid \mathcal{O}\r) \geq \l(1 - \frac{1}{\sqrt{n}}\r) \cdot \mathbf{1}_{\mathcal{O}\in\mathcal{E}_1\cap \mathcal{E}_2}.
\end{align*}
Taking the expectation over the observed data $\mathcal{O}$ on both sides, we get
\begin{align*}
    \P\l( \exists 1\leq m\leq M: |\hat{\beta}^\m - \hat{\beta}^{\rm ora}| \leq r\r) \geq \l(1 - \frac{1}{\sqrt{n}}\r) \cdot \P(\mathcal{E}_1\cap\mathcal{E}_2).
\end{align*}
Let $f(u)=u\exp(u)$. By the definition of Lambert W function, $f(W(x)) = x$. Note that $f(\log x) = x\log x$, so we get $f(\log x) > f(W(x))$ for all $x>e$. Because $f(u)$ is strictly increasing on $[-1,\infty)$, it follows that $\log x > W(x)$ whenever $x>e$. Combining this inequality with $\tilde\alpha = 2\alpha_{T_0}$ and $L^*=C\max\{L_g, L_f\}/\sqrt{n}$, we have
\begin{align*}
    r &= \frac{4L^*}{\log(2)\tilde\alpha^2} \l( \l\lfloor\frac{(1-\tilde\alpha)M}{W((1-\tilde\alpha)\sqrt{n}M)}\r\rfloor \r)^{-1} \\
    &\leq \frac{4L^*}{\log(2)\tilde\alpha^2} \l( \l\lfloor\frac{(1-\tilde\alpha)M}{\log((1-\tilde\alpha)\sqrt{n}M)}\r\rfloor \r)^{-1} \\ 
    &\lesssim \frac{\max\{L_g, L_f\}}{\log(2)}\frac{1}{\alpha_{T_0}^2\sqrt{n}} \l( \l\lfloor\frac{(1-2\alpha_{T_0})M}{\log(\sqrt{n}M)}\r\rfloor \r)^{-1}.
\end{align*}

Taking the limit as $n\to\infty$ and $M\to\infty$, we establish Theorem \ref{thm:mstar ml beta}.

\subsubsection{Coverage and length of the confidence interval}

As the statement of the theorem follows by essentially the same arguments used to prove Theorem \ref{thm: coverage lasso}, we omit certain details. Let $m^*$ denote the smallest index such that the following event holds (if it holds for at least one $1 \leq m \leq M$): 
\begin{align*}
    \mathcal{G}_1^\mstar = \left\{\left| \widehat\beta^\mstar - \widehat\beta^{\rm ora} \right| \leq \bar{C} \cdot \errml \right\},
\end{align*}
and define 
\begin{align*}
    \mathcal{G}_1 = \left\{ \|\widehat{f} - f\|_{2, \P_X} \leq R_{2, f}, \quad \|\widehat{g} - g\|_{2, \P_X} \leq R_{2,g}, \quad \|\widehat{f} - f\|_{4, \P_X} \leq R_{4, f}, \quad \|\widehat{g} - g\|_{4, \P_X} \leq R_{4, g} \right\}.
\end{align*}
Notice that for $M$ large enough, we have $\bar{C} \cdot \errml \leq 0.01$. In this light, we prove the coverage statement under the smaller filtering radius $\rho_n + \bar{C} \cdot \errml + \widehat\sigma_\beta / \sqrt{n}$. By Theorem \ref{thm:mstar ml beta} and Assumption \ref{assump:nuisance convergence}, we have $\Prob\left((\mathcal{G}_1^\mstar)^c \cup \mathcal{G}^c_1\right) \leq \alpha_0 + \tau_n$. Therefore, we have
\begin{align*}
\Prob\left(\beta \not\in \rm {CI} \right) & \leq \Prob\left(\left\{\left| \widehat\beta^\mstar - \widehat\beta\right| > (R_{2, g} + R_{2,f}) R_{2, f} + \bar{C} \cdot \errml + \frac{\widehat\sigma_\beta}{\sqrt{n}}\right\} \cap \mathcal{G}_1^\mstar \cap \mathcal{G}_1\right) \\
& \hphantom{\leq}  + \Prob\left(\left\{\frac{\left| \widehat\beta^\mstar - \beta\right|}{\widehat\sigma_\beta / \sqrt{n}} > z_{\alpha'/2} \right\} \cap \mathcal{G}_1^\mstar \cap \mathcal{G}_1\right)
+ \alpha_0 + \tau_n.
\end{align*}
On the event $\mathcal{G}_1^\mstar$, we have
\begin{align*}
\left| \widehat\beta^\mstar - \widehat\beta\right| & \leq    \left|\widehat\beta^\mstar - \widehat\beta^{\rm ora} \right| + \left|\widehat\beta - \widehat\beta^{\rm ora} \right| \leq \bar{C} \cdot \errml + \left|\widehat\beta - \widehat\beta^{\rm ora} \right|. 
\end{align*}
By the same reasoning as in the Proof of Lemma \ref{lemma:tau_5}, we can find a sequence of constants $t_5(n) \to 0$ such that, for $\mathcal{G}_5 = \left\{|\widehat\sigma_\beta / \sigma_\beta - 1 | \leq t_5(n)\right\}$, it holds that $\Prob\left(\mathcal{G}_5^c \cap \mathcal{G}_1\right) \leq t_0(n)$, where $t_0(n) \to 0$. Thus, we have
\begin{align*}
   & \Prob\left(\left\{\left| \widehat\beta^\mstar - \widehat\beta\right| > (R_{2, g} + R_{2,f}) R_{2, f} + \bar{C} \cdot \errml + \frac{\widehat\sigma_\beta}{\sqrt{n}}\right\} \cap \mathcal{G}_1^\mstar \cap \mathcal{G}_1\right) \\
   & \leq \Prob\left(\left\{\left| \widehat\beta^{\rm ora} - \widehat\beta\right| > (R_{2, g} + R_{2,f}) R_{2, f} + \frac{\sigma_\beta\{1 - t_5(n)\}}{\sqrt{n}} \right\} \cap \mathcal{G}_1^\mstar \cap \mathcal{G}_1 \cap \mathcal{G}_5 \right) + t_0(n).
\end{align*}
Reasoning as in the proof of Lemma \ref{lem: connect beta m with eta m gamma m}, we have 
\begin{align*}
    \left| \widehat\beta - \widehat\beta^{\rm ora}\right| & \lesssim \left|\widehat\psi_1 - \widehat\psi_1^{\rm ora}\right| + \left|\widehat\psi_2 - \widehat\psi_2^{\rm ora}\right| \\
    & = \left|\frac{1}{n}\sum_{i \in \Ical} \epsilon_i\{\widehat{f}(X_i) - f(X_i)\} + \delta_i\{\widehat{g}(X_i) - g(X_i)\} + \{\widehat{f}(X_i) - f(X_i)\}\{\widehat{g}(X_i) - g(X_i)\}\right| \\
    & \hphantom{=} + \left|\frac{2}{n}\sum_{i \in \Ical} \delta_i\{\widehat{f}(X_i) - f(X_i)\} + \{\widehat{f}(X_i) - f(X_i)\}^2\right|.
\end{align*}
Next, notice that
\begin{align*}
    & \Prob\left(\left\{\left|\frac{1}{n}\sum_{i \in \Ical} \{\widehat{f}(X_i) - f(X_i)\}\{\widehat{g}(X_i) - g(X_i)\}\right| > R_{2, g} \cdot R_{2, f} + \frac{\widehat\sigma_\beta}{5\sqrt{n}}\right\} \cap \mathcal{G}_1 \mid \Ical^c \right) \\
    & \leq \Prob\left(\left\{\left|\frac{1}{n}\sum_{i \in \Ical} \{\widehat{f}(X_i) - f(X_i)\}\{\widehat{g}(X_i) - g(X_i)\} \right.\right. \right. \\
    & \hphantom{\Prob\left(\left\{\left|\frac{1}{n}\sum_{i \in \Ical} \right.\right.\right. }   \left.\left.\left. \vphantom{\frac{1}{n}} - \E\left[ \{\widehat{f}(X_i) - f(X_i)\}\{\widehat{g}(X_i) - g(X_i)\} \mid \Ical^c\right]\right| >  \frac{\widehat\sigma_\beta}{5\sqrt{n}}\right\} \cap \mathcal{G}_1 \mid \Ical^c\right) \\
    & \lesssim \sqrt{\E\left[ \{\widehat{f}(X_i) - f(X_i)\}^2\{\widehat{g}(X_i) - g(X_i)\}^2 \mid \Ical^c\right]} \\
    & \leq \|\widehat{f} - f\|_{4, \P_X} \cdot \|\widehat{g} - g\|_{4, \P_X} \to 0.
\end{align*}
Similar inequalities can be derived for the other terms, using the fact that $\E(\epsilon_i \mid X_i) = \E(\delta_i \mid X_i) = 0$. We thus have that
\begin{align*}
    \liminf_{n \to \infty} \Prob\left(\left\{\left| \widehat\beta^\mstar - \widehat\beta\right| > (R_{2, g} + R_{2,f}) R_{2, f} + \bar{C} \cdot \errml + \frac{\widehat\sigma_\beta}{\sqrt{n}}\right\} \cap \mathcal{G}_1^\mstar \cap \mathcal{G}_1\right) = 0.
\end{align*}
Next, we have
\begin{align*}
    & \Prob\left(\left\{\frac{\left| \widehat\beta^\mstar - \beta\right|}{\widehat\sigma_\beta / \sqrt{n}} > z_{\alpha'/2} \right\} \cap \mathcal{G}_1^\mstar \cap \mathcal{G}_1\right) \\
    & \lesssim \Prob\left(\left\{\sqrt{n} \left| \widehat\beta^\mstar - \widehat\beta^{\rm ora} \right| > z_{\alpha'/2} \cdot \sigma_\beta \cdot \{1 - t_5(n)\} \right\} \cap \mathcal{G}_1^\mstar \cap \mathcal{G}_1\right) \\
    & \hphantom{\leq} + \Prob\left(\left\{\sqrt{n} \left| \widehat\beta^{\rm ora} - \beta \right| > z_{\alpha'/2} \cdot \sigma_\beta \cdot \{1 - t_5(n)\} \right\} \cap \mathcal{G}_1^\mstar \cap \mathcal{G}_1\right)  + t_0(n).
\end{align*}
Notice that, on $\mathcal{G}_1^*$, $|\widehat\beta^\mstar - \widehat\beta^{\rm ora}| \lesssim \errml$, where $\sqrt{n} \cdot \errml \to 0$, for a fixed $n$ and $M \to \infty$. Therefore, for a fixed $n$, there exists $M$ large enough, so that the first probability is zero. Finally, we have
\begin{align*}
    & \Prob\left(\left\{\sqrt{n} \left| \widehat\beta^{\rm ora} - \beta \right| > z_{\alpha'/2} \cdot \sigma_\beta \cdot \{1 - t_5(n)\} \right\} \cap \mathcal{G}_1^\mstar \cap \mathcal{G}_1\right) \\
    & = \Prob\left(\left\{\sqrt{n} \left|\frac{1}{n} \sum_{i \in \Ical} \varphi_\beta(O_i) \right| > z_{\alpha'/2} \cdot \sigma_\beta \cdot \{1 - t_5(n)\} \cdot \frac{\widehat\psi_2^{\rm ora}}{\psi_2} \right\} \cap \mathcal{G}_1^\mstar \cap \mathcal{G}_1\right). 
\end{align*}
Following the reasoning of Lemma \ref{lem: connect beta m with eta m gamma m}, we can find a sequence of constants $t_4(n) \to 0$ such that $\Prob\left(\left\{|\widehat\psi_2^{\rm ora}/ \psi_2 - 1| > t_4(n) \right\} \cap \mathcal{G}_1\right) \lesssim t_0(n)$, therefore the right-hand-side converges to $\alpha'$ by the CLT as $n \to \infty$. This concludes our proof that
\begin{align*}
    \liminf_{n \to \infty} \liminf_{M \to \infty} \Pb(\beta \in \rm CI) \geq 1-\alpha' - \alpha_0 = 1-\alpha.
\end{align*}
The statement regarding the length follows as in Section \ref{appendix:length_CI_lasso}.

\section{Auxiliary lemmas}

\subsection{Proof of Lemma \ref{lem: supp prob of event 1}}

\begin{proof}
Define the event 
\begin{align*}
    \mathcal{B}_1 = \l\{ \max_{1\leq j\leq p}|\hat{\Sigma}_{j,j} - \Sigma_{j,j}| \leq B(n,p,s_\eta) \r\}. 
\end{align*}
On event $\mathcal{B}_1$, we have for all $1\leq j\leq p$,
\begin{gather*}
    \min_{1\leq j\leq p}\Sigma_{j,j} - B(n,p,s_\eta) \leq \Sigma_{j,j} - B(n,p,s_\eta) \leq \hat{\Sigma}_{j,j} \leq \Sigma_{j,j} + B(n,p,s_\eta) \leq \max_{1\leq j\leq p}\Sigma_{j,j} + B(n,p,s_\eta), \\ 
    \min_{1\leq j\leq p}\Sigma_{j,j} - B(n,p,s_\eta) \leq \nu = \min_{1\leq j\leq p} \hat{\Sigma}_{j,j} \leq \max_{1\leq j\leq p}\Sigma_{j,j} + B(n,p,s_\eta).
\end{gather*}
Adding the above two inequalities together, we get, on the event $\mathcal{B}_1$,
\begin{align*}
    2\min_{1\leq j\leq p}\Sigma_{jj} - 2B(n,p,s_\eta) \leq (\widehat\Sigma + \nu I)_{j, j} \leq 2\max_{1\leq j\leq p}\Sigma_{j,j} + 2B(n,p,s_\eta).
\end{align*}

We next prove the event $\mathcal{B}_1$ holds with high probability, i.e., 
\begin{align*}
    \P(\mathcal{B}_1) = \P\l( \max_{1\leq j\leq p} |\hat{\Sigma}_{j,j} - \Sigma_{j,j}| \lesssim \log (np)\frac{s_\eta \log p}{n} + \frac{(\log n)^{5/2}}{\sqrt{n}} + \frac{1}{\sqrt{n}}\r) \geq 1 - (np)^{-c} - p^{-c}.
\end{align*}
We have the decomposition
\begin{align}
    \hat{\Sigma}_{j,j} - \Sigma_{j,j} &= \frac{1}{n}\sum_{i\in\Ical^c} \hat{\epsilon}_i^2X_{i,j}^2 - \E[\epsilon_i^2X_{i,j}^2] \nonumber \\
    &= \frac{1}{n}\sum_{i\in\Ical^c}(\hat{\epsilon}_i^2 - \epsilon_i^2)X_{i,j}^2  + \l(\frac{1}{n}\sum_{i\in\Ical^c} \epsilon_i^2X_{i,j}^2 - \E[\epsilon_i^2 X_{i,j}^2]\r). \label{eq:supp Sigma diagonal decomp}
\end{align}
For the first term, note that $\frac{1}{n}\sum_{i\in\Ical^c}(\hat{\epsilon}_i^2 - \epsilon_i^2)X_{i,j}^2 \leq \l(\max_{1\leq j\leq p}X_{i,j}^2\r) \cdot \frac{1}{n}\sum_{i\in\Ical^c}(\hat{\epsilon}_i^2 - \epsilon_i^2)$. Since $X_{i,j}$ is subgaussian and we can control the maximum $X_{i,j}$ for $1\leq i\leq n$ and $1\leq j\leq p$ by 
\begin{align*}
    &\quad \P\l(\max_{1\leq i\leq n,1\leq j\leq p} |X_{ij}| \geq \max_{1\leq j\leq p} \E[X_{i,j}] + C\sqrt{\log (np)}\r) \\
    &\leq \P\l(\max_{1\leq i\leq n,1\leq j\leq p} |X_{ij} - \E[X_{i,j}]| \geq C\sqrt{\log (np)}\r) \\
    & \leq 2np \exp(-c\sqrt{\log (np)}^2) = 2(np)^{-c},
\end{align*}
where the first inequality follows from $\max_{1\leq j\leq p} |X_{i,j} - \E[X_{i,j}]| \geq \max_{1\leq j\leq p} |X_{i,j}| - \max_{1\leq j\leq p} |\E[X_{i,j}]|$.
This implies that $\max_{1\leq j\leq p} X_{i,j}^2 \leq C\log(np)$ with probability $1-2(np)^{-c}$.
Meanwhile, we have
\begin{align}
    \frac{1}{n}\sum_{i\in\Ical^c}(\hat{\epsilon}_i^2 - \epsilon_i^2) = \frac{1}{n}\sum_{i\in\Ical^c} (X_i^\T \eta - X_i^\T \hat{\eta})^2 + \frac{2}{n} \sum_{i\in\Ical^c} \epsilon_iX_i^\T(\eta - \hat{\eta}). \label{eq:supp epsilonHat square decomp}
\end{align}
By the standard Lasso theory (Theorem 7.2 in \citeapp{bickel2009simultaneous}), we have 
\begin{align*}
    \P\l(\l| \frac{1}{n}\sum_{i\in\Ical^c}(X_i^\T\eta - X_i^\T\hat{\eta})^2 \r| \geq C\frac{s_\eta\log p}{n}\r) \leq p^{-c}.
\end{align*}
For the second term in \eqref{eq:supp epsilonHat square decomp}, we apply the H\"{o}lder's inequality to get
\begin{align}
    &\quad \P\l( \l|\frac{2}{n}\sum_{i\in\Ical^c}\epsilon_i(X_i^\T\eta - X_i^\T\hat{\eta}) \r| \geq C\frac{s_\eta \log p}{n}\r) \nonumber \\
    &\leq \P\l( 2\l\|\frac{1}{n}\sum_{i\in\Ical^c}X_i\epsilon_i \r\|_\infty \|\hat{\eta} - \eta\|_1 \geq C\frac{s_\eta \log p}{n}\r) \nonumber \\
    &\leq \P\l(2\l\|\frac{1}{n}\sum_{i\in\Ical^c}X_i\epsilon_i \r\|_\infty \geq C\sqrt{\frac{\log p}{n}}\r) + \P\l(\|\hat{\eta} - \eta\|_1 \geq Cs_\eta\sqrt{\frac{\log p}{n}}\r). \label{eq:supp epsilon x(etaHat-eta) decomp}
\end{align}
Note that $X_i\epsilon_i$ is a mean-zero product of sub-Gaussian random variables, we then use Corollary 5.17 in \citeapp{vershynin2010introduction} with $\epsilon=\sqrt{\log p/n}$ to bound it by
\begin{align}
    \P\l(\l\|\frac{1}{n}\sum_{i\in\Ical^c}X_i\epsilon_i \r\|_\infty \geq C\sqrt{\frac{\log p}{n}}\r) \leq 2p\exp\l(-c\min\l\{{\frac{\log p}{n}} , \sqrt{\frac{\log p}{n}} \r\}n\r) = 2p^{-c'}. \label{eq:supp x epsilon subexponential}
\end{align}
Again by Theorem 7.2 in \citeapp{bickel2009simultaneous}, we have 
\begin{align}
    \P\l(\|\hat{\eta} - \eta\|_1 \geq Cs_\eta\sqrt{\frac{\log p}{n}}\r) \leq p^{-c'}. \label{eq:supp etaHat 1 norm}
\end{align}
Plugging \eqref{eq:supp x epsilon subexponential} and \eqref{eq:supp etaHat 1 norm} to \eqref{eq:supp epsilon x(etaHat-eta) decomp}, we get 
\begin{align*}
    \P\l( \l|\frac{2}{n}\sum_{i\in\Ical^c}\epsilon_i(X_i^\T\eta - X_i^\T\hat{\eta}) \r| \geq C\frac{s_\eta \log p}{n}\r) \leq p^{-c}.
\end{align*}
Given the above inequalities, we bound the first term in \eqref{eq:supp Sigma diagonal decomp} by
\begin{align}
    \P\l(\max_{1\leq j \leq p} \l| \frac{1}{n}\sum_{i\in\Ical^c}(\hat{\epsilon}_i^2 - \epsilon_i^2)X_{i,j}^2 \r| \gtrsim \log(np)\frac{s_\eta \log p}{n} \r) \lesssim (np)^{-c}+p^{-c}. \label{eq:supp bound fourth power epsilon2X2}
\end{align}

We next bound the variation of the quartic term, $\frac{1}{n}\sum_{i\in\Ical^c} \epsilon_i^2X_{i,j}^2 - \E[\epsilon_i^2 X_{i,j}^2]$ in \eqref{eq:supp Sigma diagonal decomp}. Let $A_{ij} = \epsilon_i^2X_{i,j}^2$. Define the truncated variable $\bar{A}_{ij} = A_{ij} \mathbf{1}_{|\epsilon_i|\leq C\sqrt{\log p} \textrm{ and } |X_{i,j}| \leq C\sqrt{\log p}}$ and $\tilde{A}_{ij} = A_{ij} \mathbf{1}_{|\epsilon_i|> C\sqrt{\log p} \textrm{ or } |X_{i,j}| > C\sqrt{\log p}}$. Then we have 
\begin{align*}
    \frac{1}{n}\sum_{i\in\Ical^c} (A_{ij} - \E A_{ij}) = \frac{1}{n}\sum_{i\in\Ical^c} (\bar{A}_{ij} - \E \bar{A}_{ij}) + \frac{1}{n}\sum_{i\in\Ical^c} (\tilde{A}_{ij} - \E \tilde{A}_{ij}).
\end{align*}
For the second term $\frac{1}{n}\sum_{i\in\Ical^c} (\tilde{A}_{ij} - \E \tilde{A}_{ij})$, we first bound the $\E\tilde{A}_{ij}$ by applying the Markov inequality,
\begin{align*}
    \P\l(\l| \frac{1}{n}\sum_{i\in\Ical^c} (\tilde{A}_{ij} - \E \tilde{A}_{ij}) \r| \gtrsim \frac{1}{\sqrt{n}}\r) & \lesssim \frac{\sqrt{\E\l(\frac{1}{n}\sum_{i\in\Ical^c} \tilde{A}_{ij} - \E \tilde{A}_{ij}\r)^2}}{1/\sqrt{n}} = \Var(\tilde{A}_{ij}) \leq \E[\tilde{A}_{ij}^2].
\end{align*}
We then bound $\E[\tilde{A}_{ij}^2]$ by Cauchy-Shwarz inequality,
\begin{align*}
    \E\tilde{A}_{ij}^2  &= \E\epsilon_i^4X_{i,j}^4\mathbf{1}_{|\epsilon_i|> C\sqrt{\log p} \textrm{ or } |X_{i,j}| > C\sqrt{\log p}} \\
    & \leq \sqrt{\E\epsilon_i^8X_{i,j}^8}\sqrt{\P(|\epsilon_i|> C\sqrt{\log p} \textrm{ or } |X_{i,j}| > C\sqrt{\log p})} \\
    & \lesssim \sqrt{\P(|\epsilon_i|> C\sqrt{\log p}) + \P(|X_{i,j}| > C\sqrt{\log p})} \lesssim p^{-c}.
\end{align*}
In the above expression, the second inequality is by the finite moments of subgaussian $\epsilon_i$ and $X_{i,j}$, as $\E\epsilon_i^8 X_{i,j}^8 \leq \sqrt{\E\epsilon_i^{16} \cdot \E X_{i,j}^{16}} \leq C$. The last inequality is by the tail probability of subgaussian variables, {where the mean is included in $C\sqrt{\log p}$}. This implies 
\begin{align*}
    \P\l(\l| \frac{1}{n}\sum_{i\in\Ical^c} (\tilde{A}_{ij} - \E \tilde{A}_{ij}) \r| \gtrsim \frac{1}{\sqrt{n}}\r) \lesssim p^{-c}.
\end{align*}

We introduce the following lemma to bound the first term, $\frac{1}{n}\sum_{i\in\Ical^c} (\bar{A}_{ij} - \E \bar{A}_{ij})$.

\begin{lemma}\label{lem:bound quartic term}
    (From Lemma 1, \citeapp{cai2011adaptive}) Let $\xi_1,\dots,\xi_n$ be independent random variables with mean zero. Suppose that there exists some $\eta>0$ and $M_n$ such that $\sum_{i=1}^n \E\xi_i^2\exp(\eta|\xi_i|) \leq M_n^2$. Then for $0<t\leq M_n$,
    \begin{align*}
        \P\l(\sum_{i=1}^n \xi_i \geq C_\eta M_n t\r)\leq \exp(-t^2),
    \end{align*}
    where $C_\eta = \eta + \eta^{-1}$.
\end{lemma}

For any given $j$, taking $\xi_i = \bar{A}_{ij} - \E \bar{A}_{ij}$ and $\eta = (C\log p)^{-2}$, we verify the condition of Lemma \ref{lem:bound quartic term} is satisfied. Note that 
\begin{align*}
    \sum_{i\in\Ical^c} \E(\bar{A}_{ij} - \E \bar{A}_{ij})^2\exp(\eta |\bar{A}_{ij} - \E \bar{A}_{ij}|) \lesssim \sum_{i\in\Ical^c} \E(\bar{A}_{ij} - \E \bar{A}_{ij})^2 \lesssim n,
\end{align*}
where the first inequality follows from $|\bar{A}_{ij} - \E \bar{A}_{ij}| \leq |\bar{A}_{ij}| + |\E \bar{A}_{ij}| \leq (C\log p)^2$ by the construction of $\bar{A}_{ij}$, and the second inequality follows from $\Var(\bar{A}_{ij})\leq \E A_{ij}^2 \leq C$ by finite moments of subgaussian variables.  Therefore, taking $M_n = \sqrt{Cn}$ and $t = \sqrt{C\log p}$, we apply Lemma \ref{lem:bound quartic term} and get 
\begin{align*}
    \P\l(\frac{1}{n}\sum_{i\in\Ical^c} \bar{A}_{ij} - \E \bar{A}_{ij}\geq C\frac{(\log p)^{5/2}}{\sqrt{n}}\r) \lesssim p^{-c}.
\end{align*}

Combining the above bounds together with \eqref{eq:supp bound fourth power epsilon2X2}, we get
\begin{align*}
    \P\l(|\hat{\Sigma}_{j,j} - \Sigma_{j,j}| \lesssim \log (np)\frac{s_\eta \log p}{n} + \frac{(\log p)^{5/2}}{\sqrt{n}} + \frac{1}{\sqrt{n}}\r) \geq 1- (np)^{-c} - p^{-c}.
\end{align*}
Taking the union bound over $j$, we have
\begin{align*}
    \P\l( \max_{1\leq j\leq p} |\hat{\Sigma}_{j,j} - \Sigma_{j,j}| \lesssim \log (np)\frac{s_\eta \log p}{n} + \frac{(\log p)^{5/2}}{\sqrt{n}} + \frac{1}{\sqrt{n}}\r) \geq 1 - (np)^{-c} - p^{-c}.
\end{align*}

\end{proof}

\subsection{Proof of Lemma \ref{lem: supp prob of event 2}}

\begin{proof}
By Jensen's inequality, we have, for $r \geq 2$
\begin{align*}
    \E \|\xi\|_2^r \leq p^{r/2-1}\sum_{j=1}^p \E |\xi_j|^r \leq p^{r/2} \max_{1\leq j\leq p}\E |\xi_j|^r \implies (\E \|\xi\|_2^r)^{1/r} \leq \sqrt{p} \max_{1 \leq j \leq p} (\E(|\xi_j|^r)^{1/r}.
\end{align*}
Note that for $1\leq j\leq p$, the expectation is upper bounded by 
\begin{align*}
    \E|\xi_j|^r = \int_{0}^{\infty} P(|\xi_j| \geq s) r s^{r-1} ds & \leq 2 r \left\{ \int_0^{\sqrt{n}} e^{-\overline{c} s^2} s^{r-1} ds + \int_{\sqrt{n}}^\infty e^{-\overline{c} s \sqrt{n}} s^{r-1} ds\right\} \\
    & = 2r \left\{\frac{1}{2\overline{c}^{r/2}} \Gamma\left(\frac{r}{2}\right) + (\overline{c} \sqrt{n})^{-r} \Gamma(r)\right\}.
\end{align*}
Using the inequality $\Gamma(x) \leq 3 x^x$ for $x \geq 1/2$, we conclude that there exists a constant $C'$ such that
\begin{align*}
    ( \E|\xi_j|^r)^{1/r} \lesssim \sqrt{r} + \frac{r}{\sqrt{n}} \leq C' r.
\end{align*}
Therefore, we have $\max_{1 \leq j \leq p} (\E(|\xi_j|^r)^{1/r} \lesssim r$; hence, $(\E \|\xi\|_2^r)^{1/r} \lesssim \sqrt{p} \cdot r$. Then, by Markov's inequality, there exists a constant $c_{2, \xi}$ such that
\begin{align*}
\P\left(\|\xi\|_2 \geq c_{2, \xi} \cdot \sqrt{p} \cdot \log (2/\alpha_0) \right) \leq \frac{\alpha_0}{2},
\end{align*}
for any $\alpha_0 \leq 2 e^{-2}$. 
\end{proof}

\subsection{Proof of Lemma \ref{lem:Lemma11_Zijian}}

\begin{proof}
    The proof of \eqref{eq:supp zijian cnetered} directly follows from the proof of Lemma 11 in \citeapp{tony2020semisupervised}, so we only show the proof of \eqref{eq:supp zijian mean noncentered} in the below. 
    
    Let $\hat{\Sigma}_X = \frac{1}{n}\sum_{i=1}^n X_i X_i^\T$. By \eqref{eq:supp zijian cnetered} and triangle inequality, we have
    \begin{align*}
        \P\l(|w^\T \hat\Sigma_X v| \gtrsim t\frac{\|\Sigma_X^{1/2}w\|_2 \|\Sigma_X^{1/2}v\|_2}{\sqrt{n}} - |w^\T \Sigma_X v|\r) \leq 2\exp(-ct^2).
    \end{align*}
    Since $|w^\T \Sigma_X v| \leq \|\Sigma_X^{1/2}w\|_2 \|\Sigma_X^{1/2}v\|_2$, taking $t = \sqrt{t_0(n)}$, we further have 
    \begin{align*}
        \P\l(|w^\T \hat\Sigma_X v| \gtrsim \l(1+\sqrt{\frac{t_0(n)}{n}}\r){\|\Sigma_X^{1/2}w\|_2 \|\Sigma_X^{1/2}v\|_2} \r) \leq 2\exp(-ct_0(n)).
    \end{align*}
    Because $\|\Sigma_X^{1/2}w\|_2 \|\Sigma_X^{1/2}v\|_2 \leq \|\Sigma_X\|_\op \|w\|_2 \|v\|_2$ and $\sqrt{t_0(n/n)} \lesssim 1$, we have
    \begin{align*}
        \P\l(|w^\T \hat\Sigma_X v| \gtrsim \|\Sigma_X\|_\op \|w\|_2 \|v\|_2\r) \leq 2\exp(-ct_0(n)).
    \end{align*}
\end{proof}

\subsection{Proof of Lemma \ref{lemma:xi_subgaussian}}

\begin{proof}
By Lemma 2.8.6 in \citeapp{vershynin2009high}, $X_{ji} \epsilon_i$ is sub-Exponential with Orlicz norm $\|X_{ij} \epsilon_i\|_{\psi_1} \leq \|X_{ij}\|_{\psi_2} \|\epsilon_i\|_{\psi_2} \lesssim 1$. By Corollary 2.9.2 in \citeapp{vershynin2009high} applied with $a = n^{-1/2}$ (see also Remark 2.9.4), there exists a constant $\overline{c}$ such that
\begin{align*}
   \Prob\left(|\xi_j| \geq t\right) =  \Prob\left(n^{-1/2} \left| \sum_{i=1}^n X_{ji} \epsilon_i\right| \geq t \right) \leq \begin{cases} 2 \exp(-\overline{c} t^2), & \text{ if } t \leq \sqrt{n}; \\
    2 \exp(-\overline{c} t\sqrt{n}), & \text{ if } t \geq \sqrt{n}. 
    \end{cases}
\end{align*}
Under the condition that $(\log p) / n \to 0$, for $n$ and $p$ large enough and by a union bound, we have:
\begin{align*}
    \Prob\left(\max_{1 \leq j \leq p} |\xi_j| \geq C \sqrt{\log p} \right) \leq \exp\{-\overline{c}C^2\log p + \log(2p)\}.
\end{align*}
Therefore, one can choose $C$ large enough so that the right-hand-side is upper bounded by $p^{-c}$ for some $c > 0$.  
\end{proof}

\subsection{Proof of Lemma \ref{lemma:tau_5}}

\begin{proof}
Notice that
\begin{align*}
    \l|\hat\sigma_\beta^2 - \sigma^2_\beta \r|< \sigma^2_\beta \cdot \tau_5' & \implies \frac{\hat{\sigma}_\beta}{\sigma_\beta} - 1 \in \left[\sqrt{1-\tau_5'}-1, \sqrt{1+\tau_5'}-1\right] \\
    & \implies \l| \frac{\hat{\sigma}_\beta}{\sigma_\beta} - 1 \r| \leq 1-\sqrt{1-\tau_5'}
\end{align*}
because 
\begin{align*}
  \sqrt{1 + \tau_5'} - 1 \leq 1 - \sqrt{1 - \tau_5'}  \iff \frac{\sqrt{1+\tau_5'} + \sqrt{1-\tau_5'}}{2} \leq 1 
\end{align*}
and the second inequality holds by Jensen's. We have
\begin{align*}
    \hat\sigma_\beta^2 - \sigma^2_\beta & = \left\{\left(\frac{\psi_2^2}{\widehat\psi_2^2}- 1\right) + 1 \right\} \cdot \psi_2^{-2} \cdot\left(\frac{1}{n}\sum_{i=1}^{n} \{\widehat\varphi_1(O_i) - \widehat\beta\widehat\varphi_2(O_i)\}^2 - \E[\{\varphi_1(O_i) - \beta\varphi_2(O_i)\}^2]\right) \\
    & \hphantom{=} \quad + \left(\frac{\psi_2^2}{\widehat\psi_2^2}- 1\right) \cdot \psi_2^{-2} \cdot \E[\{\varphi_1(O_i) - \beta\varphi_2(O_i)\}^2].
\end{align*}
From analogous arguments as the ones in Section \ref{sec:proof beta in ci*}, we have
\begin{align*}
    \Prob\left(\left\{\left| \frac{\widehat\psi_2}{\psi_2} - 1\right| \geq \frac{\overline\tau_4(n, M, p)}{\psi_2}\right\} \cap \mathcal{G}_1 \right) \lesssim \tau_0(n, p)
\end{align*}
where 
\begin{align*}
    \overline\tau_4(n, p) = \sqrt{\frac{\Var\{\varphi_2(O)\}}{n \cdot t_0(n, p)}} + \frac{c \cdot \|\Lambda\|^{1/2}_{\text{op}} \cdot \sqrt{s_\gamma \log p}}{n \cdot \sqrt{t_0(n, p)}} + \left(c + \frac{c}{\sqrt{ n \cdot t_0(n, p)}}\right) \cdot \|\Sigma_X\|_{\text{op}} \cdot \frac{ s_\gamma \log p}{n}
\end{align*}
is simply $\tau_4(n, M, p)$ with $\err$ replaced by $\sqrt{\log p}$ and $t_0(n, M, p)$ replaced by $t_0(n, p)$. This implies that
\begin{align*}
    & \Prob\left(\left\{\left|\frac{\psi_2^2}{\widehat\psi_2^2} - 1 \right\} \cap \mathcal{G}_1 \right| \leq \overline\tau_4'(n, p) \right) \geq 1 - c\tau_0(n, p), \quad \text{ where } \\
    & \tau_4'(n, M) =  \frac{\overline\tau_4(n, p)\{\overline\tau_4(n, p)+2\}}{1 - \overline\tau_4(n, p)\{\overline\tau_4(n, p)+2\}},
\end{align*}
because
\begin{align*}
   \left| \frac{\widehat\psi_2}{\psi_2} - 1 \right| \leq \overline\tau_4(n, M) \implies \left|\frac{\widehat{\psi}_2^2}{\psi_2^2} - 1\right| \leq \overline\tau_4(n, M)\{\overline\tau_4(n, M) + 2\} \implies \left|\frac{\psi_2^2}{\widehat\psi_2^2} - 1\right| \leq \overline\tau_4'(n, M). 
\end{align*}
Similarly to the derivations from Section \ref{sec:proof m* in M}, we have
\begin{align*}
    \widehat\beta - \beta & = \left\{\frac{1}{n}\sum_{i \in \Ical} \varphi_\beta(O_i) + \Delta\right\} \cdot \left\{\left(\frac{\psi_2}{\widehat\psi_2} - 1\right) + 1\right\},
\end{align*}
where $\Delta = (R_1 + R_2) / \psi_2$ and
\begin{align*}
R_1 & = \frac{1}{n}\sum_{i \in \Ical} \left\{2\beta X_i^\intercal\delta_i(\widehat\gamma - \gamma) - X_i^\intercal \epsilon_i (\widehat\gamma-\gamma) - X_i^\intercal \delta_i (\widehat\eta-\eta) \right\}, \\
R_2 & = \left\{(\widehat\eta - \eta) - \beta\cdot (\widehat\gamma - \gamma)\right\}^\intercal \cdot \left(\frac{1}{n}\sum_{i \in \Ical} X_iX_i^\intercal\right)\cdot (\widehat\gamma - \gamma).
\end{align*}
We have
\begin{align*}
    & \Prob\left(\left|\frac{1}{n}\sum_{i \in \Ical} \varphi_\beta(O_i)\right| \geq \sqrt{\frac{\Var\{\varphi_\beta(O)\}}{n \cdot t_0(n, p)}} \right) \leq t_0(n, p), \\
    & \Prob\left(\left\{|R_1| \geq \frac{\overline\tau_2(n, p)}{\sqrt{n}}\right\} \cap \mathcal{G}_1\right) \leq t_0(n, p), \\
    & \Prob\left(\left\{|R_2| \geq \frac{\overline\tau_3(n, p)}{\sqrt{n}}\right\} \cap \mathcal{G}_1\right) \leq t_0(n, p), \\
    & \Prob\left(\left\{\left| \left(\frac{\psi_2}{\widehat\psi_2} - 1\right) + 1\right| \geq 1 + \frac{\overline\tau_4(n, p)}{1-\overline\tau_4(n, p)}\right\} \cap \mathcal{G}_1\right) \lesssim t_0(n, p),
\end{align*}
where $\overline\tau_2(n, p)$ and $\overline\tau_3(n, p)$ are equal to $\tau_2(n, M, p)$ and $\tau_3(n, M, p)$ with $\err$ replaced by $\sqrt{\log p}$. Therefore, it holds that
\begin{align*}
& \Prob\left(\left|\widehat\beta - \beta\right| \gtrsim \tau_{5,1}(n, M) \right) \lesssim \tau_0(n, M), \quad \text{ where } \\
& \tau_{5, 1}(n, M) = \frac{1}{\sqrt{n}} \cdot \left\{1 + \frac{\overline\tau_4(n, M)}{1-\overline\tau_4(n, M)}\right\}\left\{ \sqrt{\frac{\Var\{\varphi_\beta(O)\}}{\tau_0(n, M)}} + \frac{\overline\tau_2(n, M)}{\psi_2} + \frac{\overline\tau_3(n, M)}{\psi_2}\right\}.
\end{align*}
Next, consider the decomposition:
\begin{align*}
  & \{\widehat\varphi_1(O_i) - \widehat\beta \widehat\varphi_2(O_i)\} - \{\varphi_i(O_i) - \beta\varphi_2(O_i)\} \\
  & = \{\widehat\varphi_1(O_i) - \varphi_1(O_i)\} - \beta\{\widehat\varphi_2(O_i) - \varphi_2(O_i)\} - (\widehat\beta - \beta)\widehat\varphi_2(O_i) \\
  & = R_1(O_i) + R_2(O_i) - (\widehat\beta - \beta)\widehat\varphi_2(O_i), 
\end{align*}
where
\begin{align*}
& R_1(O_i) = 2\beta X_i^\intercal\delta_i(\widehat\gamma - \gamma) - X_i^\intercal \epsilon_i (\widehat\gamma-\gamma) - X_i^\intercal \delta_i (\widehat\eta-\eta), \\
& R_2(O_i) = \left\{(\widehat\eta - \eta) - \beta\cdot (\widehat\gamma - \gamma)\right\}^\intercal \cdot \left(X_iX_i^\intercal\right)\cdot (\widehat\gamma - \gamma).
\end{align*}
Notice that
\begin{align*}
    & R^2_1(O_i) \lesssim \beta^2(\widehat\gamma - \gamma)^\T X_i X_i^\T \delta_i^2 (\widehat\gamma - \gamma) + (\widehat\gamma - \gamma)^\T X_i X_i^\T \epsilon_i^2 (\widehat\gamma - \gamma) + (\widehat\eta - \eta)^\T X_i X_i^\T \delta_i^2 (\widehat\eta - \eta), \\
    & R_2^2(O_i) \lesssim \{X_i^\T(\widehat\eta - \eta)\}^2  \{X_i^\T(\widehat\gamma - \gamma)\}^2 +  \{X_i^\T(\widehat\gamma - \gamma)\}^4. 
\end{align*}
Since $X_1, \ldots, X_n$ are sub-Gaussian random vectors, independent of $\widehat\eta$, and with parameter $\sigma_X$, we have that $X_i^\T(\widehat\eta - \eta)$ is sub-Gaussian with parameter $\sigma_X \|\widehat\eta - \eta\|$. Similarly, $X_i^\T(\widehat\gamma - \gamma)$ is sub-Gaussian with parameter $\sigma_X \|\widehat\gamma - \gamma\|$. Thus, for some constant $C$, we have
\begin{align*}
    & \Prob\left(\max_{i} |X_i^\T(\widehat\gamma - \gamma)| > \left\{\lambda^{1/2}_{\max}(\Sigma_X) + C \sqrt{\log n}\right\} \cdot \|\widehat\gamma - \gamma\|_2 \mid \Ical^c\right) \\
    & \leq  \Prob\left(\max_{i} |X_i^\T(\widehat\gamma - \gamma)| > \E\left\{\left|X_i^\T(\widehat\gamma - \gamma)\right| \mid \Ical^c\right\}+ C \sqrt{\log n} \cdot \|\widehat\gamma - \gamma\|_2 \mid \Ical^c\right) \\
    & \leq \Prob\left(\max_{i} \left| X_i^\T(\widehat\gamma - \gamma) -  \E\left\{X_i^\T(\widehat\gamma - \gamma) \mid \Ical^c\right\}\right| >  C \sqrt{\log n} \cdot \|\widehat\gamma - \gamma\|_2 \mid \Ical^c\right) \\
    & \lesssim n^{-c} \lesssim t_0(n, p),
\end{align*}
where the last inequality follows by a union bound and sub-Gaussianity of $X_i^\T(\widehat\gamma - \gamma)$ (see, e.g., Proposition  2.6.6 in \citeapp{vershynin2009high}). In addition,
\begin{align*}
\Prob\left( \frac{1}{n}\sum_{i \in \Ical} \{X_i^\T(\widehat\gamma - \gamma)\}^2 \gtrsim \frac{\lambda_{\max}(\Sigma_X) \cdot \|\widehat\gamma - \gamma\|_2^2}{t_0(n, p)} \mid \Ical^c \right) \lesssim t_0(n, p)
\end{align*}
by Markov's inequality. 
Therefore, we have
\begin{align*}
    \Prob\left(\frac{1}{n}\sum_{i \in \Ical} \{X_i^\T(\widehat\gamma - \gamma)\}^4 \gtrsim \frac{(\log n)^2\cdot\|\widehat\gamma - \gamma\|_2^4}{t_0(n, p)} \mid \Ical^c \right) \lesssim t_0(n, p).
\end{align*}
We may similarly derive
\begin{align*}
    \Prob\left(\frac{1}{n}\sum_{i \in \Ical} \{X_i^\T(\widehat\gamma - \gamma)\}^2\{X_i^\T(\widehat\eta - \eta)\}^2 \gtrsim \frac{(\log n)^2 \cdot \|\widehat\eta - \eta\|_2^2\cdot \|\widehat\gamma - \gamma\|_2^2}{t_0(n, p)} \mid \Ical^c \right) \lesssim t_0(n, p).
\end{align*}
Therefore, we conclude that
\begin{align*}
    \Prob\left(\frac{1}{n}\sum_{i \in \Ical} R_2^2(O_i) \gtrsim \frac{(\log n)^2}{t_0(n, p)}\cdot (\|\widehat\eta - \eta\|^2_2  + \|\widehat\gamma - \gamma\|_2^2) \cdot \|\widehat\gamma - \gamma\|_2^2 \mid \Ical^c \right) \lesssim t_0(n, p).
\end{align*}
Because $\E\{R_1^2(O)\mid \Ical^c\} \lesssim (\|\widehat\gamma - \gamma\|_2^2 + \|\widehat\eta - \eta\|_2^2) \cdot \lambda_{\max}(\Lambda) + \|\widehat\gamma - \gamma\|_2^2 \cdot \lambda_{\max}(\Sigma)$, we have
\begin{align*}
    & \Prob\left(\left\{\frac{1}{n}\sum_{i \in \Ical} \left\{R_1^2(O_i) + R_2^2(O_i)\right\} \gtrsim \tau_{5, 2}(n, p) \right\} \cap \mathcal{G}_1 \right) \lesssim t_0(n, p), \quad \text{where} \\
    & \tau_{5, 2}(n, p) = \frac{1}{t_0(n, p)}\left[\frac{\log p}{n} \{ (s_\gamma + s_\eta)  \cdot \lambda_{\max} (\Lambda) + s_\gamma \cdot \lambda_{\max} (\Sigma) \} +   \frac{\log^2 p \cdot \log^2 n}{n^2} \cdot (s_\eta + s_\gamma) \cdot s_\gamma\right].
\end{align*}
Next, we have
\begin{align*}
\widehat\varphi^2_2(O_i) & = \left\{\delta_i^2 + 2\delta_i X_i^\T(\widehat\gamma - \gamma) + (\widehat\gamma - \gamma)^\T X_i X_i^\T(\widehat\gamma - \gamma)\right\}^2 \\
& \lesssim \delta_i^4 + (\widehat\gamma - \gamma)^T X_i X_i^\T \delta_i^2(\widehat\gamma - \gamma) + \{X_i^\T(\widehat\gamma - \gamma)\}^4, 
\end{align*}
so that
\begin{align*}
& \Prob\left(\left\{\frac{1}{n}\sum_{i \in \Ical} \widehat\varphi^2_2(O_i) \gtrsim \tau_{5, 3}(n, M)\right\} \cap \mathcal{G}_1 \right) \lesssim t_0(n, p), \quad \text{ where } \\
& \tau_{5, 3}(n, p) =  \E(\delta^4) + \sqrt{\frac{\Var(\delta^4)}{n \cdot t_0(n, p)}} + \frac{\lambda_{\max}(\Lambda) \cdot s_\gamma \cdot \log p}{n \cdot t_0(n, p)} + \frac{\log^2 p \cdot \log^2 n \cdot s_\gamma^2}{n^2 \cdot t_0(n, p)}.
\end{align*}
Similarly, 
\begin{align*}
    & \Prob\left(\frac{1}{n}\sum_{i \in \Ical} \left\{\varphi_1(O_i) - \beta \varphi_2(O_i)\right\}^2 \geq \tau_{5, 4}(n, p) \right) \leq t_0(n, p), \quad \text{ where } \\
    &\tau_{5, 4}(n, p) = \var\{\varphi_\beta(O_i)\} + \sqrt{\frac{\Var[\{\varphi_1(O_i) - \beta \varphi_2(O_i)\}^2]}{n \cdot \tau_0(n, p)}}
\end{align*}
Thus, using $a^2 - b^2 = 2b(a - b) + (a-b)^2$, we have arrived
\begin{align*}
    & \Prob\left(\left\{\frac{1}{n}\sum_{i \in \Ical} \left[\{\widehat\varphi_1(O_i) - \widehat\beta \widehat\varphi_2(O_i)\}^2 - \{\varphi_i(O_i) - \beta\varphi_2(O_i)\}^2 \right]\right\} \cap \mathcal{G}_1 \gtrsim \tau_{5, 5}(n, p)\right) \lesssim t_0(n, p),
\end{align*}
where
\begin{align*}
    & \tau_{5, 5}(n, p) \\
    & = \{\tau_{5,4}(n, p)\}^{\frac{1}{2}} \{\tau_{5, 2}(n, p) + \tau^2_{5, 1}(n, M) \tau_{5, 3}(n, p)\}^{\frac{1}{2}} + \tau_{5, 2}(n, p) + \tau^2_{5, 1}(n, p) \tau_{5, 3}(n, p).
\end{align*}
Finally, 
\begin{align*}
    & \Prob\left(\left| \frac{1}{n}\sum_{i \in \Ical} \{\varphi_1(O_i) - \beta \varphi_2(O_i)\}^2 - \E[\{\varphi_1(O) - \beta \varphi_2(O)\}^2]\right| \geq \tau_{5, 6}(n, M) \right) \leq t_0(n, p), \text{ where } \\
    & \tau_{5, 6}(n, M) = \sqrt{\frac{\Var[\{\varphi_1(O) - \beta \varphi_2(O)\}^2]}{n \cdot t_0(n, p)}}.
\end{align*}
Thus, we have arrived at
\begin{align}
    & \Prob\left(|\widehat\sigma^2_\beta - \sigma^2_\beta| \gtrsim \tau'_{5}(n,p)\right) \lesssim t_0(n, p), \text{ where for some constant } C \nonumber \\
    & \tau'_5(n, p) = \overline\tau_4'(n, p) \cdot \frac{\E[\{\varphi_1(O) - \beta\varphi_2(O)\}^2]}{\psi_2^2}  + \{1 + \overline\tau_4'(n, p)\} \cdot \frac{\tau_{5, 5}(n, p) + \tau_{5, 6}(n, p)}{\psi_2^2}, \label{eq:supp tau5'}
\end{align}
which yields $\Prob\left(\mathcal{G}^c_5 \cap \mathcal{G}_1\right) \lesssim t_0(n, p)$, with $\tau_5(n, p) = 1 - \sqrt{1 - \sigma^2_\beta \cdot \tau_5'(n, p)}$. Notice that
\begin{align*}
\tau'_5(n, p) & \lesssim \tau_4'(n, p) + \sqrt{\tau_{5, 2}(n, M, p)} + \tau_{5, 1}(n, M, p) \lesssim \overline\tau_4(n, p) + \sqrt{\frac{(s_\gamma + s_\eta) \log p}{n \cdot \tau_0(n, p)}}.
\end{align*}
\end{proof}

\subsection{Proof of Lemma \ref{lem:lipschitz continuity}}

In this proof, we adopt the sample splitting scheme where the observations in fold $\Ical^c$ are used to fit the nuisance models $\hat{g}^\m,\hat{f}^\m$ while those in fold $\Ical$ are used to compute the nuisance predictions and $\hat{\beta}^\m$. In particular, we use $\hat{g}^\m(X)$ and $\hat{f}^\m(X)$ to denote the out-of-sample prediction vectors $(\hat{g}^\m(X_1) \;\cdots\; \hat{g}^\m(X_n))^\T \in \RR^n$ and $(\hat{f}^\m(X_1) \;\cdots\; \hat{f}^\m(X_n))^\T \in \RR^n$ where $X_1,\dots,X_n$ are covariates from fold $\Ical$.

We prove the Lipschitz continuity of the mapping $\psi$ conditional on the observed data by its composited mappings. In this proof, define $\psi$'s composited mappings using the following notations: 
\begin{align*}
    \psi(z) &= \psi_3 \odot \psi_2 \odot \psi_1(z)
\end{align*}
with 
\begin{align*}
    \psi_1 &: \RR^{2n} \to \RR^{2n}, \quad \psi_1(z) = (I_n \otimes \hat{\Pi}^{1/2})z := {e}; \\
    \psi_2 &: \RR^{2n} \to \RR^{2n}, \quad \psi_2({e}) = \bigg(
    \begin{array}{c}
         \hat{g}^\m(X) \\
         \hat{f}^\m(X)
    \end{array}
    \bigg) := {u}; \\
    \psi_3 &: \RR^{2n} \to \RR, \quad \psi_3({u}) = \frac{\sum_{i\in\Ical} (Y_i - \hat{g}^\m(X_i)) (D_i - \hat{f}^\m(X_i))}{\sum_{i\in\Ical} (D_i - \hat{f}^\m(X_i))^2}.
\end{align*}
where all sub-mappings are dependent on the observed data. Specifically, the perturbed nuisance models $\hat{g}^\m$ and $\hat{f}^\m$ are fitted with injected noise $e$. For notation simplicity, we omit the superscript $\m$ in $e$ and $u$ to indicate their perturbed nature. 

By Assumption \ref{assump:ml lipschitz}, the mapping $\psi_2$ is Lipschitz continuous with probability $1-\tau_n$ since the variation in the outcome vector $Y$ and treatment vector $D$ can be translated as the variation in stacking noises ${e}$. It remains to show the rest of mappings are Lipschitz continuous. In Steps 1-2 below, we will show that with probability at least $1-c(1/t_0(n)+\tau_n)$,
\begin{align}
    \l\| \psi_1(z_1) - \psi_1(z_2) \r\|_2 &\leq L_1 \|z_1 - z_2\|_2, \label{eq:supp first step lipschitz} \\
    \l|\psi_3(u_1) - \psi_3(u_2)\r| &\leq L_2\|u_1 - u_2\|_2. \label{eq:supp third step lipschitz}
\end{align}
with 
\begin{align*}
    L_1 = C, \quad L_2 = \frac{C}{\sqrt{n}},
\end{align*}
where $t_0(n)$ is a slowly increasing rate in $n$, for example, $t_0(n)=\log\log n$. 

\textbf{Step 1. } In this step, we establish \eqref{eq:supp first step lipschitz}. The transformation from $z$ to $e$ is linear and can be expressed as
\begin{align*}
\psi_1(z) = (I_n \otimes \hat{\Pi}^{1/2})z \quad \textrm{with } \hat\Pi = \frac{1}{n}\sum_{i\in\Ical_0} \hat{o}_i \hat{o}_i^\T \textrm{ and } \hat{o}_i^\T = \bigg(
    \begin{array}{c}
         Y_i - \hat{g}(X_i) \\
         D_i - \hat{f}(X_i)
    \end{array}
    \bigg),
\end{align*}
where $\hat{g}$ and $\hat{f}$ are unperturbed nuisance models. 
Then we have, for any $z_1$ and $z_2$ in $\RR^{2n}$
\begin{align*}
    \l\| \psi_1(z_1) - \psi_1(z_2) \r\|_2 \leq \|I_n \otimes \hat{\Pi}^{1/2}\|_\op \|z_1 - z_2\|_2.
\end{align*}
The matrix $I_n \otimes \hat{\Pi}^{1/2}$ is a block diagonal matrix and it satisfies $\|I_n \otimes \hat{\Pi}^{1/2}\|_\op = \|\hat{\Pi}^{1/2}\|_\op$. Note that $\|\hat{\Pi}^{1/2}\|_\op = \sqrt{\lambda_{\rm max}(\hat{\Pi})}$, so it suffices to bound $\lambda_{\rm max}(\hat{\Pi})$. Note that 
\begin{align*}
    \hat{e}_i = o_i + r_i, \quad \textrm{with } o_i = \bigg(
    \begin{array}{c}
        Y_i - g(X_i) \\
        D_i - f(X_i)
    \end{array}
    \bigg) \textrm{ and }r_i = \bigg(
    \begin{array}{c}
        g(X_i) - \hat{g}(X_i) \\
        f(X_i) - \hat{f}(X_i) 
    \end{array}
    \bigg).
\end{align*}
Plugging the above into the construction of $\hat{\Pi}$, we get
\begin{align*}
    \lambda_{\rm max}(\hat{\Pi}) = \|\hat\Pi\|_\op \leq \l\|\frac{1}{n}\sum_{i\in\Ical_0}o_io_i^\T\r\|_\op + 2\l\|\frac{1}{n}\sum_{i\in\Ical_0}o_ir_i^\T\r\|_\op + \l\|\frac{1}{n}\sum_{i\in\Ical_0}r_ir_i^\T\r\|_\op.
\end{align*}
Let $o\in\RR^{n\times 2}$ be the stacking matrix of $o_i$ and $r\in\RR^{n\times 2}$ be the stacking matrix of $r_i$ for $i\in\Ical_0$. The first term is upper bounded by $\|o^\T o/n\|_\op \leq \|o^\T o/n - \Pi\|_\op + \|\Pi\|_\op$. By Remark 4.7.3 in \citeapp{vershynin2018high} with $u\asymp n$ , we define the event
\begin{align*}
    \mathcal{B}_1 = \{\|o^\T o/n - \Pi\|_\op \leq C\}, \quad \textrm{which satisfies } \P(\mathcal{B}_1) \geq 1 - e^{-cn} \geq 1 - \frac{c'}{t_0(n)}.
\end{align*}
The third term can be bounded by $\|r^\T r/n\|_\op \leq \textrm{tr}(r^\T r/n) = \sum_{i\in\Ical_0} \|r_i\|_2^2/n$. We next show $\sum_{i\in\Ical_0} \|r_i\|_2^2/n$ concentrates around its expectation with the rate or $n^{-1/2}$. Define the event 
\begin{align*}
    \mathcal{B}_2 = \l\{ \l|\frac{1}{n}\sum_{i\in\Ical_0} r_i^\T r_i - \E[r_i^\T r_i \mid \Ical^c]\r| \leq C\sqrt{\frac{ t_0(n)}{{n}}}\r\}, \quad \textrm{which satisfies } \P(\mathcal{B}_2 \mid \Ical^c) \geq 1-\frac{c}{t_0(n)}.
\end{align*}
This finite-sample bound is obtained by Chebyshev inequality with converging fourth moment condition in Assumption \ref{assump:nuisance convergence}. Meanwhile, the above conditional probability inequality implies the $\P(\mathcal{B}_2) \geq 1 - c/t_0(n)$ by taking the expectation over $\Ical^c$. Note that $\E[r_i^\T r_i \mid \Ical^c] = \|\hat{g} - g\|_{2,\P_X}^2 + \|\hat{f} - f\|_{2,\P_X}^2$. By Assumption \ref{assump:nuisance convergence}, we define the following event with probability $1-\tau_n$,
\begin{align*}
    \mathcal{B}_3 = \bigg\{\|\hat{g} - g\|_{2,\P_X} \lesssim R_{2,g}, \quad \|\hat{f} - f\|_{2,\P_X}\lesssim R_{2,f}, \quad \|\hat{g} - g\|_{4,\P_X} \leq R_{4,g}, \quad \|\hat{f} - f\|_{4,\P_X} \leq R_{4,f} \bigg\}.
\end{align*}
On the event $\mathcal{B}_3$, $\E[r_i^\T r_i \mid \Ical^c] $ is bounded by the rate $R_{2,g}^2 + R_{2,f}^2$ and the third term $\frac{1}{n}\sum_{i\in\Ical_0} r_i^\T r_i$ is thus bounded by the rate of $R_{2,g}^2 + R_{2,f}^2 + \sqrt{t_0(n)/n}$.
To bound the second term, by Cauchy-Schwarz, we have
\begin{align*}
    2\l\|\frac{1}{n}\sum_{i\in\Ical_0}o_ir_i^\T\r\|_\op \leq 2\sqrt{\frac{1}{n}\sum_{i\in\Ical_0}\|o_i\|_2^2} \sqrt{\frac{1}{n}\sum_{i\in\Ical_0}\|r_i\|_2^2}.
\end{align*}
It suffices to bound $\frac{1}{n}\sum_{i\in\Ical_0} \|o_i\|_2^2$. Note that $\frac{1}{n}\sum_{i\in\Ical_0}\|o_i\|_2^2 = \textrm{tr}(o^\T o/n)\leq 2\|o^\T o/n\|_\op$. Hence, on the event $\mathcal{B}_1\cap\mathcal{B}_2\cap\mathcal{B}_3$, we have 
\begin{align*}
    \lambda_{\rm max}(\hat{\Pi}) &\leq C\l(1+\sqrt{R_{2,g}^2 + R_{2,f}^2+\sqrt{\frac{ t_0(n)}{{n}}}} + \l(R_{2,g}^2 + R_{2,f}^2+\sqrt{\frac{ t_0(n)}{{n}}}\r)\r) \\
    & \leq C'.
\end{align*}

\textbf{Step 2.} In this step, we establish \eqref{eq:supp third step lipschitz}. 

We denote the nuisance predictions $\hat{g}^\m(X)$ and $\hat{f}^\m(X)$ as a projection of the stacking vector $u$:
\begin{align*}
    \hat{g}^\m(X) &= P_g u \quad \textrm{with } P_g = \l( I_n \;\; \mathbf{0}_n \r) \in \RR^{n\times2n} \\
    \hat{f}^\m(X) &= P_f u \quad \textrm{with } P_f = \l( \mathbf{0}_n \;\; I_n \r)\in \RR^{n\times2n}.
\end{align*}
With this notation, the mapping $\psi_3$ is written as
\begin{align*}
    \psi_3(u) = \frac{n^{-1}(Y - P_g u)^\T(D - P_f u^\m)}{n^{-1}\|D - P_f u\|^2_2} =: \frac{a(u)}{b(u)}.
\end{align*}
For any $u_1$ and $u_2$, the distance $|\psi_3(u_1) - \psi_3(u_2)|$ can be decomposed as 
\begin{align}
    \l|\psi_3(u_1) - \psi_3(u_2)\r| & = \l| \frac{a(u_1) - a(u_2)}{b(u_1)} - \frac{a(u_2)\{b(u_1) - b(u_2)\}}{b(u_1)b(u_2)} \r| \nonumber \\
    & \leq \frac{|a(u_1) - a(u_2)|}{b(u_1)} + \frac{|a(u_2)| \cdot |b(u_1) - b(u_2)|}{b(u_1)b(u_2)}. \label{eq:supp decomp betam distance}
\end{align}

We first bound the distance $|a(u_1) - a(u_2)|$ and $|b(u_1) - b(u_2)|$ in \eqref{eq:supp decomp betam distance}. Notice that for some $\tilde{u}$ between $u_1$ and $u_2$, we have 
\begin{align*}
    \l|a(u_1) - a(u_2)\r| &= \l|\nabla a(\tilde{u}) ^\T (u_1 - u_2)\r| \leq \|\nabla a(\tilde{u})\|_2\cdot \|u_1 - u_2\|_2, \\
    \l|b(u_1) - b(u_2)\r| &= \l|\nabla b(\tilde{u}) ^\T (u_1 - u_2)\r| \leq \|\nabla b(\tilde{u})\|_2\cdot \|u_1 - u_2\|_2.
\end{align*}
It suffices to bound the gradients' norms $\|\nabla a(\tilde{u})\|_2$ and $\|\nabla b(\tilde{u})\|_2$. The same definition applies to all function notations including $\tilde{f}$ and $\tilde{g}$. By the definition of $a(\cdot)$ and triangle inequality, we have
\begin{align*}
    \|\nabla a(\tilde{u})\|_2 
    &= \l\|\frac{1}{n}( - P_g^\T D - P_f^\T Y + 2 P_g^\T P_f \tilde{u})\r\|_2 \\
    &\leq \l\|\frac{1}{n}P_g^\T D\r\|_2 + \l\|\frac{1}{n}P_f^\T Y\r\|_2 + 2\l\|\frac{1}{n}P_g^\T P_f\r\|_\op\|\tilde{u}\|_2 \\
    &= \frac{1}{n}\l\|D\r\|_2 + \frac{1}{n}\l\|Y\r\|_2 + \frac{2}{n}\|\tilde{u}\|_2, 
\end{align*}
where the last equality is derived by the construction of the projection matrices $P_g$ and $P_f$. Note that $\tilde{u} = tu_1 + (1-t)u_2$ for $t\in[0,1]$, so we have $\|\tilde{u}\|_2 \leq \max\{\|u_1\|_2, \|u_2\|_2\}$. It suffices to bound $\|u\|_2$ based on universal perturbed nuisance models. Note that $\frac{1}{n}\|u\|_2^2 = \frac{1}{n}\sum_{i\in\Ical} \{\hat{g}^\m(X_i)^2 + \hat{f}^\m(X_i)^2\}$, so, by Assumption \ref{assump:nuisance convergence}, we have $\frac{1}{n}\|u^\m\|_2^2 \leq C$ for some constant $C>0$. 
\begin{align*}
    \mathcal{B}_4 &= \l\{\l|\frac{1}{n}\|Y\|_2^2 - \E[Y_i^2]\r| \leq \sqrt{t_0(n)\frac{\Var(Y_i^2)}{{n}}}\r\}, \quad
    \mathcal{B}_5 = \l\{\l| \frac{1}{n}\|D\|_2^2 - \E[D_i^2] \r| \leq \sqrt{t_0(n)\frac{\Var(D_i^2)}{{n}}}\r\}. 
\end{align*}
In $\mathcal{B}_4$, note that $Y_i^2 = (g(X_i)+\epsilon_i)^2 \leq 2g(X_i)^2 + 2\epsilon_i^2$. Since $g(\cdot)$ is upper bounded by a positive constant $C$ as stated in Assumption \ref{assump:nuisance convergence}, this implies $\E(Y_i^2)$ and $\Var(Y_i^2)$ are bounded.
The same arguments carry over to bound $\frac{1}{n}\|D\|_2^2$ in the event $\mathcal{B}_5$. 
Therefore, for these events, we have 
\begin{align*}
    \min_{j=4,5} \P(\mathcal{B}_j) \geq 1 - \frac{c}{t_0(n)}.
\end{align*}
On the event $\mathcal{B}_4\cap\mathcal{B}_5$, it implies that 
\begin{align*}
    \frac{1}{n}\|Y\|_2\leq \frac{C}{\sqrt{n}}, \quad \frac{1}{n}\|D\|_2\leq \frac{C}{\sqrt{n}}.
\end{align*}
Therefore, combining the above bounds together, on the event $\cap_{j=4,5}\mathcal{B}_j$, we bound $\|\nabla a(\tilde{u})\|_2$ as
\begin{align*}
    \|\nabla a(\tilde{u})\|_2 
    &\leq \frac{C}{\sqrt{n}}.
\end{align*}
Following the similar derivation, from the construction of $b(\cdot)$ we get
\begin{align*}
    \|\nabla b(\tilde{u})\|_2 \leq \frac{C}{\sqrt{n}}.
\end{align*}

Next we bound the term $|a(u_2)|$ in \eqref{eq:supp decomp betam distance}. By the definition of $a(\cdot)$, we have 
\begin{align*}
    |a(u_2)| &= \frac{1}{n} \l| Y^\T D - D^\T P_g u_2 - Y^\T P_fu_2 + u_2^\T P_f^\T P_f u_2 \r| \\
    &\leq \frac{1}{n}\l(|Y^\T D| + \l(\|P_g^\T D\|_2 + \|P_f^\T Y\|_2\r)\|u_2\|_2 + \|P_g^\T P_f\|_\op \|u_2\|_2^2\r) \\
    &= \l|\frac{1}{n}Y^\T D\r| + \l(\frac{1}{\sqrt{n}}\|D\|_2 + \frac{1}{\sqrt{n}}\|Y\|_2\r)\frac{1}{\sqrt{n}}\|u_2\|_2 + \frac{1}{n}\|u_2\|_2^2.
\end{align*}
By Cauchy-Schwarz, the first term $|Y^\T D/n|$ can be bounded by
\begin{align*}
    \l|\frac{Y^\T D}{n}\r| &\leq \frac{1}{\sqrt{n}}\|Y\|_2 \cdot \frac{1}{\sqrt{n}}\|D\|_2\leq C.
\end{align*}
Hence, on the event $\cap_{j=4,5}\mathcal{B}_j$, the first two terms are bounded since $\frac{1}{\sqrt{n}}\|u_2\|_2$ is bounded by a constant following the previous reasoning based on Assumption \ref{assump:nuisance convergence}.
Therefore, on the event $\cap_{j=4,5}\mathcal{B}_j$, we have
\begin{align*}
    |a(u_2)| \leq C.
\end{align*}

We next show that the denominator $b(u)$ is bounded away from zero for large $n$. 
\begin{align*}
    b(u) &= \frac{1}{n}\sum_{i\in\Ical}(D_i - \hat{f}^\m(X_i))^2 \\
    &= \frac{1}{n}\sum_{i\in\Ical}\delta_i^2 + \frac{1}{n}\sum_{i\in\Ical} \l(\hat{f}^\m(X_i) - f(X_i)\r)^2 - \frac{2}{n}\sum_{i\in\Ical} \delta_i\l(\hat{f}^\m(X_i) - f(X_i)\r) \\
    &\geq \frac{1}{n}\sum_{i\in\Ical}\delta_i^2 - \l|\frac{2}{n}\sum_{i\in\Ical} \delta_i\l(\hat{f}^\m(X_i) - f(X_i)\r)\r|. 
\end{align*}
Note that $\frac{1}{n}\sum_{i\in\Ical}\delta_i^2 - \E[\delta_i^2] \indist N(0,\Var(\delta_i^2))$, so we can define the high probability event
\begin{align*}
    \mathcal{B}_6 = \l\{\l|\frac{1}{n}\sum_{i\in\Ical}\delta_i^2 - \E[\delta_i^2]\r| \leq C\sqrt{\frac{t_0(n)}{n}}\r\} \quad \textrm{with } \P(\mathcal{B}_6) \geq 1- \frac{c}{t_0(n)}.
\end{align*}
To bound the abstract value in the second term, we define 
\begin{align*}
    \mathcal{B}_7 = \l\{ \l| \frac{1}{n}\sum_{i\in\Ical}\delta_i(\hat{f}^\m(X_i) - f(X_i)) \r| \leq C\sqrt{\frac{t_0(n)}{n}} \r\}.
\end{align*}
Note that $\E[\delta_i(\hat{f}^\m(X_i) - f(X_i)) \mid \Ical^c] = 0$ by $\E[\delta_i \mid X_i]=0$ and $\hat{f}^\m$ being independent of $\{X_i,\delta_i\}_{i\in\Ical}$, then by Chebyshev inequality, we have 
\begin{align*}
    \P(\mathcal{B}_7^c \mid \Ical^c) \lesssim \frac{\Var\{\delta_i(\hat{f}^\m(X_i) - f(X_i) \mid \Ical^c\}}{t_0(n)} \lesssim \frac{1}{t_0(n)}.
\end{align*}
This holds because $\Var\{\delta_i(\hat{f}^\m(X_i) - f(X_i) \mid \Ical^c\} = \E[\delta_i^2(\hat{f}^\m(X_i) - f(X_i))^2 \mid \Ical^c]$ is bounded by a constant by Assumption \ref{assump:nuisance convergence} and the subgaussian property of $\delta_i$.
Therefore, on the event $\mathcal{B}_6 \cap \mathcal{B}_7$, we have 
\begin{align}\label{eq:supp cn}
    b(u) \geq \E[\delta_i^2] - C\sqrt{\frac{t_0(n)}{n}} =: C_n
\end{align}
where $C_n \to \E[\delta_i^2]$ as $n\to\infty$.

Given the bounds for gradients, $|a(\cdot)|$ and $|b(\cdot)|$, we can derive the Lipschitz constant for \eqref{eq:supp decomp betam distance}. We get, on the event $\cap_{1\leq j\leq 7}\mathcal{B}_j$,
\begin{align}
    \l|\psi_3(u_1) - \psi_3(u_2)\r|
    & \leq \frac{\|\nabla a(\tilde{u})\|_2\cdot \|u_1 - u_2\|_2}{C_n} + \frac{|a(u_2)| \cdot \|\nabla b(\tilde{u})\|_2\cdot \|u_1 - u_2\|_2}{C_n^2} \nonumber \\
    &\leq \frac{C}{\sqrt{n}}\|u_1 - u_2\|_2. \nonumber
\end{align}

Combining the inequalities in \eqref{eq:supp first step lipschitz} and \eqref{eq:supp third step lipschitz} with Assumption \ref{assump:ml lipschitz}, we get, with probability at least $1-c(1/t_0(n)+\tau_n)$, 
\begin{align*}
    & \quad \l|\psi(z_1) - \psi(z_2)\r| \leq L L_1L_2\|z_1 - z_2\|_2.
\end{align*}
Since $LL_1L_2 = CL\frac{1}{\sqrt{n}}$, we establish Lemma \ref{lem:lipschitz continuity}.

\bibliographystyleapp{plainnat}
\bibliographyapp{refs}

\end{document}

%% file: myoperator.tex
\def\indist{\rightsquigarrow}

\def\T{{ \mathrm{\scriptscriptstyle T} }}

\newcommand{\var}{\text{var}}

\newcommand{\Pb}{\mathbb{P}}

\newcommand{\E}{\mathbb{E}}
\newcommand{\R}{\mathbb{R}}